\documentclass[seceq]{ptptex}

\usepackage{amsmath}
\usepackage{amsfonts}
\usepackage{latexsym}
\usepackage{amssymb} 
\usepackage{verbatim}
\usepackage{amsthm}
\usepackage{graphicx}
\usepackage{wrapft}

\newcommand{\MR}{{{\mathbb R}}}

\title{%        %You can use \\ for explicit line-break
  Second-Order Gauge Invariant Cosmological Perturbation Theory 
}

\subtitle{%      %use this when you want a subtitle 
  Einstein Equations in Terms of Gauge Invariant Variables
}

\author{%       %Use \scshape  for the family name
  Kouji \textsc{Nakamura}%
}

\inst{%         %Affiliation, neglected when [addenda] or [errata]
  Department of Astronomical Science,
  the Graduate University for Advanced Studies,
  Mitaka 181-8588, Japan
}

\recdate{May 22, 2006}%            %Editorial Office will fill in this.

\abst{%         %this abstract is neglected when [addenda] or [errata]
  Following the general framework of the gauge invariant
  perturbation theory developed in the papers [K.~Nakamura,
  Prog.~Theor.~Phys.\ {\bf 110} (2003), 723; {\it ibid}.
  {\bf 113} (2005), 481], we formulate second-order gauge
  invariant cosmological perturbation theory in a
  four-dimensional homogeneous isotropic universe. 
  We consider perturbations both in the universe dominated
  by a single perfect fluid and in that dominated by a single
  scalar field.
  We derive all the components of the Einstein equations in the
  case that the first-order vector and tensor modes are
  negligible.
  All equations are derived in terms of gauge invariant
  variables without any gauge fixing.
  These equations imply that second-order vector and tensor
  modes may be generated due to the mode-mode coupling of the
  linear-order scalar perturbations.
  We also briefly discuss the main progress of this work through
  comparison with previous works.
}

\begin{document}

\maketitle

%%%%%%%%%%%%%%%%%%%%%%%%%%%%%%%%%%%%%%%%%%%%%%%%%%%%%%%%%%%%%%%%%%%%%%
\section{Introduction}
\label{sec:intro}
%%%%%%%%%%%%%%%%%%%%%%%%%%%%%%%%%%%%%%%%%%%%%%%%%%%%%%%%%%%%%%%%%%%%%%

%****************************************************************

The general relativistic cosmological {\it linear} perturbation
theory has been developed to a high degree of sophistication
during the last 25
years\cite{Bardeen-1980,Kodama-Sasaki-1984,Mukhanov-Feldman-Brandenberger-1992}.
One of the motivations of this development is to clarify the
relation between the scenarios of the early universe and
cosmological data, such as the cosmic microwave background (CMB) 
anisotropies.
Recently, the first-order approximation of our universe from a
homogeneous isotropic one was revealed through the observation
of the CMB by the Wilkinson Microwave Anisotropy Probe
(WMAP)\cite{WMAP}.
This observation suggests that the fluctuations are adiabatic
and Gaussian at least to a first order approximation.
One of the next important theoretical studies is to clarify the
accuracy of these results by studying the non-Gaussian behavior,
non-adiabatic behaviors, and so on.  
These will be goals of future satellite missions.
To estimate the accuracy of the first-order approximation,
theoretically, it is necessary to investigate {\it second-order}
cosmological perturbations.
From the observational point of view, also, with the increase of
precision of the CMB data, the study of relativistic
cosmological perturbations beyond linear order is becoming a
topical subject, especially in regard to studying the generation
of the primordial non-Gaussian behavior in inflationary
scenarios\cite{Non-Gaussianity-inflation} and the non-Gaussian
component in the CMB anisotropy\cite{Non-Gaussianity-in-CMB}.

%********************************************************************

In the literature, the second-order general relativistic
perturbation theory has been investigated by many
researchers.
For the pioneering work, Tomita\cite{Tomita-1967} investigated
general relativistic second-order perturbations in the
Einstein-de Sitter model (vanishing $\Lambda$ model), and his
treatment is in the synchronous gauge. 
His second-order perturbation theory was later extended to the
general relativistic Zel'dovich approximation\cite{rela-Zel-ap}.
Recently, nonlinear gauge transformations and the concept of
gauge invariance have been studied by Bruni et
al.\cite{M.Bruni-S.Soonego-CQG1997}.
As the by-product of their research, Sonego and
Bruni\cite{S.Sonego-M.Bruni-CMP1998} obtained a representation
of the higher-order Taylor expansion of tensors on a manifold in
a quite generic form, and the second-order gauge transformation
from the synchronous gauge to the Poisson gauge has been
performed by Matarrese et
al.\cite{Matarrese-Mollerach-Bruni-1998}.
More recently, Noh and Hwang\cite{Noh-Hwang-2004} derived
second-order perturbation equations in the Friedmann universe. 
Further, Tomita\cite{Tomita-2005} also extended his original
works to second-order perturbations of nonzero-$\Lambda$
cosmological models and studied the CMB anisotropy and its
non-Gaussian nature.
However, their treatments of the general relativistic
second-order perturbation is very complicated. 
Hence, to avoid this complicated formulation, there have also
been several attempts to investigate the nonlinear effects of
general relativistic perturbations\cite{other-branch}.

%********************************************************************

In this paper, we present a very clear formulation of the
general relativistic second-order cosmological perturbations in
a homogeneous isotropic universe.
This paper is the complete version of the previous short
paper\cite{kouchan-cosmo-second-let} by the present author.
Our formulation in this paper is one of the applications of the
gauge invariant formulation of the second-order perturbation
theory on a generic background spacetime developed in two
papers by the present
author\cite{kouchan-gauge-inv,kouchan-second}.
These papers are referred to in this paper as
KN2003\cite{kouchan-gauge-inv} and KN2005\cite{kouchan-second}.
This formulation is a by-product of investigations of the
oscillatory behavior of a self-gravitating Nambu-Goto
membrane\cite{kouchan-papers} and was first applied to a
comparison between the oscillatory behavior of a gravitating
Nambu-Goto string and that of a test string\cite{kouchan-flat}.
This was a trivial application of the general formulation
developed in KN2003 and KN2005, while the second-order
cosmological perturbation carried out in this paper is the first
non-trivial application of the formulation presented in KN2003
and KN2005.

%********************************************************************

The formulation developed in KN2003 and KN2005 is an extension
of the works of Bruni et al.\cite{M.Bruni-S.Soonego-CQG1997}.
The gauge transformation rules of the perturbations formulated
by Bruni et al.\cite{M.Bruni-S.Soonego-CQG1997} are extended to
those in multi-parameter perturbation
theory\cite{Bruni-Gualtieri-Sopuerta,kouchan-gauge-inv}.
Based on these gauge transformation rules, in KN2003, we
proposed a procedure to find gauge invariant variables on a
generic background spacetime to third-order perturbations, 
assuming that we already know the procedure to find gauge
invariant variables for the linear-order metric
perturbations.
We also showed in KN2005 that the proposal of the gauge
invariant variables in KN2003 provides a self-consistent second 
order perturbation theory in the generic background spacetime.
It is straightforward to apply this general formulation to
cosmological perturbations.
In the cosmological perturbation case, there are some proposals
of gauge invariant formulations of the second-order
perturbation. 
For example, Mukhanov et
al.\cite{Mukhanov-Abramo-Brandenberger-1997} proposed a gauge
invariant second-order perturbation to evaluate the back
reaction effect of the inhomogeneities in the universe on the
effective expansion law of the universe.
However, we should distinguish our formulation presented in
this paper from the proposal of Mukhanov et
al\cite{Mukhanov-Abramo-Brandenberger-1997}. 
These are quite different approaches.

%********************************************************************

To develop the gauge invariant perturbation theory, we start
by explaining the concept of the ``gauge'' in general
relativistic perturbation theory to avoid any misunderstanding
of our formulation. 
General relativity is based on the concept of general
covariance.
Intuitively, the principle of general covariance states that
there is no preferred coordinate system in nature, though the
notion of general covariance is mathematically included in the
definition of a spacetime manifold in a trivial way.
This is based on the philosophy that coordinate systems are
originally chosen by us, and natural phenomena have nothing to do
with our coordinate systems.
If we apply a peculiar coordinate system to investigate
natural phenomena, we will see peculiar behavior of that
natural phenomena due to this peculiar coordinate system.
This is an intuitive explanation of general covariance.
Due to this general covariance, the 
{\it gauge degree of freedom}, which is the unphysical degree of 
freedom of perturbations, arises in general relativistic
perturbations.
To obtain physically meaningful results, we have to fix this
gauge degree of freedom or to extract the 
{\it gauge invariant part of perturbations}.

%********************************************************************

As reviewed in detail in
\S\ref{sec:General-framework-of-the-gauge-invariant-perturbation-theory}
of this paper, the developments in
KN2003 and KN2005 are based on the understanding of the
``gauge'' in the perturbation theory which was first proposed by
Stewart et al.\cite{J.M.Stewart-M.Walker11974} and developed by
Bruni et
al.\cite{Bruni-Gualtieri-Sopuerta,M.Bruni-S.Soonego-CQG1997}.
Based on this formulation, we define the complete set of gauge
invariant variables of the second-order cosmological 
perturbations in the Friedmann-Robertson-Walker universe.
We consider two cases of the Friedmann-Robertson-Walker
universe, one in which the universe is filled with a single
perfect fluid and one in which the universe is filled with a
single scalar field.
We also derive the second-order Einstein equations of
cosmological perturbations in terms of these gauge invariant
variables without any gauge fixing in these two cases.

%********************************************************************

The organization of this paper is as follows. In
\S\ref{sec:General-framework-of-the-gauge-invariant-perturbation-theory},
we review the general framework of the second-order gauge
invariant perturbation theory developed in KN2003 and
KN2005.
This review also includes additional explanation not given in
those papers. 
In \S\ref{sec:Cosmological-Background-spacetime}, we summarize
the Einstein equations in the case of a background homogeneous
isotropic universe, which are used in the derivation of the
first- and second-order Einstein equations. 
In \S\ref{sec:Gauge-Invariant-Variables-of-Cosmological-Perturbations},
we define the first- and second-order gauge invariant
variables for the cosmological perturbations.
The first-order perturbation of the Einstein equations is
reviewed in \S\ref{sec:First-order-Einstein-equations}.
This perturbation is used in the derivation of the
second-order Einstein equations.
Then, the derivation of the second-order Einstein equation is
given in \S\ref{sec:Secnd-order-Einstein-equations}.
The final section, \S\ref{sec:summary}, is devoted to a
summary and discussions concerning the relation between this and
previous works.

%********************************************************************

We employ the notation of KN2003 and KN2005 and use abstract
index notation\cite{Wald-book}. 
We also employ natural units in which Newton's gravitational 
constant is denoted by $G$ and the velocity of light satisfies
$c=1$.

%****************************************************************

%%%%%%%%%%%%%%%%%%%%%%%%%%%%%%%%%%%%%%%%%%%%%%%%%%%%%%%%%%%%%%%%%%%%%%
\section{General framework of the gauge invariant perturbation theory}
\label{sec:General-framework-of-the-gauge-invariant-perturbation-theory}
%%%%%%%%%%%%%%%%%%%%%%%%%%%%%%%%%%%%%%%%%%%%%%%%%%%%%%%%%%%%%%%%%%%%%%

%****************************************************************

In this section, we briefly review the general framework of the
gauge invariant perturbation theory developed in KN2003 and
KN2005 by the present author.
To explain the {\it gauge degree of freedom} in perturbation
theories, we have to recall what we are doing when we consider
perturbations.
Further, we comment on the Taylor expansion of tensors on a
manifold, at first, in 
\S\ref{sec:Taylor-expansion-of-tensors-on-a-manifold}, because
any perturbation theory is based on the Taylor expansion.
Next, in
\S\ref{sec:Gauge-degree-of-freedom-in-perturbation-theory}, we
review the basic understanding of the gauge degree of freedom in 
perturbation theory based on the work of Stewart et
al.\cite{J.M.Stewart-M.Walker11974} and Bruni et
al.\cite{M.Bruni-S.Soonego-CQG1997}.
When we consider perturbations in the theory with general
covariance, we have to exclude these gauge degrees of freedom in
the perturbations.
To accomplish this, gauge invariant variables of perturbations
are useful, and these are regarded as physically meaningful
quantities.
In \S\ref{sec:gauge-invariant-variables}, we review the
procedure for finding gauge invariant variables of
perturbations, which was developed by the present author in
KN2003. 
Then, in
\S\ref{sec:Perturbation-of-the-Einstein-tensor-and-the-Einstein-equations},
we briefly review the general issue of the gauge invariant
formulation for the second-order perturbation of the Einstein
equation developed in KN2005.
We emphasize that the ingredients of this section do not
depend on the background spacetime, and they are applicable not
only to cosmological perturbations but also to any other general
relativistic perturbations.

%****************************************************************

%%%%%%%%%%%%%%%%%%%%%%%%%%%%%%%%%%%%%%%%%%%%%%%%%%%%%%%%%%%%%%%%%%%%%%
\subsection{Taylor expansion of tensors on a manifold}
\label{sec:Taylor-expansion-of-tensors-on-a-manifold}

%****************************************************************

Here, we comment on the general form of the Taylor expansion of
tensors on a manifold ${\cal M}$.
We first consider the Taylor expansion of a scalar function
$f:{\cal M}\mapsto\MR$, which can be extended to any tensor
field on a manifold.

%****************************************************************

The Taylor expansion of a function $f$ is an approximated form
of $f(q)$ at $q\in{\cal M}$ in terms of the variables at
$p\in{\cal M}$, where $q$ is in the neighborhood of $p$.
To consider the Taylor expansion of a function $f$, we introduce
a one-parameter family of diffeomorphisms 
$\Phi_{\lambda}:{\cal M}\mapsto{\cal M}$, where 
$\Phi_{\lambda}(p)=q$ and $\Phi_{\lambda=0}(p)=p$.
One example of a diffeomorphisms $\Phi_{\lambda}$ is
an exponential map. 
However, we consider a more general class of diffeomorphisms.

%****************************************************************

In terms of the diffeomorphism $\Phi_{\lambda}$, the Taylor
expansion of the function $f(q)$ is given by
\begin{equation}
  \label{eq:symbolic-Taylor-expansion-of-f}
  f(q)
  = f(\Phi_{\lambda}(p))
  = (\Phi^{*}_{\lambda}f)(p)
  =
  f(p)
  +
  \left.\frac{\partial}{\partial\lambda}(\Phi^{*}_{\lambda}f)\right|_{p}
  \lambda 
  +
  \frac{1}{2}
  \left.\frac{\partial^{2}}{\partial\lambda^{2}}(\Phi^{*}_{\lambda}f)\right|_{p}
  \lambda^{2} 
  + O(\lambda^{3}).
\end{equation}
Since this expression hold for an arbitrary smooth function
$f$, we may regard the Taylor expansion to be the expansion of 
the pull-back $\Phi_{\lambda}^{*}$ of the diffeomorphism
$\Phi_{\lambda}$, rather than the expansion of the function $f$.

%****************************************************************

Further, as shown by Sonego and
Bruni\cite{S.Sonego-M.Bruni-CMP1998}, there exist vector fields
$\xi_{1}^{a}$ and $\xi_{2}^{a}$ such that the expansion
(\ref{eq:symbolic-Taylor-expansion-of-f}) is given by
\begin{equation}
  \label{eq:Taylor-expansion-of-f}
  f(q)
  = (\Phi^{*}_{\lambda}f)(p)
  = f(p)
  + \left.\left({\pounds}_{\xi_{1}}f\right)\right|_{p} \lambda 
  + \frac{1}{2}
  \left.\left({\pounds}_{\xi_{2}}+{\pounds}_{\xi_{1}}^{2}\right)f\right|_{p}
  \lambda^{2} 
  + O(\lambda^{3}),
\end{equation}
without loss of generality.
In the representation (\ref{eq:Taylor-expansion-of-f}) of the
Taylor expansion, $\xi_{1}^{a}$ and $\xi_{2}^{a}$ are the
generators of the one-parameter family of diffeomorphisms
$\Phi_{\lambda}$ and these represent the direction along which
the Taylor expansion is carried out.
The generator $\xi_{1}^{a}$ is the first-order approximation of
the flow of the diffeomorphism $\Phi_{\lambda}$, and the
generator $\xi_{2}^{a}$ is the second-order correction to this
flow.

%****************************************************************

When the generator $\xi_{2}^{a}$ is proportional to the
generator $\xi_{1}^{a}$, the representation
(\ref{eq:Taylor-expansion-of-f}) of the Taylor expansion reduces
to that of the pull-back of an exponential map.
Therefore, we may regard that the Taylor expansion
(\ref{eq:Taylor-expansion-of-f}) is the generalization of an
exponential map (one-parameter group of diffeomorphisms).
However, as shown by Sonego and
Bruni\cite{S.Sonego-M.Bruni-CMP1998}, the Taylor expansion of a 
$C^{n}$ one-parameter family of diffeomorphisms can always be 
represented in the form of (\ref{eq:Taylor-expansion-of-f}).
In general, the generator $\xi_{2}^{a}$ may not be proportional
to the generator $\xi_{1}^{a}$.
Hence, we regard the generators $\xi_{1}^{a}$ and $\xi_{2}^{a}$
to be independent.

%****************************************************************

Further, $\Phi_{\lambda}$ can be extended to diffeomorphisms
acting on tensor fields of all types.
Thus, the Taylor expansion of a tensor field $Q$ of any type on a
manifold ${\cal M}$ is given by
\begin{equation}
  \label{eq:Taylor-expansion-of-Q}
  Q(q)
  = (\Phi^{*}_{\lambda}Q)(p)
  = Q(p)
  + \left.\left({\pounds}_{\xi_{1}}Q\right)\right|_{p} \lambda 
  + \frac{1}{2}
  \left.\left({\pounds}_{\xi_{2}}+{\pounds}_{\xi_{1}}^{2}\right)Q\right|_{p}
  \lambda^{2} 
  + O(\lambda^{3}),
\end{equation}
and we conclude that the representation of the Taylor
expansion (\ref{eq:Taylor-expansion-of-Q}) is quite general.

%****************************************************************

%%%%%%%%%%%%%%%%%%%%%%%%%%%%%%%%%%%%%%%%%%%%%%%%%%%%%%%%%%%%%%%%%%%%%%
\subsection{Gauge degree of freedom in perturbation theory}
\label{sec:Gauge-degree-of-freedom-in-perturbation-theory}

%****************************************************************

Now, we explain the concept of gauge in general relativistic
perturbation theory.
To explain this, we first point out that, in any perturbation
theory, we always treat two spacetime manifolds. 
One is the physical spacetime ${\cal M}$, which we attempt to
describe in forms of perturbations, and the other is the
background spacetime ${\cal M}_{0}$, which is a fictitious
manifold which we prepare for perturbative analyses.
We emphasize that these two spacetime manifolds
${\cal M}$ and ${\cal M}_{0}$ are distinct.
Let us denote the physical spacetime by $({\cal M},\bar{g}_{ab})$
and the background spacetime by $({\cal M}_{0},g_{ab})$, where 
$\bar{g}_{ab}$ is the metric on the physical spacetime manifold,
${\cal M}$, and $g_{ab}$ is the metric on the background
spacetime manifold, ${\cal M}_{0}$.
Further, we formally denote the spacetime metric and the other
physical tensor fields on the physical spacetime by $Q$ and its
background value on the background spacetime by $Q_{0}$.

%****************************************************************

Second, in any perturbation theories, we always write equations
for the perturbation of the physical variable $Q$ in the form
\begin{equation}
  \label{eq:variable-symbolic-perturbation}
  Q(``p\mbox{''}) = Q_{0}(p) + \delta Q(p).
\end{equation}
Usually, this equation is simply regarded as a relation
between the physical variable $Q$ and its background value
$Q_{0}$, or as the definition of the deviation $\delta Q$ of
the physical variable $Q$ from its background value $Q_{0}$.
However, Eq.~(\ref{eq:variable-symbolic-perturbation}) has
deeper implications.
Keeping in our mind that we always treat two different
spacetimes, $({\cal M},\bar{g}_{ab})$ and $({\cal M}_{0},g_{ab})$,
in perturbation theory,
Eq.~(\ref{eq:variable-symbolic-perturbation}) is a rather 
strange equation in the following sense: 
The variable on the left-hand side of
Eq.~(\ref{eq:variable-symbolic-perturbation}) is a variable on
the physical spacetime $({\cal M},\bar{g}_{ab})$, while the
variables on the right-hand 
side of Eq.~(\ref{eq:variable-symbolic-perturbation}) are
varibles on the background spacetime, $({\cal M}_{0},g_{ab})$.
Hence, Eq.~(\ref{eq:variable-symbolic-perturbation}) gives a
relation between variables on two different manifolds.

%****************************************************************

Further, through Eq.~(\ref{eq:variable-symbolic-perturbation}),
we have implicitly identified points in these two different
manifolds.
More specifically, $Q(``p\mbox{''})$ on the left-hand side of
Eq.~(\ref{eq:variable-symbolic-perturbation}) is a field on the
physical spacetime, ${\cal M}$, and $``p\mbox{''}\in{\cal M}$.
Similarly, we should regard the background value
$Q_{0}(p)$ of $Q(``p\mbox{''})$ and its deviation $\delta Q(p)$
of $Q(``p\mbox{''})$ from $Q_{0}(p)$, which are on the
right-hand side of
Eq.~(\ref{eq:variable-symbolic-perturbation}), as fields on the 
background spacetime, ${\cal M}_{0}$, and $p\in{\cal M}_{0}$.
Because Eq.~(\ref{eq:variable-symbolic-perturbation}) is
regarded as an equation for field variables, it implicitly
states that the points $``p\mbox{''}\in{\cal M}$ and 
$p\in{\cal M}_{0}$ are same.
This represents the implicit assumption of the existence of a map 
${\cal M}_{0}\rightarrow{\cal M}$ $:$ $p\in{\cal M}_{0}\mapsto
``p\mbox{''}\in{\cal M}$, which is usually called a 
{\it gauge choice} in perturbation
theory\cite{J.M.Stewart-M.Walker11974}.

%****************************************************************

It is important to note that the correspondence between points
on ${\cal M}_{0}$ and ${\cal M}$, which is established by such a
relation as Eq.~(\ref{eq:variable-symbolic-perturbation}), is
not unique to the theory in which general covariance is
imposed.
Rather, Eq.~(\ref{eq:variable-symbolic-perturbation}) involves
the degree of freedom corresponding to the choice of the map
${\cal X}$ $:$ ${\cal M}_{0}\mapsto{\cal M}$.
This is called the {\it gauge degree of freedom}.
Such a degree of freedom always exists in perturbations of a
theory in which we impose general covariance.
``General covariance'' intuitively means that there is no
preferred coordinate system in the theory.
If general covariance is not imposed on the theory, there is a
preferred coordinate system, and we naturally introduce this
coordinate system onto both ${\cal M}_{0}$ and 
${\cal M}$.
Then, we can choose the identification map ${\cal X}$ using this
coordinate system.
However, there is no such coordinate system in general
relativity, due to its general covariance, and we have no
guiding principle to choose the identification map ${\cal X}$.
Indeed, we could identify $``p\mbox{''}\in{\cal M}$ with 
$q\in{\cal M}_{0}$ ($q\neq p$) instead of $p\in{\cal M}_{0}$.
In the above understanding of the concept of ``gauge'' in
general relativistic perturbation theory, a gauge transformation is
simply a change of the map ${\cal X}$.

%****************************************************************

These are the basic ideas necessary to understand 
{\it gauge degree of freedom} in the general relativistic
perturbation theory proposed by Stewart and
Walker\cite{J.M.Stewart-M.Walker11974}.
This understanding has been developed by Bruni et
al.\cite{M.Bruni-S.Soonego-CQG1997}, and by the present
author\cite{kouchan-gauge-inv,kouchan-second}.
We briefly review this development.

%****************************************************************

To formulate the above understanding in more detail, we
introduce an infinitesimal parameter $\lambda$ for the
perturbation.
Further, we consider the $4+1$-dimensional manifold 
${\cal N}={\cal M}\times\MR$, where $4=\dim{\cal M}$ and
$\lambda\in\MR$.
The background spacetime 
${\cal M}_{0}=\left.{\cal N}\right|_{\lambda=0}$ and the
physical spacetime 
${\cal M}={\cal M}_{\lambda}=\left.{\cal N}\right|_{\MR=\lambda}$ are
also submanifolds embedded in the extended manifold ${\cal N}$.
Each point on ${\cal N}$ is identified by a pair, $(p,\lambda)$,
where $p\in{\cal M}_{\lambda}$, and each point in the background
spacetime ${\cal M}_{0}$ in ${\cal N}$ is identified by
$\lambda=0$.

%****************************************************************

Through this construction, the manifold ${\cal N}$ is foliated by
four-dimensional submanifolds ${\cal M}_{\lambda}$ of each
$\lambda$, and these are diffeomorphic to the physical
spacetime ${\cal M}$ and the background spacetime 
${\cal M}_{0}$. 
The manifold ${\cal N}$ has a natural differentiable structure
consisting of the direct product of ${\cal M}$ and $\MR$.
Further, the perturbed spacetimes ${\cal M}_{\lambda}$ for each
$\lambda$ must have the same differential structure with this
construction.
In other words, we require that perturbations be continuous in
the sense that $({\cal M},\bar{g}_{ab})$ and 
$({\cal M}_{0},g_{ab})$ are connected by a continuous curve
within the extended manifold ${\cal N}$.
Hence, the changes of the differential structure resulting from
the perturbation, for example the formation of singularities and
singular perturbations in the sense of fluid mechanics, are
excluded from consideration in this paper.

%****************************************************************

Let us consider the set of field equations 
\begin{equation}
  \label{eq:field-eq-for-Q}
  {\cal E}[Q_{\lambda}] = 0
\end{equation}
on the physical spacetime ${\cal M}_{\lambda}$ for the physical
variables $Q_{\lambda}$ on ${\cal M}_{\lambda}$.
The field equation (\ref{eq:field-eq-for-Q}) formally represents
the Einstein equation for the metric on ${\cal M}_{\lambda}$ and
the equations for matter fields on ${\cal M}_{\lambda}$.
If a tensor field $Q_{\lambda}$ is given on each 
${\cal M}_{\lambda}$, $Q_{\lambda}$ is automatically extended to a
tensor field on ${\cal N}$ by $Q(p,\lambda):=Q_{\lambda}(p)$,
where $p\in{\cal M}_{\lambda}$.
In this extension, the field equation (\ref{eq:field-eq-for-Q})
is regarded as an equation on the extended manifold ${\cal N}$.
Thus, we have extended an arbitrary tensor field and the field
equations (\ref{eq:field-eq-for-Q}) on each ${\cal M}_{\lambda}$
to those on the extended manifold ${\cal N}$.

%****************************************************************

Tensor fields on ${\cal N}$ obtained through the above construction
are necessarily ``tangent'' to each ${\cal M}_{\lambda}$, i.e.,
their normal component to each ${\cal M}_{\lambda}$ identically
vanishes.
To consider the basis of the tangent space of ${\cal N}$, we
introduce the normal form and its dual, which are normal to each
${\cal M}_{\lambda}$ in ${\cal N}$.
These are denoted by $(d\lambda)_{a}$ and
$(\partial/\partial\lambda)^{a}$, respectively, and they satisfy
\begin{equation}
  (d\lambda)_{a} \left(\frac{\partial}{\partial\lambda}\right)^{a} = 1.
  \label{eq:depsilon-normalization}
\end{equation}
The form $(d\lambda)_{a}$ and its dual,
$(\partial/\partial\lambda)^{a}$, are normal to any tensor field
extended from the tangent space on each ${\cal M}_{\lambda}$
through the above construction. 
The set consisting of $(d\lambda)_{a}$,
$(\partial/\partial\lambda)^{a}$ and the basis of the tangent
space on each ${\cal M}_{\lambda}$ is regarded as the basis of
the tangent space of ${\cal N}$.

%****************************************************************

To define the perturbation of an arbitrary tensor field $Q$, we
compare $Q$ on the physical spacetime ${\cal M}_{\lambda}$ with
$Q_{0}$ on the background spacetime, and it is necessary to
identify the points of ${\cal M}_{\lambda}$ with those of 
${\cal M}_{0}$.
This point identification map is the so-called 
{\it gauge choice} in the context of perturbation theories,
as mentioned above.
The gauge choice is made by assigning a diffeomorphism
${\cal X}_{\lambda}$ $:$ ${\cal N}$ $\rightarrow$ ${\cal N}$
such that ${\cal X}_{\lambda}$ $:$ ${\cal M}_{0}$ $\rightarrow$
${\cal M}_{\lambda}$.
Following the paper of Bruni et
al.\cite{M.Bruni-S.Soonego-CQG1997}, we introduce a gauge choice
${\cal X}_{\lambda}$ as one of the one-parameter groups of
diffeomorphisms, i.e., an exponential map, for simplicity.
We denote the generator of this exponential map by 
${}^{{\cal X}}\!\eta^{a}$.
This generator ${}^{{\cal X}}\!\eta^{a}$ is decomposed by the 
basis on the tangent space of ${\cal N}$ which are constructed
above.
Though the generator ${}^{{\cal X}}\!\eta^{a}$ should satisfy
some appropriate conditions, the arbitrariness of the gauge
choice ${\cal X}_{\lambda}$ is represented by the tangential
component of the generator ${}^{{\cal X}}\!\eta^{a}$ to the
tangent space of ${\cal M}_{\lambda}$.

%****************************************************************

The pull-back ${\cal X}_{\lambda}^{*}Q$, which is induced by the
exponential map ${\cal X}_{\lambda}$, maps a tensor field $Q$
on the physical manifold ${\cal M}_{\lambda}$ to a tensor field
${\cal X}_{\lambda}^{*}Q$ on the background spacetime.
In terms of this generator ${}^{{\cal X}}\!\eta^{a}$, the
pull-back ${\cal X}_{\lambda}^{*}Q$ is represented by the Taylor
expansion
\begin{eqnarray}
  Q(r)
  =
  Q({\cal X}_{\lambda}(p))
  =
  {\cal X}_{\lambda}^{*}Q(p)
  =
  Q(p)
  + \lambda \left.{\pounds}_{{}^{{\cal X}}\!\eta}Q \right|_{p}
  + \frac{1}{2} \lambda^{2} 
  \left.{\pounds}_{{}^{{\cal X}}\!\eta}^{2}Q\right|_{p}
  + O(\lambda^{3}),
  \label{eq:Taylor-expansion-of-calX-org}
\end{eqnarray}
where $r={\cal X}_{\lambda}(p)\in{\cal M}_{\lambda}$.
Because $p\in{\cal M}_{0}$, we may regard the equation
\begin{eqnarray}
  {\cal X}_{\lambda}^{*}Q(p)
  =
  Q_{0}(p)
  + \lambda \left.{\pounds}_{{}^{{\cal X}}\!\eta}Q\right|_{{\cal M}_{0}}(p)
  + \frac{1}{2} \lambda^{2} 
  \left.{\pounds}_{{}^{{\cal X}}\!\eta}^{2}Q\right|_{{\cal M}_{0}}(p)
  + O(\lambda^{3})
  \label{eq:Taylor-expansion-of-calX}
\end{eqnarray}
as an equation on the background spacetime ${\cal M}_{0}$,
where $Q_{0}=\left.Q\right|_{{\cal M}_{0}}$ is the background
value of the physical variable of $Q$.
Once the definition of the pull-back of the gauge choice 
${\cal X}_{\lambda}$ is given, the perturbation 
$\Delta^{\cal X}Q_{\lambda}$ of a tensor field $Q$ under the
gauge choice ${\cal X}_{\lambda}$ is simply defined as 
\begin{equation}
  \label{eq:Bruni-34}
  \Delta^{\cal X}Q_{\lambda} :=
  \left.{\cal X}^{*}_{\lambda}Q\right|_{{\cal M}_{0}} - Q_{0}.
\end{equation}
We note that all variables in this definition are defined on
${\cal M}_{0}$.
Expanding the first term on the right-hand side of
(\ref{eq:Bruni-34}) as
\begin{equation}
  \label{eq:Bruni-35}
  \left.{\cal X}^{*}_{\lambda}Q_{\lambda}\right|_{{\cal M}_{0}}
  =
  Q_{0}
  + \lambda {}^{(1)}_{\;\cal X}\!Q
  + \frac{1}{2} \lambda^{2} {}^{(2)}_{\;\cal X}\!Q
  + O(\lambda^{3}),
\end{equation}
we define the first- and the second-order perturbations of a
physical variable $Q_{\lambda}$ under the gauge choice 
${\cal X}_{\lambda}$ by 
\begin{eqnarray}
  {}^{(1)}_{\;\cal X}\!Q = 
  \left.{\pounds}_{{}^{\cal X}\!\eta} Q\right|_{{\cal M}_{0}},
  \quad
  {}^{(2)}_{\;\cal X}\!Q = 
  \left.{\pounds}_{{}^{\cal X}\!\eta}^{2} Q\right|_{{\cal M}_{0}}.
  \label{eq:representation-of-each-order-perturbation}
\end{eqnarray}

%****************************************************************

Now, we consider two {\it different gauge choices} based on
the above understanding of the gauge choice in the perturbation
theory.
Suppose that ${\cal X}_{\lambda}$ and ${\cal Y}_{\lambda}$ are
two exponential maps with the generators ${}^{\cal X}\eta^{a}$
and ${}^{\cal Y}\eta^{a}$ on ${\cal N}$, respectively.
In other words, ${\cal X}_{\lambda}$ and ${\cal Y}_{\lambda}$
are two gauge choices.
Then, the integral curves of each ${}^{\cal X}\!\eta^{a}$ and
${}^{\cal Y}\!\eta^{a}$ in ${\cal N}$ are the orbits of the
actions of the gauge choices ${\cal X}_{\lambda}$ and 
${\cal Y}_{\lambda}$, respectively.
Since we choose the generators ${}^{\cal X}\!\eta^{a}$ and
${}^{\cal Y}\!\eta^{a}$ so that these are transverse to each
${\cal M}_{\lambda}$ everywhere on ${\cal N}$, the integral
curves of these vector fields intersect with each 
${\cal M}_{\lambda}$.
Therefore, points lying on the same integral curve of either of
the two are to be regarded as {\it the same point} within the
respective gauges.
When these curves are not identical, i.e., the tangential
components to each ${\cal M}_{\lambda}$ of 
${}^{\cal X}\!\eta^{a}$ and ${}^{\cal Y}\!\eta^{a}$ are
different, these point identification maps ${\cal X}_{\lambda}$
and ${\cal Y}_{\lambda}$ are regarded as
{\it two different gauge choices}.

%****************************************************************

We next introduce the concept of {\it gauge invariance}.
Following the paper by Bruni et
al.\cite{Bruni-Gualtieri-Sopuerta}, we consider the concept of
{\it gauge invariance up to order $n$}.
Suppose that ${\cal X}_{\lambda}$ and ${\cal Y}_{\lambda}$ are
two different gauge choices which are generated by the vector
fields ${}^{\cal X}\!\eta^{a}$ and ${}^{\cal Y}\!\eta^{a}$,
respectively.
These gauge choices also pull back a generic tensor field $Q$ on
${\cal N}$ to two other tensor fields, ${\cal X}_{\lambda}^{*}Q$
and ${\cal Y}_{\lambda}^{*}Q$, for any given value of $\lambda$.
In particular, on ${\cal M}_{0}$, we now have three tensor
fields associated with a tensor field $Q$; one is the background
value $Q_{0}$ of $Q$, and the other two are the pulled-back
variables of $Q$ from ${\cal M}_{\lambda}$ to ${\cal M}_{0}$ by
the two different gauge choices,
\begin{eqnarray}
  {}^{\cal X}Q_{\lambda} &:=&
  \left.{\cal X}^{*}_{\lambda}Q\right|_{{\cal M}_{0}}
  = 
  Q_{0}
  + \lambda {}^{(1)}_{\;{\cal X}}\!Q
  + \frac{1}{2} \lambda^{2} {}^{(2)}_{\;{\cal X}}\!Q
  + O(\lambda^{3})
  = Q_{0} + \Delta^{\cal X}Q_{\lambda},
  \label{eq:Bruni-39-one}
  \\
  {}^{\cal Y}Q_{\lambda} &:=&
  \left.{\cal Y}^{*}_{\lambda}Q\right|_{{\cal M}_{0}}
  = 
  Q_{0}
  + \lambda {}^{(1)}_{\;{\cal Y}}\!Q
  + \frac{1}{2} \lambda^{2} {}^{(2)}_{\;{\cal Y}}\!Q
  + O(\lambda^{3})
  = Q_{0} + \Delta^{\cal Y}Q_{\lambda}.
  \label{eq:Bruni-40-one}
\end{eqnarray}
Here, we have used Eqs.~(\ref{eq:Bruni-34}) and
(\ref{eq:Bruni-35}).
Because ${\cal X}_{\lambda}$ and ${\cal Y}_{\lambda}$ are gauge
choices that map the background spacetime ${\cal M}_{0}$ into
the physical spacetime ${\cal M}_{\lambda}$, 
${}^{\cal X}Q_{\lambda}$ and ${}^{\cal Y}Q_{\lambda}$ are the
representations on ${\cal M}_{0}$ of the perturbed tensor field
$Q$ in the two different gauges.
The quantities ${}^{(k)}_{\;\cal X}\!Q$ and 
${}^{(k)}_{\;\cal Y}\!Q$ in Eqs.~(\ref{eq:Bruni-39-one}) and
(\ref{eq:Bruni-40-one}) are the perturbations of $O(k)$ in the
gauges ${\cal X}_{\lambda}$ and ${\cal Y}_{\lambda}$, respectively.
We say that $Q$ is {\it gauge invariant up to order $n$} iff
for any two gauges ${\cal X}_{\lambda}$ and ${\cal Y}_{\lambda}$
the following holds:
\begin{equation}
  {}^{(k)}_{\;\cal X}\!Q = {}^{(k)}_{\;\cal Y}\!Q \quad 
  \forall k, \quad \mbox{with} \quad k<n.
\end{equation}
From this definition, we can prove that the $n$th-order
perturbation of a tensor field $Q$ is gauge invariant up to
order $n$ iff in a given gauge ${\cal X}_{\lambda}$ we have
${\pounds}_{\xi} {}^{(k)}_{\;\cal X}\!Q = 0$ for any vector
field $\xi^{a}$ defined on ${\cal M}_{0}$ and for any $k<n$.
As a consequence, the $n$th-order perturbation of a tensor field
$Q$ is gauge invariant up to order $n$ iff $Q_{0}$ and all its
perturbations of lower than $n$th order are, in any gauge,
either vanishing or constant scalars, or a combination of
Kronecker deltas with constant
coefficients\cite{Bruni-Gualtieri-Sopuerta,J.M.Stewart-M.Walker11974,M.Bruni-S.Soonego-CQG1997}.  
Further, even if its lower order perturbations are not
trivial, we can decompose any perturbation of $Q$ into the gauge
invariant and gauge variant parts, as shown in KN2003. 
This will be also explained in the next subsection.

%****************************************************************

Now, we consider the {\it gauge transformation rules}
between different gauge choices.
In general, the representation ${}^{\cal X}Q_{\lambda}$ on
${\cal M}_{0}$ of the perturbed variable $Q$ on 
${\cal M}_{\lambda}$ depends on the gauge choice 
${\cal X}_{\lambda}$.
If we employ a different gauge choice, the representation of
$Q_{\lambda}$ on ${\cal M}_{0}$ may change.
Suppose that ${\cal X}_{\lambda}$ and ${\cal Y}_{\lambda}$ are
different gauge choices, which are the point identification maps
from ${\cal M}_{0}$ to ${\cal M}_{\lambda}$, and the generators
of these gauge choices are given by ${}^{\cal X}\!\eta^{a}$ and
${}^{\cal Y}\!\eta^{a}$, respectively.
Then, the change of the gauge choice from ${\cal X}_{\lambda}$
to ${\cal Y}_{\lambda}$ is represented by the diffeomorphism
\begin{equation}
  \label{eq:diffeo-def-from-Xinv-Y}
  \Phi_{\lambda} :=
  ({\cal X}_{\lambda})^{-1}\circ{\cal Y}_{\lambda}.
\end{equation}
This diffeomorphism $\Phi_{\lambda}$ is the map $\Phi_{\lambda}$
$:$ ${\cal M}_{0}$ $\rightarrow$ ${\cal M}_{0}$ for each value
of $\lambda\in\MR$.
The diffeomorphism $\Phi_{\lambda}$ does change the point
identification, as expected from the understanding of the gauge
choice discussed above.
Therefore, the diffeomorphism $\Phi_{\lambda}$ is regarded as
the gauge transformation $\Phi_{\lambda}$ $:$
${\cal X}_{\lambda}$ $\rightarrow$ ${\cal Y}_{\lambda}$.

%****************************************************************

The gauge transformation $\Phi_{\lambda}$ induces a pull-back
from the representation ${}^{\cal X}Q_{\lambda}$ of the
perturbed tensor field $Q$ in the gauge choice 
${\cal X}_{\lambda}$ to the representation 
${}^{\cal Y}Q_{\lambda}$ in the gauge choice 
${\cal Y}_{\lambda}$.
Actually, the tensor fields ${}^{\cal X}Q_{\lambda}$ and
${}^{\cal Y}Q_{\lambda}$, which are defined on ${\cal M}_{0}$,
are connected by the linear map $\Phi^{*}_{\lambda}$ as
\begin{eqnarray}
  {}^{\cal Y}Q_{\lambda}
  &=&
  \left.{\cal Y}^{*}_{\lambda}Q\right|_{{\cal M}_{0}}
  =
  \left.\left(
      {\cal Y}^{*}_{\lambda}
      \left({\cal X}_{\lambda}
      {\cal X}_{\lambda}^{-1}\right)^{*}Q\right)
  \right|_{{\cal M}_{0}}
  \nonumber\\
  &=&
  \left.
    \left(
      {\cal X}^{-1}_{\lambda}
      {\cal Y}_{\lambda}
    \right)^{*}
    \left(
      {\cal X}^{*}_{\lambda}Q
    \right)
  \right|_{{\cal M}_{0}}
  =  \Phi^{*}_{\lambda} {}^{\cal X}Q_{\lambda}.
  \label{eq:Bruni-45-one}
\end{eqnarray}
According to generic arguments concerning the Taylor expansion
of the pull-back of a tensor field on the same manifold, given
in \S\ref{sec:Taylor-expansion-of-tensors-on-a-manifold}, it
should be possible to express the gauge transformation
$\Phi^{*}_{\lambda} {}^{\cal X}Q_{\lambda}$ in the form
\begin{eqnarray}
  \Phi^{*}_{\lambda}{}^{\cal X}\!Q = {}^{\cal X}\!Q
  + \lambda {\pounds}_{\xi_{1}} {}^{\cal X}\!Q
  + \frac{\lambda^{2}}{2} \left\{
    {\pounds}_{\xi_{2}} + {\pounds}_{\xi_{1}}^{2}
  \right\} {}^{\cal X}\!Q
  + O(\lambda^{3}),
  \label{eq:Bruni-46-one} 
\end{eqnarray}
where the vector fields $\xi_{1}^{a}$ and $\xi_{2}^{a}$ are the
generators of the gauge transformation $\Phi_{\lambda}$.

%****************************************************************

Comparing the representation (\ref{eq:Bruni-46-one}) of the
Taylor expansion in terms of the generators $\xi_{1}^{a}$ and
$\xi_{2}^{a}$ of the pull-back 
$\Phi_{\lambda}^{*}{}^{\cal X}\!Q$ and that in terms of the
generators ${}^{\cal X}\eta^{a}$ and ${}^{\cal Y}\eta^{a}$ of
the pull-back 
${\cal Y}^{*}_{\lambda}\circ\left({\cal X}_{\lambda}^{-1}\right)^{*}\;{}^{{\cal X}}\!Q$ 
($=\Phi_{\lambda}^{*}{}^{\cal X}\!Q$), we readily obtain explicit
expressions for the generators $\xi_{1}^{a}$ and $\xi_{2}^{a}$
of the gauge transformation 
$\Phi={\cal X}^{-1}_{\lambda}\circ{\cal Y}_{\lambda}$ in terms
of the generators ${}^{\cal X}\eta^{a}$ and 
${}^{\cal Y}\eta^{a}$ of the gauge choices as follows: 
\begin{eqnarray}
  \xi_{1}^{a}
  =
  {}^{\cal Y}\eta^{a}
  -
  {}^{\cal X}\eta^{a},
  \quad
  \xi_{2}^{a}
  = 
  \left[
    {}^{\cal Y}\eta
    ,
    {}^{\cal X}\eta
  \right]^{a}.
  \label{eq:relation-between-xi-eta}
\end{eqnarray}
Further, because the gauge transformation $\Phi_{\lambda}$ is a
map within the background spacetime ${\cal M}_{0}$, the
generator should consist of vector fields on ${\cal M}_{0}$.
This can be satisfied by imposing some appropriate conditions on
the generators ${}^{\cal Y}\eta^{a}$ and ${}^{\cal X}\eta^{a}$.

%****************************************************************

We can now derive the relation between the perturbations in the
two different gauges.
Up to second order, these relations are derived by substituting
(\ref{eq:Bruni-39-one}) and (\ref{eq:Bruni-40-one}) into
(\ref{eq:Bruni-46-one}):
\begin{eqnarray}
  \label{eq:Bruni-47-one}
  {}^{(1)}_{\;{\cal Y}}\!Q - {}^{(1)}_{\;{\cal X}}\!Q &=& 
  {\pounds}_{\xi_{1}}Q_{0}, \\
  \label{eq:Bruni-49-one}
  {}^{(2)}_{\;\cal Y}\!Q - {}^{(2)}_{\;\cal X}\!Q &=& 
  2 {\pounds}_{\xi_{(1)}} {}^{(1)}_{\;\cal X}\!Q 
  +\left\{{\pounds}_{\xi_{(2)}}+{\pounds}_{\xi_{(1)}}^{2}\right\} Q_{0}.
\end{eqnarray}
These results are, of course, consistent with the concept of
gauge invariance up to order $n$, as introduced above.
Further, inspecting these gauge transformation rules, we can
define the gauge invariant variables at each order, as shown
below.

%****************************************************************

Here, we should comment on the gauge choice in the above
explanation.
We have introduced an exponential map ${\cal X}_{\lambda}$ (or
${\cal Y}_{\lambda}$) as the gauge choice, for simplicity.
However, this simplified introduction of the gauge choice 
${\cal X}_{\lambda}$ as an exponential map is not essential to
the gauge transformation rules (\ref{eq:Bruni-47-one}) and
(\ref{eq:Bruni-49-one}).
Indeed, we can generalize the diffeomorphism 
${\cal X}_{\lambda}$ from an exponential map.
If we generalize the diffeomorphism ${\cal X}_{\lambda}$, the
representation (\ref{eq:Taylor-expansion-of-calX}) of the
pulled-back variable ${\cal X}_{\lambda}^{*}Q(p)$, the
representations of the perturbations
(\ref{eq:representation-of-each-order-perturbation}), and the
relations (\ref{eq:relation-between-xi-eta}) between generators
of $\Phi_{\lambda}$, ${\cal X}_{\lambda}$, and 
${\cal Y}_{\lambda}$ will be changed.
However, the gauge transformation rules (\ref{eq:Bruni-47-one})
and (\ref{eq:Bruni-49-one}) are direct consequences of the
Taylor expansion (\ref{eq:Bruni-46-one}) of $\Phi_{\lambda}$.
As commented in
\S\ref{sec:Taylor-expansion-of-tensors-on-a-manifold}, the
representation of the Taylor expansion (\ref{eq:Bruni-46-one})
of $\Phi_{\lambda}$ is quite
general\cite{S.Sonego-M.Bruni-CMP1998}.
Therefore, the gauge transformation rules
(\ref{eq:Bruni-47-one}) and (\ref{eq:Bruni-49-one}) do not
change, even if we generalize the choice of ${\cal X}_{\lambda}$.
Further, the relations (\ref{eq:relation-between-xi-eta})
between generators also imply that, even if we consider a simple
exponential map as the gauge choice, both of the generators
$\xi_{1}^{a}$ and $\xi_{2}^{a}$ are naturally induced by the
generators of the original gauge choices.
Hence, we conclude that the gauge transformation rules
(\ref{eq:Bruni-47-one}) and (\ref{eq:Bruni-49-one}) are quite
general and irreducible.
In this paper, we develop a second-order gauge invariant
cosmological perturbation theory based on the above
understanding of the gauge degree of freedom only through the
gauge transformation rules (\ref{eq:Bruni-47-one}) and
(\ref{eq:Bruni-49-one}).
Hence, the development of the cosmological perturbation theory
presented below is not changed if we generalize the gauge choice
${\cal X}_{\lambda}$ from a simple exponential map.

%****************************************************************

%%%%%%%%%%%%%%%%%%%%%%%%%%%%%%%%%%%%%%%%%%%%%%%%%%%%%%%%%%%%%%%%%%%%%%
\subsection{Gauge invariant variables}
\label{sec:gauge-invariant-variables}

%****************************************************************

Inspecting the gauge transformation rules
(\ref{eq:Bruni-47-one}) and (\ref{eq:Bruni-49-one}), we can define
the gauge invariant variables for a metric perturbation and for
arbitrary matter fields.
Employing the idea of gauge invariance up to order $n$ for
$n$th-order perturbations\cite{M.Bruni-S.Soonego-CQG1997}, we
proposed a procedure to construct gauge invariant variables of
higher-order perturbations\cite{kouchan-gauge-inv}.
This proposal is as follows.
First, we decompose a linear-order metric perturbation into its
gauge invariant and variant parts. 
The procedure for decomposing linear-order metric perturbations
is extended to second-order metric perturbations, and we can
decompose the second-order metric perturbations through a
procedure similar to that for the linear-order metric
perturbation. 
Then, we define the gauge invariant variables for the first-
and second-order perturbations of an arbitrary field other
than the metric by using the gauge variant parts of the first-
and second-order metric perturbations.
Though the procedure for finding gauge invariant variables for
linear-order metric perturbations is highly non-trivial, once we
know this procedure, we can easily find the gauge invariant part
of a higher-order perturbation through a simple extension of the
procedure for the linear-order perturbations.

%****************************************************************

To consider a metric perturbation, we expand the metric on the
physical spacetime ${\cal M}$, which is pulled back to the
background spacetime ${\cal M}_{0}$ using a gauge choice in the
form given in (\ref{eq:Bruni-35}),
\begin{eqnarray}
  {\cal X}^{*}_{\lambda}\bar{g}_{ab}
  &=&
  g_{ab} + \lambda {}_{{\cal X}}\!h_{ab} 
  + \frac{\lambda^{2}}{2} {}_{{\cal X}}\!l_{ab}
  + O^{3}(\lambda),
  \label{eq:metric-expansion}
\end{eqnarray}
where $g_{ab}$ is the metric on the background spacetime 
${\cal M}_{0}$.
Of course, the expansion (\ref{eq:metric-expansion}) of the
metric depends entirely on the gauge choice 
${\cal X}_{\lambda}$.
Nevertheless, henceforth, we do not explicitly express the index
of the gauge choice ${\cal X}_{\lambda}$ in an expression if
there is no possibility of confusion.

%****************************************************************

Our starting point to construct gauge invariant variables is the
assumption that 
{\it we already know the procedure for finding gauge invariant
  variables for the linear metric perturbations.}
Then, a linear metric perturbation $h_{ab}$ is decomposed as
\begin{eqnarray}
  h_{ab} =: {\cal H}_{ab} + {\pounds}_{X}g_{ab},
  \label{eq:linear-metric-decomp}
\end{eqnarray}
where ${\cal H}_{ab}$ and $X^{a}$ are the gauge invariant and
variant parts of the linear-order metric
perturbations\cite{kouchan-gauge-inv}, i.e., under the gauge
transformation (\ref{eq:Bruni-47-one}), these are transformed as
\begin{equation}
  {}_{{\cal Y}}\!{\cal H}_{ab} - {}_{{\cal X}}\!{\cal H}_{ab} =  0, 
  \quad
  {}_{\quad{\cal Y}}\!X^{a} - {}_{{\cal X}}\!X^{a} = \xi^{a}_{(1)}. 
  \label{eq:linear-metric-decomp-gauge-trans}
\end{equation}

%****************************************************************

As emphasized in KN2003 and KN2005, the above assumption is
quite strong and it is not simple to carry out the systematic
decomposition (\ref{eq:linear-metric-decomp}) on an arbitrary
background spacetime, since this procedure depends completely on
the background spacetime $({\cal M}_{0},g_{ab})$.
However, we show that this procedure exists in the case of
cosmological perturbations of a homogeneous and isotropic
universe in \S\ref{sec:First-order-metric-perturbations}.

%****************************************************************

Once we accept this assumption for linear-order metric
perturbations, we can always find gauge invariant variables for 
higher-order perturbations\cite{kouchan-gauge-inv}.
As shown in KN2003, at second order, the metric perturbations
are decomposed as 
\begin{eqnarray}
  \label{eq:H-ab-in-gauge-X-def-second-1}
  l_{ab}
  =:
  {\cal L}_{ab} + 2 {\pounds}_{X} h_{ab}
  + \left(
      {\pounds}_{Y}
    - {\pounds}_{X}^{2} 
  \right)
  g_{ab},
\end{eqnarray}
where ${\cal L}_{ab}$ and $Y^{a}$ are the gauge invariant and
variant parts of the second order metric perturbations, i.e.,
\begin{eqnarray}
  {}_{{\cal Y}}\!{\cal L}_{ab} - {}_{{\cal X}}\!{\cal L}_{ab} = 0,
  \quad
  {}_{{\cal Y}}\!Y^{a} - {}_{{\cal X}}\!Y^{a}
  = \xi_{(2)}^{a} + [\xi_{(1)},X]^{a}.
\end{eqnarray}
The details of the derivation of this gauge invariant part of
the second-order metric perturbation are explained in the
context of cosmological perturbations in
\S\ref{sec:Second-order-metric-perturbations}.

%****************************************************************

Furthermore, as shown in KN2003, using the first- and
second-order gauge variant parts, $X^{a}$ and $Y^{a}$, of the
metric perturbations, the gauge invariant variables for an arbitrary
field $Q$ other than the metric are given by
\begin{eqnarray}
  \label{eq:matter-gauge-inv-def-1.0}
  {}^{(1)}\!{\cal Q} &:=& {}^{(1)}\!Q - {\pounds}_{X}Q_{0}
  , \\ 
  \label{eq:matter-gauge-inv-def-2.0}
  {}^{(2)}\!{\cal Q} &:=& {}^{(2)}\!Q - 2 {\pounds}_{X} {}^{(1)}Q 
  - \left\{ {\pounds}_{Y} - {\pounds}_{X}^{2} \right\} Q_{0}
  .
\end{eqnarray}
It is straightforward to confirm that the variables
${}^{(p)}\!{\cal Q}$ defined by
(\ref{eq:matter-gauge-inv-def-1.0}) and
(\ref{eq:matter-gauge-inv-def-2.0}) are gauge invariant under
the gauge transformation rules (\ref{eq:Bruni-47-one}) and
(\ref{eq:Bruni-49-one}), respectively.

%****************************************************************

Equations (\ref{eq:matter-gauge-inv-def-1.0}) and
(\ref{eq:matter-gauge-inv-def-2.0}) have several more important 
implications.
To see this, we represent these equations as
\begin{eqnarray}
  \label{eq:matter-gauge-inv-decomp-1.0}
  {}^{(1)}\!Q &=& {}^{(1)}\!{\cal Q} + {\pounds}_{X}Q_{0}
  , \\ 
  \label{eq:matter-gauge-inv-decomp-2.0}
  {}^{(2)}\!Q  &=& {}^{(2)}\!{\cal Q} + 2 {\pounds}_{X} {}^{(1)}Q 
  + \left\{ {\pounds}_{Y} - {\pounds}_{X}^{2} \right\} Q_{0}
  .
\end{eqnarray}
These equations imply that any perturbation of first and second
order can always be decomposed into gauge invariant and gauge 
variant parts as Eqs.~(\ref{eq:matter-gauge-inv-decomp-1.0}) and
(\ref{eq:matter-gauge-inv-decomp-2.0}), respectively.
In \S\ref{sec:Matter-perturbations}, we see that these formulae
for the decomposition into gauge invariant and 
variant parts of each order perturbation are very important.

%****************************************************************

%%%%%%%%%%%%%%%%%%%%%%%%%%%%%%%%%%%%%%%%%%%%%%%%%%%%%%%%%%%%%%%%%%%%%%
\subsection{Perturbations of the Einstein tensor and the Einstein equations}
\label{sec:Perturbation-of-the-Einstein-tensor-and-the-Einstein-equations}

%****************************************************************

Now, we review the formulae for the perturbative Einstein tensor
at each order that are presented in KN2005.
The relation between the curvatures associated with the metrics
on the physical spacetime ${\cal M}_{\lambda}$ and the
background spacetime ${\cal M}_{0}$ is given by the relation
between the pulled-back operator 
${\cal X}_{\lambda}^{*}\bar{\nabla}_{a}\left({\cal X}^{-1}_{\lambda}\right)^{*}$
of the covariant derivative $\bar{\nabla}_{a}$ associated with
the metric $\bar{g}_{ab}$ on ${\cal M}_{\lambda}$ and the
covariant derivative $\nabla_{a}$ associated with the metric
$g_{ab}$ on ${\cal M}_{0}$.
The pulled-back covariant derivative 
${\cal X}_{\lambda}^{*}\bar{\nabla}_{a}\left({\cal X}^{-1}_{\lambda}\right)^{*}$
depends on the gauge choice ${\cal X}_{\lambda}$.
The property of the derivative operator 
${\cal X}^{*}_{\lambda}\bar{\nabla}_{a}\left({\cal X}^{-1}_{\lambda}\right)^{*}$
as the covariant derivative on the physical spacetime 
${\cal M}_{\lambda}$ is given by   
\begin{equation}
  {\cal X}^{*}_{\lambda}\bar{\nabla}_{a}
  \left(
    \left(
      {\cal X}^{-1}_{\lambda}\right)^{*}{\cal X}^{*}_{\lambda}\bar{g}_{ab}
  \right) = 0,
  \label{eq:property-as-covariant-derivative-on-phys-sp}
\end{equation}
where ${\cal X}^{*}_{\lambda}\bar{g}_{ab}$ is the pull-back of
the metric on the physical spacetime ${\cal M}_{\lambda}$, which
is expanded as Eq.~(\ref{eq:metric-expansion}).
In spite of the gauge dependence of the operator 
${\cal X}^{*}_{\lambda}\bar{\nabla}_{a}\left({\cal X}^{-1}_{\lambda}\right)^{*}$,
we simply denote this operator by $\bar{\nabla}_{a}$, because
our calculations are carried out only on the background spacetime
${\cal M}_{0}$ in the same gauge choice ${\cal X}_{\lambda}$.
Further, we denote the pulled-back metric 
${\cal X}^{*}_{\lambda}\bar{g}_{ab}$ of the physical spacetime
${\cal M}_{\lambda}$ by $\bar{g}_{ab}$, as mentioned above.
Though we have to keep in our mind that we are treating
perturbations in the single gauge choice when we treat the
derivative operator $\bar{\nabla}_{a}$ and the pulled-back
physical metric $\bar{g}_{ab}$ on the background spacetime
${\cal M}_{0}$, there is no confusion in the development of the
perturbation theory if we treat perturbations only in the single
gauge choice ${\cal X}_{\lambda}$.

%****************************************************************

Since the derivative operator $\bar{\nabla}_{a}$ 
($={\cal X}^{*}\bar{\nabla}_{a}\left({\cal X}^{-1}\right)^{*}$)
may be regarded as a derivative operator on the background
spacetime that satisfies the property
(\ref{eq:property-as-covariant-derivative-on-phys-sp}), there
exists a tensor field $C^{c}_{\;\;ab}$ on the background
spacetime ${\cal M}_{0}$ such that 
\begin{equation}
  \bar{\nabla}_{a}\omega_{b}
  = \nabla_{a}\omega_{b} - C^{c}_{\;\;ab} \omega_{c},
\end{equation}
where $\omega_{c}$ is an arbitrary one-form on the background
spacetime ${\cal M}_{0}$.
From the property
(\ref{eq:property-as-covariant-derivative-on-phys-sp}) of the
covariant derivative operator $\bar{\nabla}_{a}$ on ${\cal M}$,
the tensor field $C^{c}_{\;\;ab}$ on ${\cal M}_{0}$ is given by
\begin{equation}
  C^{c}_{\;\;ab} = \frac{1}{2} \bar{g}^{cd}
  \left(
      \nabla_{a}\bar{g}_{db}
    + \nabla_{b}\bar{g}_{da}
    - \nabla_{d}\bar{g}_{ab}
  \right),
  \label{eq:c-connection}
\end{equation}
where $\bar{g}^{ab}$ is the inverse metric of $\bar{g}_{ab}$,
i.e., $\bar{g}_{ac}\bar{g}^{cb}=\delta_{a}^{\;\;b}$.
We note that the gauge dependence of the derivative
$\bar{\nabla}_{a}$ as a derivative operator on ${\cal M}_{0}$ is
included only in this tensor field $C^{c}_{\;\;ab}$.
The Riemann curvature $\bar{R}_{abc}^{\;\;\;\;\;\;d}$ on the
physical spacetime ${\cal M}_{\lambda}$, which is pulled back to
the background spacetime ${\cal M}_{0}$, is given by the Riemann
curvature $R_{abc}^{\;\;\;\;\;\;d}$ on the background spacetime
${\cal M}_{0}$ and the tensor field $C^{c}_{\;\;ab}$ as follows:
\begin{equation}
  \bar{R}_{abc}^{\;\;\;\;\;\;d} = R_{abc}^{\;\;\;\;\;\;d}
  - 2 \nabla_{[a}^{} C^{d}_{\;\;b]c}
  + 2 C^{e}_{\;\;c[a} C^{d}_{\;\;b]e}.
  \label{eq:phys-riemann-back-riemann-rel}
\end{equation}
The perturbative expression for the curvatures are obtained from
the perturbative expansion of
Eq.~(\ref{eq:phys-riemann-back-riemann-rel}) through the
perturbative expansion of the tensor $C^{c}_{\;\;ab}$ defined by
Eq.~(\ref{eq:c-connection}).

%*********************************************************************

The first- and the-second order perturbations of the Riemann, the
Ricci, the scalar, the Weyl curvatures, and the Einstein tensors
on the general background spacetime are summarized in KN2005.
We also derived the perturbative form of the divergence of an
arbitrary tensor field of second rank to check the
perturbative Bianchi identities in KN2005.
In this paper, we only present the perturbative expression for
the Einstein tensor.

%*********************************************************************

We expand the Einstein tensor $\bar{G}_{a}^{\;\;b}$ on the
physical spacetime ${\cal M}_{\lambda}$ as
\begin{equation}
  \bar{G}_{a}^{\;\;b}
  =
  G_{a}^{\;\;b}
  + \lambda {}^{(1)}\!G_{a}^{\;\;b} 
  + \frac{1}{2} \lambda^{2} {}^{(2)}\!G_{a}^{\;\;b} 
  + O(\lambda^{3}).
\end{equation}
As shown in KN2005, each order perturbation of the Einstein
tensor is given by 
\begin{eqnarray}
  \label{eq:linear-Einstein}
  {}^{(1)}\!\bar{G}_{a}^{\;\;b}
  &=&
  {}^{(1)}{\cal G}_{a}^{\;\;b}\left[{\cal H}\right]
  + {\pounds}_{X} G_{a}^{\;\;b}
  ,\\
  \label{eq:second-Einstein-2,0-0,2}
  {}^{(2)}\!\bar{G}_{a}^{\;\;b}
  &=& 
  {}^{(1)}{\cal G}_{a}^{\;\;b}\left[{\cal L}\right]
  + {}^{(2)}{\cal G}_{a}^{\;\;b} \left[{\cal H}, {\cal H}\right]
  + 2 {\pounds}_{X} {}^{(1)}\!\bar{G}_{a}^{\;\;b}
  + \left\{ {\pounds}_{Y} - {\pounds}_{X}^{2} \right\} G_{a}^{\;\;b},
\end{eqnarray}
where
\begin{eqnarray}
  \label{eq:cal-G-def-linear}
  {}^{(1)}{\cal G}_{a}^{\;\;b}\left[A\right]
  &:=&
  {}^{(1)}\Sigma_{a}^{\;\;b}\left[A\right]
  - \frac{1}{2} \delta_{a}^{\;\;b} {}^{(1)}\Sigma_{c}^{\;\;c}\left[A\right]
  , \\
  \label{eq:cal-G-def-second}
  {}^{(2)}{\cal G}_{a}^{\;\;b}\left[A, B\right]
  &:=&
  {}^{(2)}\Sigma_{a}^{\;\;b}\left[A, B\right]
  - \frac{1}{2} \delta_{a}^{\;\;b} {}^{(2)}\Sigma_{c}^{\;\;c}\left[A, B\right]
  , \\
  \label{eq:(1)Sigma-def-linear}
  {}^{(1)}\Sigma_{a}^{\;\;b}\left[A\right]
  &:=&
  - 2 \nabla_{[a}^{}H_{d]}^{\;\;\;bd}\left[A\right]
  - A^{cb} R_{ac}
  , \\
  \label{eq:(2)Sigma-def-second}
  {}^{(2)}\Sigma_{a}^{\;\;b}\left[A, B\right]
  &:=& 
    2 R_{ad} B_{c}^{\;\;(b}A^{d)c}
  + 2 H_{[a}^{\;\;\;de}\left[A\right] H_{d]\;\;e}^{\;\;\;b}\left[B\right]
  + 2 H_{[a}^{\;\;\;de}\left[B\right] H_{d]\;\;e}^{\;\;\;b}\left[A\right]
  \nonumber\\
  &&
  + 2 A_{e}^{\;\;d} \nabla_{[a}H_{d]}^{\;\;\;be}\left[B\right]
  + 2 B_{e}^{\;\;d} \nabla_{[a}H_{d]}^{\;\;\;be}\left[A\right]
  \nonumber\\
  &&
  + 2 A_{c}^{\;\;b} \nabla_{[a}H_{d]}^{\;\;\;cd}\left[B\right]
  + 2 B_{c}^{\;\;b} \nabla_{[a}H_{d]}^{\;\;\;cd}\left[A\right]
  ,
\end{eqnarray}
and 
\begin{eqnarray}
  H_{ab}^{\;\;\;\;c}\left[A\right]
  &:=&
  \nabla_{(a}^{}A_{b)}^{\;\;\;c}
  - \frac{1}{2} \nabla^{c}_{}A_{ab}
  \label{eq:Habc-def-1}
  , \\
  H_{abc}\left[A\right] 
  &:=&
  g_{cd} H_{ab}^{\;\;\;\;d}\left[A\right]
  ,
  \quad
  H_{a}^{\;\;bc}\left[A\right] 
  := 
  g^{bd} H_{ad}^{\;\;\;\;c}\left[A\right]
  ,
  \nonumber\\
  H_{a\;\;c}^{\;\;b}\left[A\right] 
  &:=& 
  g_{cd} H_{a}^{\;\;bd}\left[A\right].
  \label{eq:Habc-def-2}
\end{eqnarray}
We note that ${}^{(1)}{\cal G}_{a}^{\;\;b}\left[*\right]$ and
${}^{(2)}{\cal G}_{a}^{\;\;b}\left[*,*\right]$ in
Eqs.~(\ref{eq:linear-Einstein}) and
(\ref{eq:second-Einstein-2,0-0,2}) are the gauge invariant parts
of the perturbative Einstein tensors, and
Eqs.~(\ref{eq:linear-Einstein}) and
(\ref{eq:second-Einstein-2,0-0,2}) have the same forms as 
Eqs.~(\ref{eq:matter-gauge-inv-def-1.0}) and
(\ref{eq:matter-gauge-inv-decomp-2.0}), respectively.

%*********************************************************************

We also note that ${}^{(1)}{\cal G}_{a}^{\;\;b}\left[*\right]$
and ${}^{(2)}{\cal G}_{a}^{\;\;b}\left[*,*\right]$ defined by
Eqs.~(\ref{eq:cal-G-def-linear})--(\ref{eq:(2)Sigma-def-second})
satisfy the identities
\begin{eqnarray}
  \nabla_{a}
  {}^{(1)}{\cal G}_{b}^{\;\;a}\left[A\right]
  &=& 
  - H_{ca}^{\;\;\;\;a}\left[A\right] G_{b}^{\;\;c}
  + H_{ba}^{\;\;\;\;c}\left[A\right] G_{c}^{\;\;a}
  \label{eq:linear-order-divergence-of-calGab}
  , \\
  \nabla_{a}{}^{(2)}{\cal G}_{b}^{\;\;a}\left[A, B\right]
  &=& 
  - H_{ca}^{\;\;\;\;a}\left[A\right]
    {}^{(1)}\!{\cal G}_{b}^{\;\;c}\left[B\right]
  - H_{ca}^{\;\;\;\;a}\left[B\right]
    {}^{(1)}\!{\cal G}_{b}^{\;\;c}\left[A\right]
  \nonumber\\
  &&
  + H_{ba}^{\;\;\;\;e}\left[A\right]
    {}^{(1)}\!{\cal G}_{e}^{\;\;a}\left[B\right]
  + H_{ba}^{\;\;\;\;e}\left[B\right]
    {}^{(1)}\!{\cal G}_{e}^{\;\;a}\left[A\right]
  \nonumber\\
  &&
  - \left(
    H_{bad}\left[B\right] A^{dc} + H_{bad}\left[A\right] B^{dc}
  \right)
  G_{c}^{\;\;a}
  \nonumber\\
  &&
  + \left(
    H_{cad}\left[B\right] A^{ad} + H_{cad}\left[A\right] B^{ad}
  \right)
  G_{b}^{\;\;c},
  \label{eq:second-div-of-calGab-1,1}
\end{eqnarray}
for arbitrary tensor fields $A_{ab}$ and $B_{ab}$, respectively.
We can directly confirm these identities, and these identities
guarantee the first-order and second-order perturbations of
the Bianchi identity $\bar{\nabla}_{b}\bar{G}_{a}^{\;\;b}=0$,
respectively, as shown in KN2005.
These identities are also useful when we check whether the
derived components of 
${}^{(1)}{\cal G}_{a}^{\;\;b}\left[*\right]$ and 
${}^{(2)}{\cal G}_{a}^{\;\;b}\left[*,*\right]$ are correct.

%*********************************************************************

Finally, we consider perturbations of the Einstein equation of
first and second order.
First, we expand the energy-momentum tensor as
\begin{equation}
  \bar{T}_{a}^{\;\;b}
  =
  T_{a}^{\;\;b}
  + \lambda {}^{(1)}\!T_{a}^{\;\;b} 
  + \frac{1}{2} \lambda^{2} {}^{(2)}\!T_{a}^{\;\;b} 
  + O(\lambda^{3}).
\end{equation}
Following the definitions (\ref{eq:matter-gauge-inv-def-1.0})
and (\ref{eq:matter-gauge-inv-def-2.0}) of gauge invariant
variables, the gauge invariant variables 
${}^{(1)}\!{\cal T}_{b}^{\;\;a}$ and 
${}^{(2)}\!{\cal T}_{b}^{\;\;a}$ for the perturbations
${}^{(1)}\!\bar{T}_{a}^{\;\;b}$ and 
${}^{(2)}\!\bar{T}_{a}^{\;\;b}$ of the energy-momentum tensor
are defined by 
\begin{eqnarray}
  \label{eq:Tab-gauge-inv-def-1.0} 
  {}^{(1)}\!{\cal T}_{b}^{\;\;a} &:=&
  {}^{(1)}\!\bar{T}_{b}^{\;\;a} - {\pounds}_{X}T_{b}^{\;\;a}
  , \\ 
  \label{eq:Tab-gauge-inv-def-2.0}
  {}^{(2)}\!{\cal T}_{b}^{\;\;a} &:=&
  {}^{(2)}\!\bar{T}_{b}^{\;\;a} 
  - 2 {\pounds}_{X} {}^{(1)}\!\bar{T}_{b}^{\;\;a} 
  - \left\{
    {\pounds}_{Y}
    -{\pounds}_{X}^{2}
  \right\} T_{b}^{\;\;a}
  .
\end{eqnarray}
In \S\ref{sec:Matter-perturbations}, we show that these
definitions of the gauge invariant part of the first- and
second-order perturbation of the energy-momentum tensor are
appropriate in the cases of both a perfect fluid and a single
scalar field. 
Further, we impose the perturbed Einstein equation of each order, 
\begin{equation}
  {}^{(1)}G_{a}^{\;\;b} = 8\pi G \;\; {}^{(1)}T_{a}^{\;\;b},
  \quad
  {}^{(2)}G_{a}^{\;\;b} = 8\pi G \;\; {}^{(2)}T_{a}^{\;\;b}.
\end{equation}
Then, the perturbative Einstein equation is given by 
\begin{eqnarray}
  \label{eq:linear-order-Einstein-equation}
  {}^{(1)}{\cal G}_{a}^{\;\;b}\left[{\cal H}\right]
  &=&
  8\pi G {}^{(1)}{\cal T}_{a}^{\;\;b}
\end{eqnarray}
at linear order and
\begin{eqnarray}
  \label{eq:second-order-Einstein-equation}
  {}^{(1)}{\cal G}_{a}^{\;\;b}\left[{\cal L}\right]
  + {}^{(2)}{\cal G}_{a}^{\;\;b}\left[{\cal H}, {\cal H}\right]
  &=&
  8\pi G \;\; {}^{(2)}{\cal T}_{a}^{\;\;b} 
\end{eqnarray}
at second order.
These explicitly show that, order by order, the Einstein
equations are necessarily given in terms of gauge invariant
variables only.
Therefore, we do not have to consider the gauge degree of
freedom, at least at the level where we concentrate only on the
perturbed Einstein equations.

%*********************************************************************

We have reviewed the general outline of the second-order gauge
invariant perturbation theory.
Within this general framework, we develop a second-order
cosmological perturbation theory in terms of the gauge invariant
variables.

%*******************************************************************

%%%%%%%%%%%%%%%%%%%%%%%%%%%%%%%%%%%%%%%%%%%%%%%%%%%%%%%%%%%%%%%%%%%%%
\section{Cosmological background spacetime}
\label{sec:Cosmological-Background-spacetime}
%%%%%%%%%%%%%%%%%%%%%%%%%%%%%%%%%%%%%%%%%%%%%%%%%%%%%%%%%%%%%%%%%%%%%

Here, we consider the background spacetime for cosmological
perturbation theory.
The background spacetime considered here is a homogeneous,
isotropic universe that is foliated by the three-dimensional 
hypersurface $\Sigma(\eta)$, which is parameterised by $\eta$.
Each hypersurface of $\Sigma(\eta)$ is a maximally symmetric
three-space\cite{Weinberg1972}, and the spacetime metric of this
universe is given by
\begin{eqnarray}
  g_{ab} = a^{2}(\eta)\left(
    - (d\eta)_{a}(d\eta)_{b}
    + \gamma_{ij}(dx^{i})_{a}(dx^{j})_{b}
  \right),
  \label{eq:background-metric}
\end{eqnarray}
where $a=a(\eta)$ is the scale factor, $\gamma_{ij}$ is the
metric on the maximally symmetric 3-space with curvature
constant $K$, and the indices $i,j,k,...$ for the spatial
components run from 1 to 3.
Depending on the behavior of the scale factor $a$, this metric
(\ref{eq:background-metric}) can represent a
Friedmann-Robertson-Walker universe or a de Sitter spacetime. 
In terms of the coordinate system in which the spacetime metric
is given by (\ref{eq:background-metric}), the components of the
Christoffel symbol of this background spacetime are given by 
\begin{eqnarray}
  \Gamma_{\eta\eta}^{\eta} = {\cal H}, \quad
  \Gamma_{i\eta}^{\eta} = 0 = \Gamma_{\eta\eta}^{i}, \quad
  \Gamma_{ij}^{\eta} = {\cal H}\gamma_{ij}, \quad
  \Gamma_{i\eta}^{j} = {\cal H} \gamma_{i}^{\;\;j}, \quad
  \Gamma_{ij}^{k} = {}^{(3)}\!\Gamma_{ij}^{k},
\end{eqnarray}
where ${\cal H} := \partial_{\eta}a/a$,
${}^{(3)}\!\Gamma_{ij}^{k}$ is the Christoffel symbol associated
with the three metric $\gamma_{ij}$, and
$\gamma_{i}^{\;\;j}=\gamma_{ik}\gamma^{kj}$ is the
three-dimensional Kronecker delta.
These components of the Christoffel symbol are useful when we
write down the components of the four-dimensional covariant
derivative of tensors in terms of the derivative with respect to
$\eta$ and the three-dimensional covariant derivative associate
with the metric $\gamma_{ij}$.

%*******************************************************************

Since $\gamma_{ij}$ is the metric on the maximally symmetric
3-space with the curvature constant $K$, the curvatures
associated with the metric $\gamma_{ij}$ are given by
\begin{eqnarray}
  \label{eq:maxsym-curv-gamma}
  {}^{(3)}\!R_{ijkl} = 2K \gamma_{k[i}\gamma_{j]l}, \quad
  {}^{(3)}\!R_{ij} = 2K \gamma_{ij}, \quad
  {}^{(3)}\!R = 6K.
\end{eqnarray}
These are useful when we calculate the components of the
perturbative curvatures in terms of the 3+1 decomposition, as in
the metric (\ref{eq:background-metric}).
The four-dimensional background curvature tensors are also
necessary to calculate the components of the perturbative
curvatures. 
These are given by 
\begin{eqnarray}
  \label{eq:ricci-ab-conformal-coordiante-calH}
  R_{ab} &=& - 3 \partial_{\eta}{\cal H}
  \left(d\eta\right)_{a}  \left(d\eta\right)_{b}
  + \left(
    \partial_{\eta}{\cal H}
    + 2 {\cal H}^{2}  
    + 2 K 
  \right) \gamma_{ab}
  , \\
  \label{eq:ricci-ab-mixed-conformal-coordiante-calH}
  R_{a}^{\;\;b} &=& \frac{3}{a^{2}} \partial_{\eta}{\cal H}
  \left(d\eta\right)_{a}  \left(\frac{\partial}{\partial\eta}\right)^{b}
  + \frac{1}{a^{2}} \left(
    \partial_{\eta}{\cal H}
    + 2 {\cal H}^{2}  
    + 2 K 
  \right) \gamma_{a}^{\;\;b}
  , \\
  \label{eq:scalar-conformal-coordiante-calH}
  R &:=& R_{a}^{\;\;a} = \frac{6}{a^{2}} 
  \left\{
    \partial_{\eta}{\cal H}
    + {\cal H}^{2}  
    + K 
  \right\}
  , \\
  G_{a}^{\;\;b}
  &:=& 
  R_{a}^{\;\;b} - \frac{1}{2} \delta_{a}^{\;\;b} R \nonumber\\ 
  \label{eq:einstein-conformal-coordiante-calH}
  &=& 
  - \frac{3}{a^{2}} 
  \left[
    {\cal H}^{2} + K
  \right]
  \left(d\eta\right)_{a}  \left(\frac{\partial}{\partial\eta}\right)^{b}
  - \frac{1}{a^{2}}
  \left[
    2 \partial_{\eta}{\cal H}
    + {\cal H}^{2}
    + K
  \right]
  \gamma_{a}^{\;\;b},
\end{eqnarray}
where $\gamma_{ab} = \gamma_{ij}(dx^{i})_{a}(dx^{j})_{b}$ and
$\gamma_{a}^{\;\;b}=\gamma_{i}^{\;\;j}(dx^{i})_{a}(\partial/\partial x^{j})^{b}$.

%*******************************************************************

To study the Einstein equation for this background spacetime,
we introduce the energy-momentum tensor for a perfect fluid,
\begin{eqnarray}
  \label{eq:energy-momentum-perfect-fluid}
  T_{a}^{\;\;b}
  &=&
  \epsilon u_{a} u^{b}
  + p (\delta_{a}^{\;\;b} + u_{a} u^{b}) 
  \\
  &=&
  - \epsilon (d\eta)_{a} \left(\frac{\partial}{\partial\eta}\right)^{b}
  + p \gamma_{a}^{\;\;b},  
  \label{eq:energy-momentum-perfect-fluid-homogeneous}
\end{eqnarray}
where we have used
\begin{eqnarray}
  u_{a} = - a (d\eta)_{a},
  \quad 
  u^{a} = \frac{1}{a} \left(\frac{\partial}{\partial\eta}\right)^{a},
  \quad 
  \delta_{a}^{\;\;b}
  = (d\eta)_{a}\left(\frac{\partial}{\partial\eta}\right)^{b}
  + \gamma_{a}^{\;\;b}.
  \label{eq:energy-momentum-for-velocity-homogeneous}
\end{eqnarray}
We also consider the energy-momentum tensor for the scalar
field, which is given by
\begin{eqnarray}
  T_{a}^{\;\;b}
  &=&
  \nabla_{a}\varphi\nabla^{b}\varphi -
  \frac{1}{2}\delta_{a}^{\;\;b}\left(\nabla_{c}\varphi\nabla^{c}\varphi +
    2V(\varphi)\right)
  \label{eq:energy-momentum-single-scalar}
  \\
  &=&
  -
  \left(
      \frac{1}{2a^{2}} (\partial_{\eta}\varphi)^{2}
    + V(\varphi)
  \right)
  (d\eta)_{a} \left(\frac{\partial}{\partial\eta}\right)^{b}
  \nonumber\\
  && \quad
  +
  \left(
    \frac{1}{2a^{2}} (\partial_{\eta}\varphi)^{2}
    - V(\varphi)
  \right)
  \gamma_{a}^{\;\;b},
  \label{eq:energy-momentum-single-scalar-homogeneous}
\end{eqnarray}
where we have assumed that the scalar field $\varphi$ is
homogeneous, i.e., $\varphi=\varphi(\eta)$.
Comparing (\ref{eq:energy-momentum-single-scalar-homogeneous})
with (\ref{eq:energy-momentum-perfect-fluid-homogeneous}), the
energy density and the pressure for the homogeneous scalar field
are given by
\begin{eqnarray}
  \label{eq:background-fluid-scalar-relations}
  \epsilon = \frac{1}{2a^{2}} (\partial_{\eta}\varphi)^{2} + V(\varphi),
  \quad
  p = \frac{1}{2a^{2}} (\partial_{\eta}\varphi)^{2} - V(\varphi).
\end{eqnarray}

%*******************************************************************

The Einstein equations $G_{a}^{\;\;b} = 8\pi G T_{a}^{\;\;b}$
for this background spacetime filled with a perfect fluid are
given by 
\begin{eqnarray}
  \label{eq:background-Einstein-equations}
  {\cal H}^{2} + K = \frac{8 \pi G}{3} a^{2} \epsilon,
  \quad
  2 \partial_{\eta}{\cal H} + {\cal H}^{2} + K = - 8 \pi G a^{2}p.
\end{eqnarray}
In the derivation of the perturbative Einstein equations, the
equation
\begin{eqnarray}
  \label{eq:background-Einstein-equations-2}
  {\cal H}^{2} + K - \partial_{\eta}{\cal H} = 4 \pi G a^{2} (\epsilon + p)
\end{eqnarray}
is also useful.
Of course, there is an equation for the energy conservation of
the matter fields, and this equation gives the behavior of the
energy density in the scale factor if we apply an appropriate
equation of state for the matter field.
This equation is consistent with the two equations in 
(\ref{eq:background-Einstein-equations}).

%*******************************************************************

Further, in the case of the single scalar field model, the
Einstein equations are given by
\begin{eqnarray}
  \label{eq:background-Einstein-equations-scalar-1}
  && 
  {\cal H}^{2} + K = \frac{8 \pi G}{3} a^{2} \left(
    \frac{1}{2a^{2}} (\partial_{\eta}\varphi)^{2} + V(\varphi)
  \right)
  ,\\
  \label{eq:background-Einstein-equations-scalar-2}
  &&
  2 \partial_{\eta}{\cal H} + {\cal H}^{2} + K = - 8 \pi G a^{2}
  \left(\frac{1}{2a^{2}} (\partial_{\eta}\varphi)^{2} - V(\varphi)\right)
  ,
\end{eqnarray}
through the relations
(\ref{eq:background-fluid-scalar-relations}).
We also note that the equations
(\ref{eq:background-Einstein-equations-scalar-1}) and
(\ref{eq:background-Einstein-equations-scalar-2}) lead to 
\begin{eqnarray}
  \label{eq:background-Einstein-equations-scalar-3}
  {\cal H}^{2} + K - \partial_{\eta}{\cal H}
  = 4\pi G (\partial_{\eta}\varphi)^{2}.
\end{eqnarray}
Equation (\ref{eq:background-Einstein-equations-scalar-3})
is also useful when we derive the perturbative Einstein
equations.
Equations (\ref{eq:background-Einstein-equations-scalar-1}) and
(\ref{eq:background-Einstein-equations-scalar-2}) are often used
to investigate the inflationary scenario.
Actually, in the situation that the potential term of the
scalar field is sufficiently larger than its kinetic term, the
spacetime is approximately a de Sitter spacetime, and this
situation may be realized in the very early
universe\cite{Mukhanov-Feldman-Brandenberger-1992}.
Hence, the background spacetime described by
Eqs.~(\ref{eq:background-Einstein-equations-scalar-1}) and
(\ref{eq:background-Einstein-equations-scalar-2}) also includes
inflationary universes, and the second-order perturbation theory
developed below is also applicable to the inflationary
universe.

%*******************************************************************

%%%%%%%%%%%%%%%%%%%%%%%%%%%%%%%%%%%%%%%%%%%%%%%%%%%%%%%%%%%%%%%%%%%%%
\section{Gauge invariant variables of cosmological perturbations}
\label{sec:Gauge-Invariant-Variables-of-Cosmological-Perturbations}
%%%%%%%%%%%%%%%%%%%%%%%%%%%%%%%%%%%%%%%%%%%%%%%%%%%%%%%%%%%%%%%%%%%%%

%*******************************************************************

Now, we develop the second-order perturbation theory with the
cosmological background spacetime in
\S\ref{sec:Cosmological-Background-spacetime} within the general
framework of the gauge invariant perturbation theory reviewed in
\S\ref{sec:General-framework-of-the-gauge-invariant-perturbation-theory}.
The important step when we apply the above general framework of
the gauge invariant perturbation theory is to confirm that the
assumption for the decomposition (\ref{eq:linear-metric-decomp})
of the linear-order metric perturbation is correct.
This confirmation is accomplished in
\S\ref{sec:First-order-metric-perturbations}.
Hence, the general framework reviewed in
\S\ref{sec:General-framework-of-the-gauge-invariant-perturbation-theory}
is applicable.
Applying this framework, we define the second-order gauge
invariant variable of the metric perturbation in
\S\ref{sec:Second-order-metric-perturbations} and of the matter
perturbations in \S\ref{sec:Matter-perturbations}.

%*******************************************************************

%%%%%%%%%%%%%%%%%%%%%%%%%%%%%%%%%%%%%%%%%%%%%%%%%%%%%%%%%%%%%%%%%%%%%
\subsection{First-order metric perturbations}
\label{sec:First-order-metric-perturbations}

%*******************************************************************

On the background spacetime discussed in
\S\ref{sec:Cosmological-Background-spacetime}, we consider the
metric perturbation to be that given in
Eq.~(\ref{eq:metric-expansion}). 
To show that the assumption for the decomposition
(\ref{eq:linear-metric-decomp}) of the linear-order metric
perturbation is correct, we first consider the components of the
linear-order metric perturbation in the coordinate system
(\ref{eq:background-metric}),
\begin{eqnarray}
  h_{ab}
  =
  h_{\eta\eta}(d\eta)_{a}(d\eta)_{b}
  + 2 h_{\eta i}(d\eta)_{(a}(dx^{i})_{b)}
  + h_{ij}(dx^{i})_{a}(dx^{j})_{b}.
\end{eqnarray}
Because components belonging to different groups are coupled
through contraction with the metric tensor and the covariant
derivatives in the Einstein equations, the grouping
$\{h_{\eta\eta}, h_{\eta i}, h_{ij}\}$ is not so useful.
Instead, using the fact that the three manifold
$(\Sigma(\eta),a^{2}\gamma_{ab})$ is maximally symmetric, we
further decompose the vector $h_{\eta i}$ and tensor $h_{ij}$ as
\begin{eqnarray}
  \label{eq:htaui-decomp}
  h_{\eta i} &=& D_{i}h_{(VL)} + h_{(V)i}, \quad D^{i}h_{(V)i} = 0, \\
  \label{eq:hij-decomp}
  h_{ij} &=& a^{2} h_{(L)} \gamma_{ij} + a^{2}h_{(T)ij}, 
  \quad {h_{(T)}}^{i}_{\;\;i} := \gamma^{ij}{h_{(T)}}_{ij} = 0, \\ 
  \label{eq:hTij-decomp}
  h_{(T)ij} &=& \left(D_{i}D_{j} - \frac{1}{3}\gamma_{ij}\Delta\right)h_{(TL)}
  + 2 D_{(i}h_{(TV)j)} + {h_{(TT)ij}}, \\
  D^{i} h_{(TV)i} &=& 0, \quad D^{i} h_{(TT)ij} = 0,
\end{eqnarray}
where $D_{i}$ is the covariant derivative associated with the
metric $\gamma_{ij}$, and $\Delta = D^{i}D_{i}$. 
It is well-known that the linearized Einstein equations
on a homogeneous isotropic universe, whose metric is given by
Eq.~(\ref{eq:background-metric}), can be decomposed into groups
that each contains only variables belonging to one
of the three sets $\{h_{\eta\eta},h_{(VL)},h_{(L)},h_{TL}\}$,
$\{{h_{(V)}}_{i},{h_{(TV)}}_{i}\}$, and
${h_{(TT)ij}}$\cite{perturbation-of-maximally-symmetric}.
Variables belonging to these sets are called scalar-type,
vector-type and tensor-type variables, respectively. 
This segregation of variables is due to the fact that the metric
tensor $a^{2}\gamma_{ab}$ is the only non-trivial tensor on the
maximally symmetric space, and as a consequence, the tensorial
operations on $h_{ab}$ to construct the linearized Einstein
tensors preserve this decomposition.

%*******************************************************************

In the linear perturbation theory, the covariant derivatives are
always combined into the Laplacian $\Delta$ in the linearized
Einstein equations after the decompositions
(\ref{eq:htaui-decomp})--(\ref{eq:hTij-decomp}), because the
metric tensor $a^{2}\gamma_{ab}$ is the only non-trivial tensor
on the maximally symmetric space.
Thus, the harmonic expansion of the perturbation variables with
respect to the Laplacian is also useful in the linear
perturbation theory.
However, in the second-order perturbation theory, mode-mode
coupling occurs due to the non-linearity of the Einstein
equations.
For this reason, we do not apply the harmonic expansion of the
perturbation variables with respect to the Laplacian in this
paper, though the harmonic expansion should be also useful after
the nonlinear terms in the second-order Einstein equations are
clarified.
Instead, we assume the existence of some Green functions, as
explained below.

%*******************************************************************

To clarify the uniqueness of the decompositions
(\ref{eq:htaui-decomp})--(\ref{eq:hTij-decomp}), we consider the
inverse relations of
Eqs.~(\ref{eq:htaui-decomp})--(\ref{eq:hTij-decomp}).
To do this, we first note that the commutator of the divergence
and the Laplacian is given by 
\begin{eqnarray}
  D^{i}\Delta t_{i} - \Delta D^{i} t_{i} 
  &=& D^{i}\left({}^{(3)}\!R_{ij} t^{j}\right)
  = 2K D^{i} t_{i}
  , \\
  D^{i}\Delta t_{ij} - \Delta D^{i} t_{ij} 
  &=& D^{i}\left(
    {}^{(3)}\!R_{i}^{\;\;e} t_{ej}
    +
    {}^{(3)}\!R_{\;\;ij}^{l\;\;\;\;k}t_{lk}
  \right)
  + 
  {}^{(3)}\!R_{ikjl} D^{k}t^{il}
  \nonumber\\
  &=& 4K \left( D^{i} t_{(ij)}
    -
    \frac{1}{2}
    D_{j} t_{i}^{\;\;i}
  \right)
\end{eqnarray}
for any tensor fields $t_{i}$ and $t_{ij}$, where we have used
Eqs.~(\ref{eq:maxsym-curv-gamma}).
These relations are also useful when we write down the
components of the perturbative Einstein tensor.
They show that on the maximally symmetric space, the Laplacian
preserves the transverse condition
\begin{equation}
  D^{j}\Delta h_{(V)j} = 
  \left(\Delta + 2K \right)D^{j}h_{(V)j} = 0.
\end{equation}
The inverse relations of the decompositions
(\ref{eq:htaui-decomp})--(\ref{eq:hTij-decomp}) are given by
\begin{eqnarray}
  \label{eq:hetai-decomp-inv-1}
  h_{(VL)} &=& \Delta^{-1} D^{j} h_{\eta j}, \\
  \label{eq:hetai-decomp-inv-2}
  h_{(V)i} &=& h_{\eta i} - D_{i} \Delta^{-1} D^{j} h_{\eta j}, \\
  \label{eq:hij-decomp-inv-1}
  h_{(L)} &=& \frac{1}{3a^{2}} h_{i}^{\;\;i}, \\
  \label{eq:hij-decomp-inv-2}
  h_{(T)ij} &=&
  \frac{1}{a^{2}} \left(h_{ij} - \frac{1}{3}h_{k}^{\;\;k}\gamma_{ij}\right), \\
  \label{eq:hij-decomp-inv-3}
  h_{(TL)} &=& \frac{3}{2} (\Delta + 3K)^{-1} \Delta^{-1} D^{i}D^{j} h_{(T)ij}
  , \\
  \label{eq:hij-decomp-inv-4}
  h_{(TV)i}
  &=&
  (\Delta + 2K)^{-1} D^{k} h_{(T)ik}
  - (\Delta + 2K)^{-1} D_{i} \Delta^{-1} D^{k}D^{l} h_{(T)kl}
  , \\
  {h_{(TT)ij}}
  &=&
  h_{(T)ij} 
  - \frac{3}{2} \left(D_{i}D_{j} - \frac{1}{3}\gamma_{ij}\Delta\right)
  (\Delta + 3K)^{-1} \Delta^{-1} D^{k}D^{l} h_{(T)kl}
  \nonumber\\
  && \quad
  - 2 D_{(i} (\Delta + 2K)^{-1} D^{k} h_{(T)j)k}
  \nonumber\\
  && \quad
  + 2 D_{(i} (\Delta + 2K)^{-1} D_{j)} \Delta^{-1} D^{k}D^{l} h_{(T)kl}
  .
  \label{eq:hTij-decomp-inv-5}
\end{eqnarray}
Equations
(\ref{eq:hetai-decomp-inv-1})--(\ref{eq:hTij-decomp-inv-5}) 
show that there should be exist Green functions of the
operators $\Delta$, $\Delta + 2K$, and $\Delta + 3K$ to
guarantee the one to one correspondence of the decompositions of
the set $\{h_{\eta\eta},h_{\eta i},h_{ij}\}$ and the sets 
$\{$
$\{h_{\eta\eta},h_{(VL)},h_{(L)},h_{TL}\}$,
$\{{h_{(V)}}_{i},{h_{(TV)}}_{i}\}$, $h_{(TT)ij}$ 
$\}$.
Actually, these Green functions exist if we specify the domain
of the perturbations, for example $L^{2}$-space on
$\Sigma(\eta)$ with appropriate boundary condition.
Therefore, we assume the existence of these Green functions in this
paper.
By this assumption, any tensor that belongs to the kernel of any
of the operators $\Delta$, $\Delta+2K$, and $\Delta+3K$ is
excluded from consideration.
For example, a Killing field $v_{i}$ on the three-dimensional
hypersurface $\Sigma(\eta)$, which satisfies the Killing equation
$D_{(i}v_{j)}=0$, belongs to the kernel of the operator
$\Delta+2K$, since we can easily confirm $(\Delta + 2K)v_{i}=0$
from the Killing equation.
If it is necessary to investigate such tensors as the
perturbative mode, separate treatments are necessary.
Because the treatment of these exceptional modes is beyond the
scope of this paper, we ignore all modes that belong to the
kernels of the operators $\Delta$, $\Delta+2K$, and $\Delta+3K$.

%*******************************************************************

Now, we consider the decomposition
(\ref{eq:linear-metric-decomp}) of the linear-order metric
perturbation and show that the decomposition
(\ref{eq:linear-metric-decomp}) is valid in the case of
cosmological perturbations, though this decomposition is merely
an assumption in the general framework reviewed in
\S\ref{sec:Gauge-degree-of-freedom-in-perturbation-theory}.

%*******************************************************************

To accomplish the decomposition (\ref{eq:linear-metric-decomp}),
we consider the gauge transformation rule
(\ref{eq:Bruni-47-one}), which is given by 
\begin{equation}
  {}_{\cal Y}\!h_{ab} - {}_{\cal X}\!h_{ab}
  = {\pounds}_{\xi} g_{ab}
  = 2\nabla_{(a}\xi_{b)}
  \label{eq:linear-gauge-trans-metric}
\end{equation}
for linear-order metric perturbations.
In Eq.~(\ref{eq:linear-gauge-trans-metric}), the generator
$\xi^{a}$ of the gauge transformation is an arbitrary vector
field on the background spacetime ${\cal M}_{0}$.
We decompose the generator $\xi^{a}$ in terms of the 3+1
decomposition as
\begin{equation}
  \xi_{a} = \xi_{\eta}(d\eta)_{a} + \xi_{i} (dx^{i})_{a},
\end{equation}
and, further, the component $\xi_{i}$ as
\begin{equation}
  \xi_{i} = D_{i}\xi_{(L)} + \xi_{(T)i}, \quad D^{i}\xi_{(T)i} = 0.
  \label{eq:xii-decomp}
\end{equation}
In terms of the 3+1 decomposition, the gauge transformation
rules (\ref{eq:linear-gauge-trans-metric}) are given by 
\begin{eqnarray}
  {}_{\cal Y}\!h_{\eta\eta} - {}_{\cal X}\!h_{\eta\eta}
  &=& 2\left(\partial_{\eta} - {\cal H}\right) \xi_{\eta}
  \label{eq:linear-gauge-trans-metric-3+1-tautau}
  , \\
  {}_{\cal Y}\!h_{\eta i} - {}_{\cal X}\!h_{\eta i}
  &=& D_{i}\xi_{\eta} + \left(\partial_{\eta} - 2 {\cal H}\right)\xi_{i} 
  \label{eq:linear-gauge-trans-metric-3+1-taui}
  , \\
  {}_{\cal Y}\!h_{ij} - {}_{\cal X}\!h_{ij}
  &=& 2 D_{(i}\xi_{j)}
  - 2 {\cal H} \gamma_{ij}\xi_{\eta} .
  \label{eq:linear-gauge-trans-metric-3+1-ij}
\end{eqnarray}
Furthermore, following the decomposition
(\ref{eq:xii-decomp}) of the component $\xi_{i}$, the gauge
transformation rules
(\ref{eq:linear-gauge-trans-metric-3+1-taui}) and
(\ref{eq:linear-gauge-trans-metric-3+1-ij}) are obtained in
terms of $\xi_{\eta}$, $\xi_{(L)}$, and $\xi_{(T)i}$ as
\begin{eqnarray}
  {}_{\cal Y}\!h_{\eta i} - {}_{\cal X}\!h_{\eta i}
  &=& 
  D_{i}\left\{
        \left(\partial_{\eta} - 2 {\cal H}\right) \xi_{(L)}
    +   \xi_{\eta}
  \right\}
  + \left(\partial_{\eta} - 2 {\cal H}\right)\xi_{(T)i}
  \label{eq:linear-gauge-trans-metric-3+1-taui-2}
  , \\
  {}_{\cal Y}\!h_{ij} - {}_{\cal X}\!h_{ij}
  &=& 
  2 \left(\frac{1}{3}\Delta\xi_{(L)} - {\cal H} \xi_{\eta} \right)\gamma_{ij}
  \nonumber\\
  &&
  +
  2 \left(D_{i}D_{j} - \frac{1}{3}\gamma_{ij}\Delta\right)\xi_{(L)} 
  + 2 D_{(i}\xi_{j)}^{(T)}.
  \label{eq:linear-gauge-trans-metric-3+1-ij-2}
\end{eqnarray}
Since the tensorial decomposition
(\ref{eq:htaui-decomp})--(\ref{eq:hTij-decomp}) has the inverse
relations
(\ref{eq:hetai-decomp-inv-1})--(\ref{eq:hTij-decomp-inv-5}), the
gauge transformation rules
(\ref{eq:linear-gauge-trans-metric-3+1-taui-2}) and
(\ref{eq:linear-gauge-trans-metric-3+1-ij-2}) lead to those for
the metric perturbations $h_{(VL)}$, $h_{(V)i}$, 
$h_{(L)}$, $h_{(TL)}$, $h_{(TV)i}$, and $h_{(TT)ij}$ as follows:
\begin{eqnarray}
  \label{eq:decomposed-gauge-trans-2}
  {}_{\cal Y}\!h_{(VL)} - {}_{\cal X}\!h_{(VL)}
  &=& \xi_{\eta} + \left(\partial_{\eta} - 2 {\cal H}\right) \xi_{(L)}, \\
  \label{eq:decomposed-gauge-trans-3}
  {}_{\cal Y}\!h_{(V)i} - {}_{\cal X}\!h_{(V)i}
  &=&
  \left( \partial_{\eta} - 2 {\cal H} \right) \xi_{(T)i}, \\
  \label{eq:decomposed-gauge-trans-4}
  a^{2} {}_{\cal Y}\!h_{(L)} - a^{2} {}_{\cal X}\!h_{(L)}
  &=&
  - 2 {\cal H} \xi_{\eta} + \frac{2}{3}\Delta\xi_{(L)}, \\
  \label{eq:decomposed-gauge-trans-5}
  a^{2} {}_{\cal Y}\!h_{(TL)} - a^{2} {}_{\cal X}\!h_{(TL)}
  &=&
  2 \xi_{(L)} , \\
  \label{eq:decomposed-gauge-trans-6}
  a^{2} {}_{\cal Y}\!h_{(TV)i} - a^{2} {}_{\cal X}\!h_{(TV)i}
  &=&
  \xi_{(T)i}, \\
  \label{eq:decomposed-gauge-trans-7}
  a^{2} {}_{\cal Y}\!h_{(TT)ij} - a^{2} {}_{\cal X}\!h_{(TT)ij}
  &=& 0.
\end{eqnarray}
Inspecting the gauge transformation rules
(\ref{eq:decomposed-gauge-trans-2})--(\ref{eq:decomposed-gauge-trans-7}), 
together with
Eq.~(\ref{eq:linear-gauge-trans-metric-3+1-tautau}), we find
gauge invariant and variant variables of the first-order metric
perturbation.
We first construct gauge invariant variables.

%*******************************************************************

First, equation (\ref{eq:decomposed-gauge-trans-7}) shows that
the transverse traceless part $h_{(TT)ij}$ is itself gauge
invariant.
We denote this transverse traceless part by
\begin{equation}
  \stackrel{(1)\;\;\;}{\chi_{ij}} :=  h_{(TT)ij}, \quad
  \stackrel{(1)\;\;\;}{\chi_{ij}} = \stackrel{(1)\;\;\;}{\chi_{ji}}, \quad
  \stackrel{(1)\;\;\;}{\chi^{i}_{\;\;i}} = 0, \quad
  D^{i}\stackrel{(1)\;\;\;}{\chi_{ij}} = 0.
\end{equation}
The transverse traceless tensor
$\stackrel{(1)\;\;\;}{\chi_{ij}}$ has two independent components
and is called the ``tensor mode'' in the context of cosmological
perturbations.
This transverse traceless part of the metric perturbations is
well known as gravitational waves on a homogeneous isotropic
universe.

%*******************************************************************

Next, we consider to the gauge transformation rules
(\ref{eq:decomposed-gauge-trans-3}) and
(\ref{eq:decomposed-gauge-trans-6}).
From these gauge transformation rules, we can easily see that
the variable defined by
\begin{eqnarray}
  a^{2} \stackrel{(1)\;\;}{\nu_{i}} &:=& h_{(V)i} 
  - \left(\partial_{\eta} - 2 {\cal H} \right) \left( a^{2}h_{(TV)i} \right) 
  \nonumber\\
  &=& h_{(V)i} - a^{2}\partial_{\eta}h_{(TV)i}
  \label{eq:vector-mode-gauge-inv-def}
\end{eqnarray}
is gauge invariant.
The gauge invariant variable $\nu_{i}$ is called a ``vector
mode'' in the context of cosmological perturbations. 
It satisfies the equation
\begin{equation}
  D^{i}\stackrel{(1)\;\;}{\nu_{i}} = 0
  \label{eq:vector-mode-divergence-free}
\end{equation}
from the divergenceless property of the variables $h_{(V)i}$ and
$h_{(TV)i}$.
Equation (\ref{eq:vector-mode-divergence-free}) implies that the
vector mode $\stackrel{(1)\;\;}{\nu_{i}}$ includes two
independent components.

%*******************************************************************

In addition to the vector and tensor mode of the perturbation,
there are two scalar modes in the linear-order metric
perturbations $h_{ab}$.
To see this, we first consider the gauge transformation rules
(\ref{eq:decomposed-gauge-trans-2}) and
(\ref{eq:decomposed-gauge-trans-5}).
From these transformation rules, the variable defined by
\begin{equation}
  \bar{X}_{\eta} :=
  h_{(VL)} 
  - \frac{1}{2}\left( \partial_{\eta} - 2 {\cal H} \right) 
  \left( a^{2} h_{(TL)} \right)
  = 
  h_{(VL)} 
  - \frac{1}{2} a^{2} \partial_{\eta} h_{(TL)} 
  \label{eq:barXeta-def}
\end{equation}
is transformed as 
\begin{eqnarray}
  {}_{\cal Y}\!\bar{X}_{\eta} - {}_{\cal X}\!\bar{X}_{\eta}
  &=&
  {}_{\cal Y}\!h_{(VL)} - {}_{\cal X}\!h_{(VL)}
  - \frac{1}{2}\left( \partial_{\eta} - 2 {\cal H} \right)
  \left(
    a^{2} \left(
      {}_{\cal Y}\!h_{(TL)} - {}_{\cal X}\!h_{(TL)}
    \right)
  \right)
  \nonumber\\ 
  &=&
  \partial_{\eta} \xi_{(L)} + \xi_{\eta} - 2 {\cal H} \xi_{(L)}
  - \frac{1}{2}\left( \partial_{\eta} - 2 {\cal H} \right)
  \left( 2 \xi_{(L)} \right).
  \nonumber\\ 
  &=&
  \xi_{\eta}.
  \label{eq:barXtau-gauge-trans}
\end{eqnarray}
Using $\bar{X}_{\eta}$ and inspecting the gauge transformation
rule (\ref{eq:linear-gauge-trans-metric-3+1-tautau}), we easily
find that the variable $\stackrel{(1)}{\Phi}$ defined by 
\begin{equation}
  - 2 a^{2} \stackrel{(1)}{\Phi} 
  := 
  h_{\eta\eta} - 2 \left( \partial_{\eta} - {\cal H} \right) \bar{X}_{\eta}
\end{equation}
is gauge invariant.
Further, from gauge transformation rules
(\ref{eq:decomposed-gauge-trans-4}),
(\ref{eq:decomposed-gauge-trans-5}), and
(\ref{eq:barXtau-gauge-trans}), the variable
$\stackrel{(1)}{\Psi}$ defined by 
\begin{equation}
  - 2 a^{2} \stackrel{(1)}{\Psi} := 
  a^{2} \left( h_{(L)} - \frac{1}{3}\Delta h_{(TL)} \right)
  + 2 {\cal H} \bar{X}_{\eta} 
\end{equation}
is gauge invariant.
The two scalar functions $\stackrel{(1)}{\Phi}$ and
$\stackrel{(1)}{\Psi}$ are called ``scalar perturbations'' in
the context of cosmological perturbations.

%*******************************************************************

Thus, we have six components of gauge invariant variables: two
components of the tensor mode, two components of the vector
mode, and two scalar modes. 
Since the metric perturbation $h_{ab}$ has ten components, there
are four remaining components, which are the components of the
gauge variant part of the metric perturbation.

%*******************************************************************

Because we already have all gauge invariant variables, we can
specify the variant part $X_{a}$ of the metric perturbation
$h_{ab}$ as in Eq.~(\ref{eq:linear-metric-decomp}).
Using the gauge invariant variables $\Phi$, $\Psi$,
$\nu_{i}$, and $\chi_{ij}$, the components of the metric
perturbation $h_{ab}$ are given by
\begin{eqnarray}
  \label{eq:htautau-gaugeinv-decomp}
  h_{\eta\eta} 
  &=& 
  - 2 a^{2} \stackrel{(1)}{\Phi}
  + 2 \left( \partial_{\eta} - {\cal H} \right) \bar{X}_{\eta}
  , \\
  h_{\eta i}
  &=& 
    a^{2} \stackrel{(1)\;\;}{\nu_{i}} 
  + a^{2} \partial_{\eta} h_{(TV)i}
  + D_{i} h_{(VL)}
  \label{eq:htaui-gaugeinv-decomp}
  ,\\
  h_{ij} 
  &=&
  - 2 a^{2} \stackrel{(1)}{\Psi} \gamma_{ij}
  + a^{2} \stackrel{(1)\;\;}{\chi_{ij}}
  + a^{2} D_{i}D_{j} h_{(TL)}
  - 2 {\cal H} \bar{X}_{\eta} \gamma_{ij}
  + 2 a^{2} D_{(i}h_{(TV)j)}.
  \label{eq:hij-gaugeinv-decomp}
\end{eqnarray}
On the other hand, in terms of the 3+1 decomposition, the
components of (\ref{eq:linear-metric-decomp}) are given by
\begin{eqnarray}
  h_{\eta\eta} &=& {\cal H}_{\eta\eta} 
  + 2\left(\partial_{\eta} - {\cal H}\right)X_{\eta}
  , \\
  h_{\eta i} &=& {\cal H}_{\eta i} 
  + D_{i}X_{\eta} + \partial_{\eta} X_{i} - 2 {\cal H} X_{i}
  , \\
  h_{ij} &=&
  {\cal H}_{ij} 
  + 2 D_{(i}X_{j)}
  - 2 {\cal H} \gamma_{ij} X_{\eta}.
\end{eqnarray}
Here, ${\cal H}_{ab}$ is gauge invariant, and its components
should be identified by
\begin{eqnarray}
  \label{eq:linear-order-metric-gauge-inv-def}
  {\cal H}_{\eta\eta} := - 2 a^{2} \stackrel{(1)}{\Phi}, \quad
  {\cal H}_{\eta i} := a^{2} \stackrel{(1)\;\;}{\nu_{i}}, \quad
  {\cal H}_{ij} := - 2 a^{2} \stackrel{(1)}{\Psi} \gamma_{ij} 
  + a^{2} \stackrel{(1)\;\;}{\chi_{ij}}.
\end{eqnarray}
Then, we obtain the equations for $X_{a}$:
\begin{eqnarray}
  \label{eq:Xa-eq-1}
  2\left(\partial_{\eta} - {\cal H}\right)X_{\eta} &=& 
  2\left(\partial_{\eta} - {\cal H} \right)\bar{X}_{\eta}, \\
  \label{eq:Xa-eq-2}
  D_{i}X_{\eta} + \left(\partial_{\eta} - 2{\cal H}\right) X_{i}
  &=&
  a^{2} \partial_{\eta} h_{(TV)i}
  + D_{i} h_{(VL)}
  , \\
  \label{eq:Xa-eq-3}
  2 D_{(i}X_{j)}
  - 2 {\cal H} \gamma_{ij} X_{\eta}
  &=&
      a^{2} D_{i}D_{j} h_{(TL)}
  - 2 {\cal H} \bar{X}_{\eta} \gamma_{ij}
  + 2 a^{2} D_{(i}h_{(TV)j)}.
\end{eqnarray}
Equation (\ref{eq:Xa-eq-1}) yields
\begin{equation}
  X_{\eta} = \bar{X}_{\eta} + a \bar{C}_{\eta},
  \label{eq:Xtau-intermediate-sol}
\end{equation}
where $\bar{C}_{\eta}$ is a scalar function satisfying the
equation $\partial_{\eta}\bar{C}_{\eta}=0$.
Substituting (\ref{eq:Xtau-intermediate-sol}) and
(\ref{eq:barXeta-def}) into (\ref{eq:Xa-eq-2}), we obtain
\begin{eqnarray}
  X_{i} 
  &=&
  a^{2} \left(
      h_{(TV)i}
    + \frac{1}{2} D_{i}h_{(TL)}
  \right)
  - D_{i}\bar{C}_{\eta} a^{2} \int \frac{d\eta}{a}
  + a^{2} \bar{C}_{i},
  \label{eq:Xi-intermediate-sol}
\end{eqnarray}
where $\bar{C}_{i}$ is the vector field satisfying the condition
$\partial_{\eta}\bar{C}_{i}=0$. 
Substituting (\ref{eq:Xtau-intermediate-sol}) and
(\ref{eq:Xi-intermediate-sol}) into (\ref{eq:Xa-eq-3}), we
obtain
\begin{equation}
    a^{2} D_{(i}\bar{C}_{j)}
  - D_{i}D_{j}\bar{C}_{\eta} a^{2} \int \frac{d\eta}{a}
  - {\cal H} \gamma_{ij} a \bar{C}_{\eta}
  =
  0.
  \label{eq:Xa-eq-3-reduced}
\end{equation}
Thus, we have found that the gauge variant part of the metric
perturbation $X_{a}$ is given by
\begin{eqnarray}
  X_{\eta} &=& \bar{X}_{\eta} + C_{\eta}
  = h_{(VL)} - \frac{1}{2} a^{2}\partial_{\tau}h_{(TL)} 
  + C_{\eta},
  \label{eq:Xeta-def}
  , \\
  X_{i} &=&  
  a^{2} \left(
      h_{(TV)i}
    + \frac{1}{2} D_{i}h_{(TL)}
  \right)
  + C_{i},
  \label{eq:Xi-def}
\end{eqnarray}
where $C_{\eta}$ and $C_{i}$ are defined by 
\begin{eqnarray}
  \label{eq:Ceta-def}
  C_{\eta} &:=& a \bar{C}_{\eta}
  , \\
  C_{i} &:=&
  - D_{i} \bar{C}_{\eta} a^{2} \int \frac{d\eta}{a}
  + a^{2} \bar{C}_{i}
  ,
  \label{eq:Ci-def}
\end{eqnarray}
and $\partial_{\eta}\bar{C}_{\eta} = 0 = \partial_{\eta}\bar{C}_{i}$.

%*******************************************************************

Through Eqs.~(\ref{eq:Ceta-def}) and (\ref{eq:Ci-def}), with the
constraint (\ref{eq:Xa-eq-3-reduced}), it is easy to confirm
that the vector field defined by 
\begin{equation}
  \label{eq:Ca-def}
  C_{a} := C_{\eta} (d\eta)_{a} + C_{i} (dx^{i})_{a}
\end{equation}
is a Killing vector on the background spacetime 
${\cal M}_{0}$.
Actually, it is readily shown that 
\begin{eqnarray}
  2\nabla_{(\eta}C_{\eta)}
  =
  2 \left(\partial_{\eta} - {\cal H}\right)C_{\eta}
  =
  0
  ,
\end{eqnarray}
due to the definition (\ref{eq:Ceta-def}) and
$\partial_{\eta}\bar{C}_{\eta} = 0$.
Further, from the definition (\ref{eq:Ci-def}), we can easily
confirm that
\begin{eqnarray}
  2\nabla_{(\eta}C_{i)} 
  = 
  \partial_{\eta}C_{i} + D_{i}C_{\eta} - 2 {\cal H} C_{i}
  =
  0.
\end{eqnarray}
Finally, the constraint (\ref{eq:Xa-eq-3-reduced}) leads to
\begin{eqnarray}
  2\nabla_{(i}C_{j)} &=& 2 D_{(i}C_{j)} - 2 {\cal H} \gamma_{ij} C_{\eta}
  = 0.
\end{eqnarray}
Thus, we have 
\begin{equation}
  \nabla_{(a}C_{b)} = 0;
\end{equation}
i.e., $C_{a}$ defined by Eq.~(\ref{eq:Ca-def}) is a Killing
vector.
Hence, we have been able to specify the gauge variant part
$X_{a}$ of the linear-order metric perturbation as
\begin{equation}
  X_{a} := X_{\eta}(d\eta)_{a} + X_{i}(dx^{i})_{a}.
  \label{eq:Xa-gauge-var-part-def}
\end{equation}
The relation between the components of the gauge variant part,
$X_{a}$, and the components of the linear-order metric
perturbation, $h_{(VL)}$, $h_{(TL)}$, $h_{(TV)i}$, is determined
up to the degree of freedom of the Killing vector field.
Since $X_{a}$ contributes to the metric perturbation as in
Eq.~(\ref{eq:linear-metric-decomp}), the 
Killing vector field $C_{a}$ in Eqs.~(\ref{eq:Xeta-def}) and
(\ref{eq:Xi-def}) does not contribute to the metric
perturbation.

%*******************************************************************

Further, because we do not consider the kernels of the operators
$\Delta$, $\Delta+2K$, and $\Delta+3K$ as the domain of the
perturbations, the gauge variant part $X_{a}$ of the metric
perturbation does not have the degree of freedom of the Killing
vector field; i.e., the gauge variant part $X_{a}$ is determined
without ambiguity. 
To satisfy Eq.~(\ref{eq:Xa-eq-3-reduced}) for any scale factor
$a$, the components $\bar{C}_{\eta}$ and $\bar{C}_{i}$ should
satisfy the equations
\begin{equation}
  \bar{C}_{\eta} = 0, \quad D_{(i}\bar{C}_{j)} = 0.
\end{equation}
This implies that $\bar{C}_{j}$ is a Killing vector on a
three-dimensional hypersurface $\Sigma(\eta)$. 
As commented just after
Eqs.~(\ref{eq:hetai-decomp-inv-1})--(\ref{eq:hTij-decomp-inv-5}),
it is easily shown that $(\Delta + 2K)\bar{C}_{i}=0$.
Therefore, the vector field $\bar{C}_{i}$ does not belong to the
domain of perturbations considered here.
Thus, we have $\bar{C}_{j}=0$.
Hence, we conclude that we can determine the gauge variant part 
$X_{a}$ without ambiguity.
Of course, it might be possible to include the Killing fields in
our consideration by extending the domain of the perturbations.
Separate treatments are necessary to do this, as mentioned above.

%*******************************************************************

Finally, we check the transformation rules for the vector field
$X_{a}$ under the gauge transformation 
${\cal X}_{\lambda}\rightarrow{\cal Y}_{\lambda}$.
Because the component $X_{\eta}$ of the vector $X_{a}$ is
transformed as
\begin{equation}
  {}_{\cal Y}\!X_{\eta} - {}_{\cal X}\!X_{\eta} = \xi_{\eta},   
  \label{eq:Xtau-gauge-trans}
\end{equation}
as noted in Eq.~(\ref{eq:barXtau-gauge-trans}).
From Eq.~(\ref{eq:Xi-def}) with $C_{i}=0$ and the gauge
transformation rules (\ref{eq:decomposed-gauge-trans-5}) and
(\ref{eq:decomposed-gauge-trans-6}), the gauge transformation
rule of the component $X_{i}$ defined by Eq.~(\ref{eq:Xi-def})
is given by
\begin{eqnarray}
  {}_{{\cal Y}}\!X_{i} - {}_{{\cal X}}\!X_{i}
  &=& 
  a^{2} {}_{{\cal Y}}\!h_{(TV)i} 
  - a^{2} {}_{{\cal X}}\!h_{(TV)i} 
  + \frac{1}{2} D_{i} \left(
    a^{2} {}_{{\cal Y}}\!h_{(TL)}
    - a^{2} {}_{{\cal X}}\!h_{(TL)}
  \right)
  \nonumber\\
  &=&
  \xi_{(T)i}
  +
  D_{i}\xi_{(L)}
  = \xi_{i}.
  \label{eq:Xi-gauge-trans}
\end{eqnarray}
Together with the transformation rules
(\ref{eq:Xtau-gauge-trans}) and (\ref{eq:Xi-gauge-trans}), the
vector field $X_{a}$ defined by
Eq.~(\ref{eq:Xa-gauge-var-part-def}) is transformed as
\begin{eqnarray}
  {}_{\cal Y}\!X_{a} - {}_{\cal X}\!X_{a} = \xi_{a}.   
\end{eqnarray}
This shows that $X_{a}$ is the gauge variant part of the
metric perturbation in the decomposition
(\ref{eq:linear-metric-decomp}).

%*******************************************************************

Thus, we know the procedure to find the gauge invariant variable
${\cal H}_{ab}$ and the gauge variant variable $X_{a}$ in the
case of cosmological perturbations. 
We have confirmed the important premise of the general framework
of the second-order perturbation theory reviewed in
\S\ref{sec:General-framework-of-the-gauge-invariant-perturbation-theory}.
Hence, we can apply this general framework for the second-order
gauge invariant perturbation theory presented in KN2003 and KN2005
to cosmological perturbations.

%*******************************************************************

%%%%%%%%%%%%%%%%%%%%%%%%%%%%%%%%%%%%%%%%%%%%%%%%%%%%%%%%%%%%%%%%%%%%%
\subsection{Second order metric perturbations}
\label{sec:Second-order-metric-perturbations}

%*******************************************************************

Here we consider second-order metric perturbations.
We expand the metric $\bar{g}_{ab}$ on the physical spacetime
${\cal M}_{\lambda}$ as Eq.~(\ref{eq:metric-expansion}).
According to the gauge transformation rule
(\ref{eq:Bruni-49-one}), the second-order metric perturbation
$l_{ab}$ is transformed as
\begin{eqnarray}
  \label{eq:second-order-gauge-trans-of-metric}
  {}_{\cal Y}\!l_{ab} - {}_{\cal X}\!l_{ab} &=& 
  2 {\pounds}_{\xi_{(1)}} {}_{\cal X}\!h_{ab} 
  +\left\{{\pounds}_{\xi_{(2)}}+{\pounds}_{\xi_{(1)}}^{2}\right\} g_{ab}
\end{eqnarray}
under the gauge transformation 
$\Phi_{\lambda}=({\cal X}_{\lambda})^{-1}\circ{\cal Y}_{\lambda}:{\cal X}_{\lambda}\rightarrow{\cal Y}_{\lambda}$.
As shown in \S\ref{sec:First-order-metric-perturbations}, the
first-order metric perturbation $h_{ab}$ is decomposed in the
form (\ref{eq:linear-metric-decomp}).
Using this important fact, as shown in KN2003, the second-order
metric perturbation $l_{ab}$ can be decomposed as
Eq.~(\ref{eq:H-ab-in-gauge-X-def-second-1}).
Here, we demonstrate this.

%*******************************************************************

Inspecting the gauge transformation rule
(\ref{eq:second-order-gauge-trans-of-metric}), we first
introduce the variable $\hat{L}_{ab}$ defined by
\begin{equation}
  \hat{L}_{ab}
  :=
  l_{ab}
  - 2 {\pounds}_{X} h_{ab}
  + {\pounds}_{X}^{2} g_{ab}.
\end{equation}
Under the gauge transformation 
$\Phi_{\lambda}=({\cal X}_{\lambda})^{-1}\circ{\cal Y}_{\lambda}:{\cal X}_{\lambda}\rightarrow{\cal Y}_{\lambda}$,
the variable $\hat{L}_{ab}$ is transformed as 
\begin{eqnarray}
  {}_{\;\cal Y}\!\hat{L}_{ab} - {}_{\;\cal X}\!\hat{L}_{ab}
  &=& 
  {}_{\;\cal Y}\!l_{ab}
  - 2 {\pounds}_{{}_{{\cal Y}}\!X} {}_{{\cal Y}}\!h_{ab}
  + {\pounds}_{{}_{{\cal Y}}\!X}^{2} g_{ab}
  \nonumber\\
  && \quad
  - {}_{\cal X}\!l_{ab}
  + 2 {\pounds}_{{}_{\cal X}\!X} {}_{\cal X}\!h_{ab}
  - {\pounds}_{{}_{\cal X}\!X}^{2} g_{ab}
  \\
  &=& 
  2 {\pounds}_{\xi_{(1)}} {}_{\cal X}\!h_{ab} 
  +\left\{{\pounds}_{\xi_{(2)}}+{\pounds}_{\xi_{(1)}}^{2}\right\} g_{ab}
  \nonumber\\
  && \quad
  - 2 {\pounds}_{{}_{{\cal X}}\!X+\xi_{(1)}} \left(
    {}_{{\cal X}}\!h_{ab} + {\pounds}_{\xi_{(1)}}g_{ab}
  \right)
  + {\pounds}_{{}_{{\cal X}}\!X+\xi_{(1)}}^{2} g_{ab}
  \nonumber\\
  && \quad
  - {}_{\cal X}\!l_{ab}
  + 2 {\pounds}_{{}_{\cal X}\!X} {}_{\cal X}\!h_{ab}
  - {\pounds}_{{}_{\cal X}\!X}^{2} g_{ab}
  \nonumber\\
  &=& 
  {\pounds}_{\sigma} g_{ab} ,
  \label{eq:4.67}
\end{eqnarray}
where
\begin{equation}
  \sigma^{a} := \xi_{(2)}^{a} + [\xi_{(1)},X]^{a}.
\end{equation}
The gauge transformation rule (\ref{eq:4.67}) is identical to
that for a linear metric perturbation. 
Therefore, we may apply the procedure to find the gauge
invariant and variant variables of linear-order metric
perturbations to the decomposition of the components of the
variable $\hat{L}_{ab}$.
Arguments completely analogous to those used in the case of
linear-order metric perturbation show that the variable
$\hat{L}_{ab}$ can be decomposed as
\begin{eqnarray}
  \hat{L}_{ab} = {\cal L}_{ab} + {\pounds}_{Y}g_{ab},
\end{eqnarray}
where ${\cal L}_{ab}$ is the gauge invariant part of the
variable $\hat{L}_{ab}$, or equivalently, of the second-order
metric perturbation $l_{ab}$, and $Y^{a}$ is the gauge variant
part of $\hat{L}_{ab}$, i.e., the gauge variant part of $l_{ab}$. 
Under the gauge transformation 
$\Phi_{\lambda}=({\cal X}_{\lambda})^{-1}\circ{\cal Y}_{\lambda}$,
the variables ${\cal L}_{ab}$ and $Y^{a}$ are transformed as  
\begin{equation}
  {}_{\;\cal Y}\!{\cal L}_{ab} - {}_{\;\cal X}\!{\cal L}_{ab} = 0, 
  \quad
  {}_{\;\cal Y}\!Y_{a} - {}_{\;\cal Y}\!Y_{a} = \sigma_{a},
\end{equation}
respectively.
Thus, we have reached the decomposition
(\ref{eq:H-ab-in-gauge-X-def-second-1}) of the second-order
metric perturbation $l_{ab}$ into the gauge variant and gauge
invariant parts. 
Following to the same argument as in the linear case, the
components of the gauge invariant variables ${\cal L}_{ab}$ are
given by
\begin{eqnarray}
  \label{eq:second-order-gauge-inv-metrc-pert-components}
  {\cal L}_{ab}
  &=& 
  - 2 a^{2} \stackrel{(2)}{\Phi} (d\eta)_{a}(d\eta)_{b}
  + 2 a^{2} \stackrel{(2)\;\;}{\nu_{i}} (d\eta)_{(a}(dx^{i})_{b)}
  \nonumber\\
  && \quad
  + a^{2} 
  \left( - 2 \stackrel{(2)}{\Psi} \gamma_{ij} 
    + \stackrel{(2)\;\;\;\;}{{\chi}_{ij}} \right)
  (dx^{i})_{a}(dx^{j})_{b},
\end{eqnarray}
where $\stackrel{(2)}{\nu}_{i}$ and $\stackrel{(2)\;\;\;\;}{\chi_{ij}}$ satisfy the equations
\begin{eqnarray}
  && D^{i}\stackrel{(2)\;\;}{\nu_{i}} = 0, \quad
  \stackrel{(2)\;\;\;\;}{\chi^{i}_{\;\;i}} = 0, \quad
   D^{i}\stackrel{(2)\;\;\;\;}{\chi_{ij}} = 0.
\end{eqnarray}
The gauge invariant variables $\stackrel{(2)}{\Phi}$ and
$\stackrel{(2)}{\Psi}$ are the scalar mode perturbations of
second order, and $\stackrel{(2)\;\;}{\nu_{i}}$ and 
$\stackrel{(2)\;\;\;\;}{\chi_{ij}}$ are the second-order vector
and tensor modes of the metric perturbations, respectively.

%*******************************************************************

%%%%%%%%%%%%%%%%%%%%%%%%%%%%%%%%%%%%%%%%%%%%%%%%%%%%%%%%%%%%%%%%%%%%%
\subsection{Matter perturbations}
\label{sec:Matter-perturbations}

Since we have obtained the first- and the second-order gauge
variant parts, $X_{a}$ and $Y_{a}$, of the metric perturbation,
we can define the gauge invariant variables for an arbitrary field
$Q$, except for the metric by following the definitions
(\ref{eq:matter-gauge-inv-def-1.0}) and
(\ref{eq:matter-gauge-inv-def-2.0})\cite{kouchan-gauge-inv}.
Here, we consider the first- and second-order gauge invariant
variables for the perturbations of the perfect fluid components,
the single scalar field, and their energy-momentum tensors.

%%%%%%%%%%%%%%%%%%%%%%%%%%%%%%%%%%%%%%%%%%%%%%%%%%%%%%%%%%%%%%%%%%%%%
\subsubsection{Perfect fluid}
\label{sec:Matter-perturbations-perfect-fluid}

%*******************************************************************

First, we consider the perturbation of a perfect fluid.
As shown in Eq.~(\ref{eq:energy-momentum-perfect-fluid}), the
total energy-momentum tenor of the fluid is characterized by the
energy density $\bar{\epsilon}$, the pressure $\bar{p}$, and the
four-velocity $\bar{u}^{a}$:
\begin{equation}
  \bar{T}_{a}^{\;\;b} = (\bar{\epsilon} + \bar{p}) \bar{u}_{a} \bar{u}^{b} 
  + \bar{p} \delta_{a}^{\;\;b}.
  \label{eq:MFB-5.2-again}
\end{equation}
Of course, this energy-momentum tensor is the representation on
the physical spacetime ${\cal M}_{\lambda}$, but we can regard
this equation to be the representation on the background
spacetime ${\cal M}_{0}$ which is pulled back by an appropriate
gauge choice ${\cal X}_{\lambda}$.
The background value of the energy momentum tensor
(\ref{eq:MFB-5.2-again}) is given by
Eqs.(\ref{eq:energy-momentum-perfect-fluid})--(\ref{eq:energy-momentum-for-velocity-homogeneous})
in \S\ref{sec:Cosmological-Background-spacetime}.
In addition to the components of Eq.~(\ref{eq:MFB-5.2-again}),
we may also include the anisotropic stress as non-diagonal
space-space components of the energy-momentum tensor, as
phenomenology. 
Because we can always extend our arguments to those including
the anisotropic stress, we ignore it here, for simplicity.

%*******************************************************************

Now, we consider perturbations of the fluid components of
the energy-momentum tensor (\ref{eq:MFB-5.2-again}):
\begin{eqnarray}
  \bar{\epsilon} &:=& \epsilon 
  + \lambda \stackrel{(1)}{\epsilon}
  + \frac{1}{2} \lambda^{2} \stackrel{(2)}{\epsilon}
  + O(\lambda^{3})
  \label{eq:kouchan-16.2}
  , \\
  \bar{p} &:=&
  p + \lambda\stackrel{(1)}{p} + \frac{1}{2}
  \lambda^{2} \stackrel{(2)}{p}
  + O(\lambda^{3})
  \label{eq:kouchan-16.3}
  , \\
  \bar{u}_{a} 
  &:=& 
  u_{a} + \lambda \stackrel{(1)}{(u_{a})}
  + \frac{1}{2} \lambda^{2} \stackrel{(2)}{(u_{a})}
  + O(\lambda^{3})
  .
  \label{eq:kouchan-16.4}
\end{eqnarray}
The fluid four-velocities $\bar{u}_{a}$ on the physical
spacetime and $u_{a}$ on the background spacetime satisfy the
normalization condition
\begin{eqnarray}
  \bar{g}^{ab}\bar{u}_{a}\bar{u}_{a} = g^{ab}u_{a}u_{b} = -1.
  \label{eq:four-velocity-norm-full}
\end{eqnarray}
The perturbative expansion of the normalization conditions
(\ref{eq:four-velocity-norm-full}) gives the constraints for the
components of the first- and second-order perturbative
four-velocities $\stackrel{(1)}{(u_{a})}$ and
$\stackrel{(2)}{(u_{a})}$.
The perturbative expansion of the normalization condition
(\ref{eq:four-velocity-norm-full}) to second order gives the
normalization condition at each order:
\begin{eqnarray}
  u^{a} \stackrel{(1)}{(u_{a})} 
  &=& \frac{1}{2}h^{ab} u_{a} u_{b}
  \label{eq:four-velocity-normalization-perturbation}
  , \\
  u^{a} \stackrel{(2)}{(u_{a})}
  &=&
      h^{ab} u_{a} \stackrel{(1)}{(u_{b})}
  -   g^{ab} \stackrel{(1)}{(u_{a})} \stackrel{(1)}{(u_{b})}
  -   h^{ac}h_{c}^{\;\;b} u_{a} u_{b} 
  + \frac{1}{2} l^{ab} u_{a} u_{b}. 
  \label{eq:second-order-four-velocity-normalization-perturbation}
\end{eqnarray}
We also consider the perturbation of the four-velocity
$\bar{u}^{a}$ as
\begin{equation}
  \bar{u}^{a} = u^{a} + \lambda \stackrel{(1)}{(u^{a})} 
  + \frac{1}{2} \lambda^{2} \stackrel{(2)}{(u^{a})}.
\end{equation}
The perturbative expansion of the equation
$\bar{u}^{a}=\bar{g}^{ab}\bar{u}_{b}$ leads to
\begin{eqnarray}
  \label{eq:four-velo-up-first}
  \stackrel{(1)}{(u^{a})}
  &=& g^{ab} \stackrel{(1)}{(u_{b})} - h^{ab} u_{b}
  , \\
  \label{eq:four-velo-up-second}
  \stackrel{(2)}{(u^{a})}
  &=&
  g^{ab} \stackrel{(2)}{(u_{b})}
  - 2 h^{ab} \stackrel{(1)}{(u_{b})}
  + ( 2 h^{ac}h_{c}^{\;\;b} - l^{ab}) u_{b} .
\end{eqnarray}
Further, the first-order perturbation
(\ref{eq:four-velocity-normalization-perturbation}) of the
normalization condition (\ref{eq:four-velocity-norm-full}) is
given by
\begin{equation}
  \stackrel{(1)}{(u^{a})}u_{a} + u^{a}\stackrel{(1)}{(u_{a})} = 0.
\end{equation}

%*******************************************************************

Next, we define the gauge invariant variable for the
perturbation of the fluid components $\bar{\epsilon}$,
$\bar{p}$, and $\bar{u}_{a}$.
Following the definitions (\ref{eq:matter-gauge-inv-def-1.0})
and (\ref{eq:matter-gauge-inv-def-2.0}) of the gauge invariant
variable for an arbitrary matter field\cite{kouchan-gauge-inv},
we define the variables
\begin{eqnarray}
  \label{eq:kouchan-016.13}
  \stackrel{(1)}{{\cal E}} 
  &:=& \stackrel{(1)}{\epsilon} - {\pounds}_{X}\epsilon
  , \\
  \label{eq:kouchan-016.14}
  \stackrel{(1)}{{\cal P}}
  &:=& \stackrel{(1)}{p} - {\pounds}_{X}p
  , \\
  \label{eq:kouchan-016.15}
  \stackrel{(1)}{{\cal U}_{a}}
  &:=& \stackrel{(1)}{(u_{a})} - {\pounds}_{X}u_{a}
  , \\
  \label{eq:kouchan-016.16}
  \stackrel{(2)}{{\cal E}} 
  &:=& \stackrel{(2)}{\epsilon} 
  - 2 {\pounds}_{X} \stackrel{(1)}{\epsilon}
  - \left\{
    {\pounds}_{Y}
    -{\pounds}_{X}^{2}
  \right\} \epsilon
  , \\
  \label{eq:kouchan-016.17}
  \stackrel{(2)}{{\cal P}}
  &:=& \stackrel{(2)}{p}
  - 2 {\pounds}_{X} \stackrel{(1)}{p}
  - \left\{
    {\pounds}_{Y}
    -{\pounds}_{X}^{2}
  \right\} p
  , \\
  \label{eq:kouchan-016.18}
  \stackrel{(2)}{{\cal U}_{a}}
  &:=& \stackrel{(2)}{(u_{a})}
  - 2 {\pounds}_{X} \stackrel{(1)}{u_{a}}
  - \left\{
    {\pounds}_{Y}
    -{\pounds}_{X}^{2}
  \right\} u_{a}
  ,
\end{eqnarray}
where the vector fields $X_{a}$ and $Y_{a}$ are the gauge
variant parts of the first- and second-order metric
perturbations, respectively, which were defined in \S\S
\ref{sec:Second-order-metric-perturbations} and
\ref{sec:First-order-metric-perturbations}.

%*******************************************************************

The first-order perturbation
(\ref{eq:four-velocity-normalization-perturbation}) of the
normalization condition (\ref{eq:four-velocity-norm-full}) is
given by
\begin{eqnarray}
  \label{eq:four-vel-norm-first-pert-gauge-inv-1}
  u^{a} \stackrel{(1)}{{\cal U}_{a}}
  &=& \frac{1}{2}{\cal H}_{ab} u^{a} u^{b}
  - {\pounds}_{X}\left(\frac{1}{2}u^{a}u_{a}\right)
  \\
  \label{eq:four-vel-norm-first-pert-gauge-inv-2}
  &=& \frac{1}{2}{\cal H}_{ab} u^{a} u^{b}
  ,
\end{eqnarray}
while the second-order perturbation
(\ref{eq:second-order-four-velocity-normalization-perturbation})
of Eq.~(\ref{eq:four-velocity-norm-full}) is given by
\begin{eqnarray}
  u^{a} \stackrel{(2)}{{\cal U}_{a}}
  &=&
             2  {\cal H}_{ab} u^{a} g^{bc} \stackrel{(1)}{{\cal U}_{c}}
  -             g^{ab} \stackrel{(1)}{{\cal U}_{a}}\stackrel{(1)}{{\cal U}_{b}}
  -             {\cal H}_{ac}{\cal H}_{db} g^{dc} u_{a} u_{b} 
  + \frac{1}{2} {\cal L}_{ab} u^{a} u^{b}
  \nonumber\\
  &&
  -          2  {\pounds}_{X} \left(
                u^{a} \stackrel{(1)}{(u_{a})} - \frac{1}{2} h_{ab} u^{a} u^{b}
                \right)
  -             \left( {\pounds}_{Y} - {\pounds}_{X}^{2} \right)
                \left(\frac{1}{2} u_{a} u^{a}\right)
  \label{eq:four-vel-norm-second-pert-gauge-inv-1}
  \\
  &=&
             2  {\cal H}_{ab} u^{a} g^{bc} \stackrel{(1)}{{\cal U}_{c}}
  -             g^{ab} \stackrel{(1)}{{\cal U}_{a}}\stackrel{(1)}{{\cal U}_{b}}
  -             {\cal H}_{ac}{\cal H}_{db} g^{dc} u_{a} u_{b} 
  + \frac{1}{2} {\cal L}_{ab} u^{a} u^{b},
  \label{eq:four-vel-norm-second-pert-gauge-inv-2}
\end{eqnarray}
where we have used Eqs.~(\ref{eq:four-velocity-norm-full}) and
(\ref{eq:four-velocity-normalization-perturbation}).
We note that (\ref{eq:four-vel-norm-first-pert-gauge-inv-1}) and
(\ref{eq:four-vel-norm-second-pert-gauge-inv-1}) have the same
forms as definitions (\ref{eq:matter-gauge-inv-def-1.0}) and
(\ref{eq:matter-gauge-inv-def-2.0}) of the first- and second-
order gauge invariant variables for an arbitrary tensor field,
respectively.
These are natural results, because
Eqs.~(\ref{eq:four-vel-norm-first-pert-gauge-inv-1}) and
(\ref{eq:four-vel-norm-second-pert-gauge-inv-1}) are the results
of the first- and second-order perturbative expansions of the
variable $(1/2)\bar{u}^{a}\bar{u}_{a}$.

%*******************************************************************

We also decompose the first- and second-order perturbations
of the four-velocity $\bar{u}^{a}$, which are given by
(\ref{eq:four-velo-up-first}) and
(\ref{eq:four-velo-up-second}), respectively, into gauge
invariant and variant parts: 
\begin{eqnarray}
  \label{eq:kouchan-16.40}
  \stackrel{(1)}{(u^{a})}
  &=&
  g^{ab} \stackrel{(1)}{{\cal U}_{b}} 
  - {\cal H}^{ab} u_{b}
  + {\pounds}_{X}u^{a}
  ,
  \\
  \label{eq:kouchan-16.41}
  \stackrel{(2)}{(u^{a})}
  &=&
    g^{ab} \stackrel{(2)}{{\cal U}_{b}}
  - 2 {\cal H}^{ab} \stackrel{(1)}{{\cal U}_{b}}
  + 2 {\cal H}^{ac}{\cal H}_{cb} u^{b}
  - {\cal L}^{a}_{\;\;b} u^{b}
  \nonumber\\
  &&
  + 2 {\pounds}_{X} \left( 
    g^{ab} \stackrel{(1)}{(u_{b})} - h^{ab} u_{b}
  \right)
  + \left({\pounds}_{Y} - {\pounds}_{X}^{2}\right) \left(u^{a}\right).
\end{eqnarray}
We note that these expressions have the same forms as
Eqs.~(\ref{eq:matter-gauge-inv-decomp-1.0}) and
(\ref{eq:matter-gauge-inv-decomp-2.0}).

%*******************************************************************

Next, we consider the expansion of the energy-momentum tensor
(\ref{eq:MFB-5.2-again}).
Substituting the expansion
(\ref{eq:kouchan-16.2})--(\ref{eq:kouchan-16.4}) of the fluid
components $\bar{\epsilon}$, $\bar{p}$, and $\bar{u}_{a}$ into
(\ref{eq:MFB-5.2-again}), we have the perturbative form of the
energy-momentum tensor:
\begin{eqnarray}
  \bar{T}_{a}^{\;\;b}
  &=:& 
  T_{a}^{\;\;b}
  + \lambda {}^{(1)}\!\left(T_{a}^{\;\;b}\right)
  + \frac{1}{2} \lambda^{2} {}^{(2)}\!\left(T_{a}^{\;\;b}\right)
  + O(\lambda^{3}),
\end{eqnarray}
where
\begin{eqnarray}
  {}^{(1)}\!\left(T_{a}^{\;\;b}\right)
  &=&
  \left( \stackrel{(1)}{\epsilon} + \stackrel{(1)}{p} \right) u_{a} u^{b}
  + \left( \epsilon + p \right) u_{a} \stackrel{(1)}{(u^{b})}
  + \left( \epsilon + p \right) \stackrel{(1)}{(u_{a})} u^{b}
  + \stackrel{(1)}{p} \delta_{a}^{\;\;b}
  , \\
  {}^{(2)}\!\left(T_{a}^{\;\;b}\right)
  &=&
  \left( 
    \stackrel{(2)}{\epsilon} + \stackrel{(2)}{p}
  \right) 
  u_{a} u^{b}
  + \left( \epsilon + p \right) u_{a} \stackrel{(2)}{(u^{b})}
  + 
  \left(
    \epsilon + p 
  \right) 
  \stackrel{(2)}{(u_{a})}
  u^{b}
  + \stackrel{(2)}{p} \delta_{a}^{\;\;b}
  \nonumber\\
  && 
  + 2 \left( 
    \stackrel{(1)}{\epsilon} + \stackrel{(1)}{p} 
  \right) 
  u_{a}
  \stackrel{(1)}{(u^{b})}
  + 2 
  \left(
    \stackrel{(1)}{\epsilon} + \stackrel{(1)}{p} 
  \right) 
  \stackrel{(1)}{(u_{a})} u^{b}
  \nonumber\\
  && 
  + 2 
  \left(
    \epsilon + p 
  \right) 
  \stackrel{(1)}{(u_{a})}
  \stackrel{(1)}{(u^{b})}.
\end{eqnarray}
Further, we consider the decomposition of the perturbed
energy-momentum tensor into its gauge invariant and variant
parts. 
Using the definitions
(\ref{eq:kouchan-016.13})--(\ref{eq:kouchan-016.15}) of the gauge
invariant variables of the first-order perturbations and
Eq.~(\ref{eq:kouchan-16.40}), the first-order perturbation of
the energy-momentum tensor of a perfect fluid is given by 
\begin{eqnarray}
  {}^{(1)}\!\left(T_{a}^{\;\;b}\right)
  &=:&
  {}^{(1)}\!{\cal T}_{a}^{\;\;b}
  + {\pounds}_{X}T_{a}^{\;\;b},
  \label{eq:first-ene-mon-tensor-decomp}
\end{eqnarray}
where the gauge invariant part of the energy-momentum tensor is
given by
\begin{eqnarray}
  {}^{(1)}\!{\cal T}_{a}^{\;\;b}
  &:=&
  \left(
    \stackrel{(1)}{{\cal E}}
    + \stackrel{(1)}{{\cal P}}
  \right) u_{a} u^{b}
  + \stackrel{(1)}{{\cal P}} \delta_{a}^{\;\;b}
  \nonumber\\
  && \quad
  + \left( \epsilon + p \right) \left(
    u_{a} \stackrel{(1)}{{\cal U}^{b}} 
    - {\cal H}^{bc} u_{c} u_{a} 
    + \stackrel{(1)}{{\cal U}_{a}} u^{b}
  \right).
  \label{eq:first-ene-mon-tensor-gauge-inv-def}
\end{eqnarray}
Similarly, using the definitions
(\ref{eq:kouchan-016.13})--(\ref{eq:kouchan-016.18}) of the gauge
invariant variables of the first- and second-order
perturbations and Eqs.~(\ref{eq:kouchan-16.40}) and
(\ref{eq:kouchan-16.41}), the second-order perturbation of the
energy-momentum tensor of a perfect fluid is given by
\begin{eqnarray}
  {}^{(2)}\!\left(T_{a}^{\;\;b}\right)
  &=:&
  {}^{(2)}\!{\cal T}_{a}^{\;\;b}
  + 2 {\pounds}_{X} {}^{(1)}\!\left(T_{a}^{\;\;b}\right)
  + \left\{
    {\pounds}_{Y}
    - {\pounds}_{X}^{2}
  \right\} T_{a}^{\;\;b},
  \label{eq:second-ene-mon-tensor-decomp}
\end{eqnarray}
where
\begin{eqnarray}
  {}^{(2)}\!{\cal T}_{a}^{\;\;b}
  &:=&
  \left( 
    \stackrel{(2)}{{\cal E}} 
    + \stackrel{(2)}{{\cal P}}
  \right) u_{a} u^{b}
  + 2 \left( 
    \stackrel{(1)}{{\cal E}} + \stackrel{(1)}{{\cal P}} 
  \right) 
  u_{a}
  \left(
    \stackrel{(1)}{{\cal U}^{b}} - {\cal H}^{bc} u_{c}
  \right)
  \nonumber\\
  &&
  + \left( \epsilon + p \right) u_{a} \left(
        g^{bc} \stackrel{(2)}{{\cal U}_{c}}
    - 2 {\cal H}^{bc} \stackrel{(1)}{{\cal U}_{c}}
    + 2 {\cal H}^{bc}{\cal H}_{cd} u^{d}
    -   g^{bc} {\cal L}_{cd} u^{d}
  \right)
  \nonumber\\
  &&
  + 2 
  \left(
    \stackrel{(1)}{{\cal E}} + \stackrel{(1)}{{\cal P}} 
  \right) 
  \stackrel{(1)}{{\cal U}_{a}} u^{b}
  + 2 \left( \epsilon + p \right)
  \stackrel{(1)}{{\cal U}_{a}} \left(
    g^{bc} \stackrel{(1)}{{\cal U}_{c}} 
    - {\cal H}^{bc} u_{c}
  \right)
  \nonumber\\
  &&
  + \left( \epsilon + p \right) \stackrel{(2)}{{\cal U}_{a}} u^{b}
  +   \stackrel{(2)}{{\cal P}} \delta_{a}^{\;\;b}.
  \label{eq:second-ene-mon-tensor-gauge-inv-def}
\end{eqnarray}
Here, again, we have seen that the perturbative expressions
(\ref{eq:first-ene-mon-tensor-decomp}) and
(\ref{eq:second-ene-mon-tensor-decomp}) of the energy-momentum
tensor have the same forms as
Eqs.~(\ref{eq:matter-gauge-inv-decomp-1.0}) and
(\ref{eq:matter-gauge-inv-decomp-2.0}), as expected.
We also note that in the derivation of the expressions
(\ref{eq:first-ene-mon-tensor-decomp})--(\ref{eq:second-ene-mon-tensor-gauge-inv-def}),
we did not explicitly use any background values of the fluid
component and the metric.
Therefore, the expressions
(\ref{eq:first-ene-mon-tensor-decomp})--(\ref{eq:second-ene-mon-tensor-gauge-inv-def})
are valid for any background spacetime.
This implies that the definitions
(\ref{eq:Tab-gauge-inv-def-1.0}) and
(\ref{eq:Tab-gauge-inv-def-2.0}) of the gauge invariant 
variables of the perturbed energy-momentum tensor in
\S\ref{sec:Perturbation-of-the-Einstein-tensor-and-the-Einstein-equations}
are appropriate in the case of a perfect fluid.

%*******************************************************************

Finally, we consider the perturbation of the equation of state
for a fluid. In the generic case, the equation of state of a
fluid is given by 
\begin{eqnarray}
  \bar{p} = \bar{p}(\bar{\epsilon},\bar{S}),
  \label{eq:equation-of-state-generic}
\end{eqnarray}
which gives the relation between the pressure $\bar{p}$, the
energy density $\bar{\epsilon}$, and the entropy $\bar{S}$.
In addition to the perturbative expansions
(\ref{eq:kouchan-16.2}) and (\ref{eq:kouchan-16.3}) for the
energy density and pressure, we consider the perturbative
expansion of the entropy: 
\begin{eqnarray}
  \bar{S} &=& S + \lambda \stackrel{(1)}{S} 
  + \frac{1}{2} \lambda^{2} \stackrel{(2)}{S}
  + O(\lambda^{3}).
  \label{eq:entropy-expansion}
\end{eqnarray}
Hence, the generic equation of state
(\ref{eq:equation-of-state-generic}) is expanded as
\begin{eqnarray}
  p + \lambda \stackrel{(1)}{p} 
  + \frac{1}{2} \lambda^{2} \stackrel{(2)}{p}
  &=&
  \bar{p}\left(
    \epsilon + \lambda \stackrel{(1)}{\epsilon} 
    + \frac{1}{2} \lambda^{2} \stackrel{(2)}{\epsilon}
    ,
    S + \lambda \stackrel{(1)}{S} 
    + \frac{1}{2} \lambda^{2} \stackrel{(2)}{S}
  \right)
  \\
  &=&
  p\left(\epsilon,S\right)
  + \lambda \left(
    \frac{\partial\bar{p}}{\partial\bar{\epsilon}} \stackrel{(1)}{\epsilon} 
    +
    \frac{\partial\bar{p}}{\partial\bar{S}} \stackrel{(1)}{S}
  \right)
  \nonumber\\
  &&
  + \frac{1}{2} \lambda^{2}
  \left(
    \stackrel{(2)}{\epsilon}
    \frac{\partial\bar{p}}{\partial\bar{\epsilon}}(\epsilon,S)
    +
    \stackrel{(1)}{\epsilon}^{2}
    \frac{\partial^{2}\bar{p}}{\partial\bar{\epsilon}^{2}}(\epsilon,S)
    + 2
    \stackrel{(1)}{\epsilon}
    \stackrel{(1)}{S}
    \frac{\partial^{2}\bar{p}}{\partial\bar{\epsilon}\partial\bar{S}}(\epsilon,S)
  \right.
  \nonumber\\
  && \quad\quad\quad\quad
  \left.
    +
    \stackrel{(2)}{S}
    \frac{\partial\bar{p}}{\partial\bar{S}}(\epsilon,S)
    +
    \stackrel{(1)}{S}^{2}
    \frac{\partial\bar{p}^{2}}{\partial\bar{S}^{2}}(\epsilon,S)
  \right).
\end{eqnarray}
Thus, we obtain the equation of state of the first- and
second-order perturbation of the fluid components: 
\begin{eqnarray}
  \label{eq:first-order-equation-of-state-bare}
  \stackrel{(1)}{p} 
  &=&
  \frac{\partial\bar{p}}{\partial\bar{\epsilon}} \stackrel{(1)}{\epsilon} 
  +
  \frac{\partial\bar{p}}{\partial\bar{S}} \stackrel{(1)}{S}
  , \\
  \label{eq:second-order-equation-of-state-bare}
  \stackrel{(2)}{p}
  &=&
  \stackrel{(2)}{\epsilon}
  \frac{\partial\bar{p}}{\partial\bar{\epsilon}}(\epsilon,S)
  +
  \stackrel{(1)}{\epsilon}^{2}
  \frac{\partial^{2}\bar{p}}{\partial\bar{\epsilon}^{2}}(\epsilon,S)
  + 2
  \stackrel{(1)}{\epsilon}
  \stackrel{(1)}{S}
  \frac{\partial^{2}\bar{p}}{\partial\bar{\epsilon}\partial\bar{S}}(\epsilon,S)
  \nonumber\\
  &&
  +
  \stackrel{(2)}{S}
  \frac{\partial\bar{p}}{\partial\bar{S}}(\epsilon,S)
  +
  \stackrel{(1)}{S}^{2}
  \frac{\partial^{2}\bar{p}}{\partial\bar{S}^{2}}(\epsilon,S).
\end{eqnarray}

%*******************************************************************

In addition to the definitions of the gauge invariant variables
(\ref{eq:kouchan-016.13})--(\ref{eq:kouchan-016.18}) for the
first- and second-order perturbations of the fluid
components, we also define the gauge invariant variables for the
entropy perturbations:
\begin{eqnarray}
  \label{eq:entropy-gauge-inv-1}
  \stackrel{(1)}{{\cal S}}
  &:=& \stackrel{(1)}{S} - {\pounds}_{X}S
  , \\
  \label{eq:entropy-gauge-inv-2}
  \stackrel{(2)}{{\cal S}}
  &:=& \stackrel{(2)}{S}
  - 2 {\pounds}_{X} \stackrel{(1)}{S}
  - \left\{
    {\pounds}_{Y}
    -{\pounds}_{X}^{2}
  \right\} S
  .
\end{eqnarray}
Substituting Eqs.~(\ref{eq:kouchan-016.13}),
(\ref{eq:kouchan-016.16}), and (\ref{eq:entropy-gauge-inv-1}) 
into the perturbations of the equation of state
(\ref{eq:first-order-equation-of-state-bare}), we obtain the
first-order perturbation of the equation of state of the fluid
in terms of the gauge invariant variables:
\begin{eqnarray}
  \stackrel{(1)}{{\cal P}} + {\pounds}_{X}p
  &=&
  \frac{\partial\bar{p}}{\partial\bar{\epsilon}}
  \left(
    \stackrel{(1)}{{\cal E}} + {\pounds}_{X}\epsilon 
  \right)
  +
  \frac{\partial\bar{p}}{\partial\bar{S}}
  \left(
    \stackrel{(1)}{{\cal S}} + {\pounds}_{X}S
  \right)
  \nonumber \\
  &=&
  \frac{\partial\bar{p}}{\partial\bar{\epsilon}} \stackrel{(1)}{{\cal E}}
  +
  \frac{\partial\bar{p}}{\partial\bar{S}} \stackrel{(1)}{{\cal S}}
  +
  {\pounds}_{X}p.
  \label{eq:first-order-equation-of-state-gauge-inv-org}
\end{eqnarray}
The right-hand side of
Eq.~(\ref{eq:first-order-equation-of-state-gauge-inv-org}) has
the same form as Eq.~(\ref{eq:matter-gauge-inv-decomp-1.0}), as
expected by considering the left-hand side of
Eq.~(\ref{eq:first-order-equation-of-state-gauge-inv-org}).
Hence, we obtain the first-order perturbation of the equation of
state in terms of the gauge invariant variables:
\begin{eqnarray}
  \stackrel{(1)}{{\cal P}}
  =
  c_{s}^{2} \stackrel{(1)}{{\cal E}}
  +
  \tau \stackrel{(1)}{{\cal S}},
  \label{eq:first-order-equation-of-state-gauge-inv}
\end{eqnarray}
where we have
\begin{equation}
  c_{s}^{2} := \frac{\partial\bar{p}}{\partial\bar{\epsilon}}, 
  \quad
  \tau := \frac{\partial\bar{p}}{\partial\bar{S}}, 
  \label{eq:sound-velocity-tau}
\end{equation}
and $c_{s}$ is interpreted as the sound velocity of the fluid.
The equation (\ref{eq:first-order-equation-of-state-gauge-inv})
is the equation of state for the gauge invariant variables of
the first-order perturbation of the fluid components.

%*******************************************************************

Next, we consider the second-order perturbation of the equation
of state of the fluid in terms of gauge invariant variables:
\begin{eqnarray}
  &&
  \stackrel{(2)}{{\cal P}} 
  + 2 {\pounds}_{X}\stackrel{(1)}{p}
  + \left({\pounds}_{Y} - {\pounds}_{X}^{2}\right)p
  \nonumber\\
  &=&
  \stackrel{(2)}{{\cal E}} 
  \frac{\partial\bar{p}}{\partial\bar{\epsilon}}(\epsilon,S)
  +
  \stackrel{(1)}{{\cal E}}^{2}
  \frac{\partial^{2}\bar{p}}{\partial\bar{\epsilon}^{2}}(\epsilon,S)
  + 2
  \stackrel{(1)}{{\cal E}}
  \stackrel{(1)}{{\cal S}}
  \frac{\partial^{2}\bar{p}}{\partial\bar{\epsilon}\partial\bar{S}}(\epsilon,S)
  +
  \stackrel{(2)}{{\cal S}} 
  \frac{\partial\bar{p}}{\partial\bar{S}}(\epsilon,S)
  +
  \stackrel{(1)}{{\cal S}}^{2}
  \frac{\partial^{2}\bar{p}}{\partial\bar{S}^{2}}(\epsilon,S)
  \nonumber\\
  && \quad
  + 
  2 {\pounds}_{X}\left(
    \stackrel{(1)}{\epsilon}
    \frac{\partial\bar{p}}{\partial\bar{\epsilon}}(\epsilon,S)
    +
    \stackrel{(1)}{S}
    \frac{\partial\bar{p}}{\partial\bar{S}}(\epsilon,S)
  \right)
  + 
  {\pounds}_{Y}p(\epsilon,S)
  - 
  {\pounds}_{X}^{2}p(\epsilon,S).
  \label{eq:second-order-equation-of-state-gauge-inv-org}
\end{eqnarray}
The right-hand side of
Eq.~(\ref{eq:second-order-equation-of-state-gauge-inv-org}) has
the same form as expected by considering the left-hand side of
Eq.~(\ref{eq:second-order-equation-of-state-gauge-inv-org}).
Then, we obtain the second-order perturbation of the equation of
state in terms of the gauge invariant variables:
\begin{eqnarray}
  \stackrel{(2)}{{\cal P}}
  &=&
  c_{s}^{2}
  \stackrel{(2)}{{\cal E}} 
  + 
  \tau
  \stackrel{(2)}{{\cal S}} 
  +
  \frac{\partial c_{s}^{2}}{\partial\epsilon}
  \stackrel{(1)}{{\cal E}}^{2}
  + 2
  \frac{\partial c_{s}^{2}}{\partial S}
  \stackrel{(1)}{{\cal E}}
  \stackrel{(1)}{{\cal S}}
  +
  \frac{\partial\tau}{\partial S}
  \stackrel{(1)}{{\cal S}}^{2},
  \label{eq:second-order-equation-of-state-gauge-inv}
\end{eqnarray}
where we have used Eqs.~(\ref{eq:sound-velocity-tau}).

%*******************************************************************

%%%%%%%%%%%%%%%%%%%%%%%%%%%%%%%%%%%%%%%%%%%%%%%%%%%%%%%%%%%%%%%%%%%%%
\subsubsection{Scalar field}
\label{sec:Matter-perturbations-scalar-field}

%*******************************************************************

Here, we consider the energy-momentum tensor of the single
scalar field $\bar{\varphi}$,
\begin{eqnarray}
  \bar{T}_{a}^{\;\;b} = 
  \nabla_{a}\bar{\varphi} \nabla^{b}\bar{\varphi} 
  - \frac{1}{2} \delta_{a}^{\;\;b}
  \left(
    \nabla_{c}\bar{\varphi}\nabla^{c}\bar{\varphi}
    + 2 V(\bar{\varphi})
  \right),
  \label{eq:MFB-6.2-again}
\end{eqnarray}
where $V(\bar{\varphi})$ is the potential of $\bar{\varphi}$.
Because we are considering a perturbation theory on a
homogeneous and isotropic universe, the scalar field must also 
be approximately homogeneous.
Hence, the scalar field $\bar{\varphi}$ can be expanded as 
\begin{eqnarray}
  \bar{\varphi} = \varphi + \lambda \hat{\varphi}_{1} +
  \frac{1}{2} \lambda^{2} \hat{\varphi}_{2}
  + O(\lambda^{3}),
  \label{eq:scalar-field-expansion-second-order}
\end{eqnarray}
where $\varphi=\varphi(\eta)$ is a homogeneous function on the
homogeneous isotropic universe.
The background field $\varphi$ is the homogeneous part of the
scalar field which drives the background homogeneous isotropic
model, and $|\hat{\varphi}_{1}|\ll |\varphi|$ and
$|\hat{\varphi}_{2}|\ll |\hat{\varphi}_{1}|$ are the first- and 
second-order perturbations of the scalar field $\varphi$,
respectively.
The energy-momentum tensor (\ref{eq:MFB-6.2-again}) can also be
decomposed into the background, the first-order perturbation,
and the second-order perturbation as 
\begin{eqnarray}
  \bar{T}_{a}^{\;\;b} = T_{a}^{\;\;b} 
  + \lambda \;{}^{(1)}\!\left(T_{a}^{\;\;b}\right)
  + \frac{1}{2} \lambda^{2} \;{}^{(2)}\!\left(T_{a}^{\;\;b}\right)
  + O(\lambda^{3}),
\end{eqnarray}
where ${}^{(1)}\!\left(T_{a}^{\;\;b}\right)$ is linear in the
metric and the matter perturbations $h_{ab}$ and
$\hat{\varphi}_{1}$, and ${}^{(2)}\!\left(T_{a}^{\;\;b}\right)$
includes the second-order metric and matter perturbations,
$l_{ab}$ and $\hat{\varphi}_{2}$, and the quadratic terms of the
first-order perturbations, $\hat{\varphi}_{1}$ and $h_{ab}$.

%*******************************************************************

Expanding the metric as in Eq.~(\ref{eq:metric-expansion}) and
the scalar field as in
Eq.~(\ref{eq:scalar-field-expansion-second-order}), the
perturbations ${}^{(1)}\!\left(T_{a}^{\;\;b}\right)$ and
${}^{(2)}\!\left(T_{a}^{\;\;b}\right)$ of the energy-momentum
tensor (\ref{eq:MFB-6.2-again}) are given by
\begin{eqnarray}
  \label{eq:first-order-energy-momentum-scalar}
  {}^{(1)}\!\left(T_{a}^{\;\;b}\right)
  &:=& 
  \nabla_{a}\varphi \nabla^{c}\hat{\varphi}_{1} 
  - \nabla_{a}\varphi h^{bc} \nabla_{c}\varphi 
  + \nabla_{a}\hat{\varphi}_{1} \nabla^{b} \varphi 
  \nonumber\\
  && \quad
  - \frac{1}{2} \delta_{a}^{\;\;b}
  \left(
    \nabla_{c}\varphi\nabla^{c}\hat{\varphi}_{1}
    - \nabla_{c}\varphi h^{dc} \nabla_{d} \varphi 
    + \nabla_{c}\hat{\varphi}_{1}\nabla^{c} \varphi
    + 2 \hat{\varphi}_{1} \frac{\partial V}{\partial\varphi}
  \right)
  , \\
  \label{eq:second-order-energy-momentum-scalar}
  {}^{(2)}\!\left(T_{a}^{\;\;b}\right)
  &:=& 
  \nabla_{a}\varphi \nabla^{b}\hat{\varphi}_{2}
  - 2 \nabla_{a}\varphi h^{bc} \nabla_{c}\hat{\varphi}_{1} 
  + \nabla_{a}\varphi\left(2h^{bd}h_{d}^{\;\;c}-l^{bc}\right)\nabla_{c}\varphi
  \nonumber\\
  && \quad
  + 2 \nabla_{a}\hat{\varphi}_{1} \nabla^{b}\hat{\varphi}_{1}
  - 2 \nabla_{a}\hat{\varphi}_{1} h^{bc} \nabla_{c}\varphi 
  + \nabla_{a}\hat{\varphi}_{2}g^{bc}\nabla_{c}\varphi 
  \nonumber\\
  && \quad
  - \frac{1}{2} \delta_{a}^{\;\;b}
  \left(
    \nabla_{c}\varphi\nabla^{c}\hat{\varphi}_{2}
    - 2 \nabla_{c}\varphi h^{dc} \nabla_{d} \hat{\varphi}_{1} 
    + \nabla_{c}\varphi \left(2 h^{de}h_{e}^{\;\;c} - l^{dc}\right)\nabla_{d}\varphi 
  \right.
  \nonumber\\
  && \quad\quad\quad\quad\quad
  \left.
    + 2 \nabla_{c}\hat{\varphi}_{1}\nabla^{c}\hat{\varphi}_{1} 
    - 2 \nabla_{c}\hat{\varphi}_{1} h^{dc} \nabla_{d} \varphi 
    + \nabla_{c}\hat{\varphi}_{2} \nabla^{c} \varphi 
  \right.
  \nonumber\\
  && \quad\quad\quad\quad\quad
  \left.
    + 2 \hat{\varphi}_{2} \frac{\partial V}{\partial\varphi}
    + 2 (\hat{\varphi}_{1})^{2} \frac{\partial^{2}V}{\partial\varphi^{2}}
  \right)
  .
\end{eqnarray}

%*******************************************************************

According to the decompositions
(\ref{eq:matter-gauge-inv-decomp-1.0}) and
(\ref{eq:matter-gauge-inv-decomp-2.0}), the perturbations
of the scalar field $\varphi$ at each order can be decomposed
into the gauge invariant and variant parts as
\begin{eqnarray}
  \label{eq:varphi-1-def}
  \hat{\varphi}_{1} &=:& \varphi_{1} + {\pounds}_{X}\varphi, \\
  \label{eq:varphi-2-def}
  \hat{\varphi}_{2} &=:& \varphi_{2} 
  + 2 {\pounds}_{X}\hat{\varphi}_{1} 
  + \left( {\pounds}_{Y} - {\pounds}_{X}^{2} \right) \varphi, 
\end{eqnarray}
where $\varphi_{1}$ and $\varphi_{2}$ are the first- order and
second-order gauge invariant perturbations of the scalar
field.
Through these gauge invariant variables, the perturbed
energy-momentum tensor at each order can also be decomposed into
the gauge invariant and variant parts.
Substituting Eqs.~(\ref{eq:varphi-1-def}) and
(\ref{eq:linear-metric-decomp}) into
Eq.~(\ref{eq:first-order-energy-momentum-scalar}), the
first-order perturbation
(\ref{eq:first-order-energy-momentum-scalar}) 
of the scalar field is given by
\begin{eqnarray}
  {}^{(1)}\!\left(T_{a}^{\;\;b}\right)
  &=:&
  {}^{(1)}\!{\cal T}_{a}^{\;\;b} + {\pounds}_{X}T_{a}^{\;\;b},
  \label{eq:first-order-energy-momentum-scalar-decomp}
\end{eqnarray}
where
\begin{eqnarray}
  {}^{(1)}\!{\cal T}_{a}^{\;\;b}
  &:=& 
  \nabla_{a}\varphi \nabla^{b}\varphi_{1} 
  - \nabla_{a}\varphi {\cal H}^{bc} \nabla_{c}\varphi 
  + \nabla_{a}\varphi_{1} \nabla^{b} \varphi 
  \nonumber\\
  &&
  - \frac{1}{2} \delta_{a}^{\;\;b}
  \left(
    \nabla_{c}\varphi\nabla^{c}\varphi_{1}
    - \nabla_{c}\varphi {\cal H}^{dc} \nabla_{d} \varphi 
    + \nabla_{c}\varphi_{1}\nabla^{c} \varphi
    + 2 \varphi_{1} \frac{\partial V}{\partial\varphi}
  \right)
  \label{eq:first-order-energy-momentum-scalar-gauge-inv}
\end{eqnarray}
is the gauge invariant part of the first-order perturbation of
the energy-momentum tensor for the single scalar field.
Through Eqs. (\ref{eq:varphi-1-def}), (\ref{eq:varphi-2-def}),
(\ref{eq:linear-metric-decomp}), and
(\ref{eq:H-ab-in-gauge-X-def-second-1}), the second-order
perturbation (\ref{eq:second-order-energy-momentum-scalar}) of
the energy-momentum tensor is given by
\begin{eqnarray}
  {}^{(2)}\!\left(T_{a}^{\;\;b}\right)
  &=:&
  {}^{(2)}\!{\cal T}_{a}^{\;\;b}
  + 2 {\pounds}_{X}{}^{(1)}\!\left(T_{a}^{\;\;b}\right)
  + \left( {\pounds}_{Y} - {\pounds}_{X}^{2}\right) T_{a}^{\;\;b},
  \label{eq:second-order-energy-momentum-scalar-decomp}
\end{eqnarray}
where
\begin{eqnarray}
  {}^{(2)}\!{\cal T}_{a}^{\;\;b}
  &=&
  \nabla_{a}\varphi \nabla^{b}\varphi_{2} 
  - 2 \nabla_{a}\varphi {\cal H}^{bc} \nabla_{c}\varphi_{1} 
  + 2 \nabla_{a}\varphi {\cal H}^{bd}{\cal H}_{dc} \nabla^{c}\varphi
  - \nabla_{a}\varphi g^{bd} {\cal L}_{dc} \nabla^{c}\varphi
  \nonumber\\
  && \quad
  + 2 \nabla_{a}\varphi_{1} \nabla^{b}\varphi_{1}
  - 2 \nabla_{a}\varphi_{1} {\cal H}^{bc} \nabla_{c}\varphi 
  + \nabla_{a}\varphi_{2} \nabla^{b}\varphi  
  \nonumber\\
  && \quad
  - \frac{1}{2} \delta_{a}^{\;\;b}
  \left(
    \nabla_{c}\varphi\nabla^{c}\varphi_{2}
    - 2 \nabla_{c}\varphi {\cal H}^{dc} \nabla_{d}\varphi_{1} 
    + 2 \nabla^{c}\varphi {\cal H}^{de}{\cal H}_{ec} \nabla_{d}\varphi 
    - \nabla^{c}\varphi {\cal L}_{dc}\nabla^{d}\varphi 
  \right.
  \nonumber\\
  && \quad\quad\quad\quad\quad
  \left.
    + 2 \nabla_{c}\varphi_{1}\nabla^{c}\varphi_{1} 
    - 2 \nabla_{c}\varphi_{1} {\cal H}^{dc} \nabla_{d} \varphi 
    + \nabla_{c}\varphi_{2}\nabla^{c} \varphi 
  \right.
  \nonumber\\
  && \quad\quad\quad\quad\quad
  \left.
    + 2 \varphi_{2}\frac{\partial V}{\partial\varphi}
    + 2 \varphi_{1}^{2}\frac{\partial^{2}V}{\partial\varphi^{2}}
  \right)
  \label{eq:second-order-energy-momentum-scalar-gauge-inv}
  .
\end{eqnarray}
The tensor ${}^{(2)}\!{\cal T}_{a}^{\;\;b}$ is the second-order
gauge invariant part of the energy-momentum tensor for the
single scalar field.
Here again, we have seen that the perturbative expressions
(\ref{eq:first-order-energy-momentum-scalar-decomp}) and
(\ref{eq:second-order-energy-momentum-scalar-decomp}) of the
energy-momentum tensor have the same forms as
Eqs.~(\ref{eq:matter-gauge-inv-decomp-1.0}) and
(\ref{eq:matter-gauge-inv-decomp-2.0}), respectively, as
expected.
We also note that in the derivation of the expressions
(\ref{eq:first-order-energy-momentum-scalar-decomp})--(\ref{eq:second-order-energy-momentum-scalar-gauge-inv}),
we did not exiplicitly use any background values of the scalar
field nor the metric.
Therefore, the expressions
(\ref{eq:first-order-energy-momentum-scalar-decomp})--(\ref{eq:second-order-energy-momentum-scalar-gauge-inv})
are valid for any background spacetime.
This implies that the definitions
(\ref{eq:Tab-gauge-inv-def-1.0}) and 
(\ref{eq:Tab-gauge-inv-def-2.0}) of the gauge invariant
variables of the perturbed energy-momentum tensor in
\S\ref{sec:Perturbation-of-the-Einstein-tensor-and-the-Einstein-equations}
are appropriate not only in the case of a perfect fluid but
also in the case of a single scalar field.

%*******************************************************************

%%%%%%%%%%%%%%%%%%%%%%%%%%%%%%%%%%%%%%%%%%%%%%%%%%%%%%%%%%%%%%%%%%%%%
\section{First-order Einstein equations}
\label{sec:First-order-Einstein-equations}
%%%%%%%%%%%%%%%%%%%%%%%%%%%%%%%%%%%%%%%%%%%%%%%%%%%%%%%%%%%%%%%%%%%%%

%*******************************************************************

In this section, we consider the perturbed Einstein equations of
linear order, (\ref{eq:linear-order-Einstein-equation}).
To derive the components of the gauge invariant part of the
linearized Einstein tensor 
${}^{(1)}{\cal G}_{a}^{\;\;b}\left[{\cal H}\right]$, which is
defined by Eqs.~(\ref{eq:cal-G-def-linear}) and
(\ref{eq:(1)Sigma-def-linear}), we first derive the components
of the tensor $H_{ab}^{\;\;\;\;c}\left[{\cal H}\right]$, which
is defined in Eq.~(\ref{eq:Habc-def-1}) with 
$A_{ab} = {\cal H}_{ab}$.
Since the components of the gauge invariant part ${\cal H}_{ab}$
of the first-order metric perturbation are given by
Eq.~(\ref{eq:linear-order-metric-gauge-inv-def}), the components
of the tensor $H_{ab}^{\;\;\;\;c}\left[{\cal H}\right]$ are as
follows:
\begin{eqnarray}
  H_{\eta\eta}^{\;\;\;\;\eta}\left[{\cal H}\right]
  &=&
  \partial_{\eta}\stackrel{(1)}{\Phi}
  \label{eq:kouchan-10.34}
  , \\
  H_{i\eta}^{\;\;\;\;\eta}\left[{\cal H}\right]
  &=&
  D_{i} \stackrel{(1)}{\Phi} + {\cal H} \stackrel{(1)\;\;}{\nu_{i}}
  \label{eq:kouchan-10.35}
  , \\
  H_{ij}^{\;\;\;\;\eta}\left[{\cal H}\right]
  &=&
  - 
  \left(
    2 {\cal H} \left(\stackrel{(1)}{\Psi} + \stackrel{(1)}{\Phi}\right)
    + \partial_{\eta} \stackrel{(1)}{\Psi}
  \right)
  \gamma_{ij}
  - D_{(i} \stackrel{(1)\;\;}{\nu_{j)}}
  + \frac{1}{2} \left(
    \partial_{\eta}
    + 2 {\cal H}
  \right) \stackrel{(1)\;\;\;\;}{\chi_{ij}}
  \label{eq:kouchan-10.36}
  , \\
  H_{\eta\eta}^{\;\;\;\;i}\left[{\cal H}\right]
  &=&
  D^{i} \stackrel{(1)}{\Phi}
  + \left( \partial_{\eta} + {\cal H} \right) \stackrel{(1)\;\;}{\nu^{i}}
  \label{eq:kouchan-10.37}
  , \\
  H_{j\eta}^{\;\;\;\;i}\left[{\cal H}\right]
  &=&
  - \partial_{\eta} \stackrel{(1)}{\Psi} \gamma_{j}^{\;\;i} 
  + \frac{1}{2} \left(
    D_{j} \stackrel{(1)\;\;}{\nu^{i}} - D^{i} \stackrel{(1)\;\;}{\nu_{j}}
  \right)
  + \frac{1}{2} \partial_{\eta} \stackrel{(1)\;\;\;\;}{\chi_{j}^{\;\;i}} 
  \label{eq:kouchan-10.38}
  , \\
  H_{jk}^{\;\;\;\;i}\left[{\cal H}\right]
  &=&
    D^{i} \stackrel{(1)}{\Psi} \gamma_{kj}
  - 2 \gamma^{i}_{\;\;(k} D_{j)} \stackrel{(1)}{\Psi}
  - {\cal H} \gamma_{kj} \stackrel{(1)\;\;}{\nu^{i}}
  + D_{(j} \stackrel{(1)\;\;\;\;}{\chi_{k)}^{\;\;i}} 
  - \frac{1}{2} D^{i} \stackrel{(1)\;\;\;\;}{\chi_{kj}}
  .
  \label{eq:kouchan-10.39}
\end{eqnarray}
The components of the tensors $H_{abc}\left[{\cal H}\right]$,
$H_{a}^{\;\;bc}\left[{\cal H}\right]$,
$H_{a\;\;c}^{\;\;b}\left[{\cal H}\right]$, and
$H^{abc}\left[{\cal H}\right]$ are also useful when we derive
the components of the gauge invariant parts 
${}^{(1)}\!{\cal G}_{a}^{\;\;b}\left[{\cal H}\right]$ and 
${}^{(2)}\!{\cal G}_{a}^{\;\;b}\left[{\cal H},{\cal H}\right]$
of the perturbative Einstein tensor.
These components are summarized in Appendix
\ref{sec:components-of-Habcs}.

%*******************************************************************

Following to the definitions (\ref{eq:cal-G-def-linear}) and
(\ref{eq:(1)Sigma-def-linear}) of the gauge invariant part
${}^{(1)}{\cal G}_{a}^{\;\;b}\left[{\cal H}\right]$ of the
first-order perturbation of the Einstein tensor, its components
are derived as follows:
\begin{eqnarray}
  {}^{(1)}{\cal G}_{\eta}^{\;\;\eta}\left[{\cal H}\right]
  &=&
  - \frac{1}{a^{2}} \left\{
    \left(
      - 6 {\cal H} \partial_{\eta}
      + 2 \Delta
      + 6 K
    \right) \stackrel{(1)}{\Psi}
    - 6 {\cal H}^{2} \stackrel{(1)}{\Phi}
  \right\}
  \label{eq:kouchan-10.120}
  , \\
  {}^{(1)}{\cal G}_{i}^{\;\;\eta}\left[{\cal H}\right]
  &=&
  - \frac{1}{a^{2}}
  \left(
    2 \partial_{\eta} D_{i} \stackrel{(1)}{\Psi}
    + 2 {\cal H} D_{i} \stackrel{(1)}{\Phi} 
    - \frac{1}{2} \left(
      \Delta
      + 2 K
    \right)
    \stackrel{(1)\;\;}{\nu_{i}}  
  \right)
  \label{eq:kouchan-10.121}
  , \\
  {}^{(1)}{\cal G}_{\eta}^{\;\;i}\left[{\cal H}\right] 
  &=&
  \frac{1}{a^{2}} \left\{
    2 \partial_{\eta} D^{i} \stackrel{(1)}{\Psi}
    + 2 {\cal H} D^{i} \stackrel{(1)}{\Phi} 
    + \frac{1}{2} \left(
      - \Delta
      + 2 K
      + 4 {\cal H}^{2}
      - 4 \partial_{\eta}{\cal H}
    \right)
    \stackrel{(1)\;\;}{\nu^{i}}
  \right\}
  \label{eq:kouchan-10.122}
  , \\
  {}^{(1)}{\cal G}_{i}^{\;\;j}\left[{\cal H}\right]
  &=& 
  \frac{1}{a^{2}} \left[
    D_{i} D^{j} \left(\stackrel{(1)}{\Psi} - \stackrel{(1)}{\Phi}\right)
  \right.
  \nonumber\\
  &&
  \left.
    + 
    \left\{
      \left(
        -   \Delta
        + 2 \partial_{\eta}^{2} 
        + 4 {\cal H} \partial_{\eta}
        - 2 K
      \right)
      \stackrel{(1)}{\Psi}
      + \left(
          2 {\cal H} \partial_{\eta}
        + 4 \partial_{\eta}{\cal H}
        + 2 {\cal H}^{2}
        + \Delta
      \right)
      \stackrel{(1)}{\Phi}
    \right\}
    \gamma_{i}^{\;\;j}
  \right.
  \nonumber\\
  &&
  \left.
    - \frac{1}{2a^{2}} \partial_{\eta} \left\{
      a^{2} \left( 
        D_{i} \stackrel{(1)\;\;}{\nu^{j}} + D^{j} \stackrel{(1)\;\;}{\nu_{i}}
      \right)
    \right\}
  \right.
  \nonumber\\
  &&
  \left.
    + \frac{1}{2} \left(
      \partial_{\eta}^{2}
      + 2 {\cal H} \partial_{\eta}
      + 2 K
      - \Delta
    \right) \stackrel{(1)\;\;\;\;}{\chi_{i}^{\;\;j}}
  \right]
  .
  \label{eq:kouchan-10.123}
\end{eqnarray}
Straightforward calculations show that these components of the
first-order gauge invariant perturbation
${}^{(1)}{\cal G}_{a}^{\;\;b}\left[{\cal H}\right]$ of the
Einstein tensor satisfies the first-order perturbation
(\ref{eq:linear-order-divergence-of-calGab}) of the Bianchi
identity.
This implies that we have derived the components
(\ref{eq:kouchan-10.120})--(\ref{eq:kouchan-10.123}) of
${}^{(1)}{\cal G}_{a}^{\;\;b}\left[{\cal H}\right]$
consistently.

%*******************************************************************

Together with the components of the gauge invariant part
${}^{(1)}{\cal T}_{a}^{\;\;b}$ of the first-order perturbation
of the energy-momentum tensor, the first-order Einstein equation
(\ref{eq:linear-order-Einstein-equation}) is given as equations
for the gauge invariant variables 
$\stackrel{(1)}{\Phi}$, $\stackrel{(1)}{\Psi}$,
$\stackrel{(1)\;\;}{\nu_{i}}$, and
$\stackrel{(1)\;\;}{\chi_{ij}}$.
We consider these equations for two cases, that in which the
energy-momentum tensor is dominated by the single perfect 
fluid and that in which it is dominated by the single scalar
field.

%*******************************************************************

%%%%%%%%%%%%%%%%%%%%%%%%%%%%%%%%%%%%%%%%%%%%%%%%%%%%%%%%%%%%%%%%%%%%%
\subsection{Perfect fluid case}
\label{sec:Perfect-fluid-case-first-order}

%*******************************************************************

Here, we consider the linearized Einstein equation of a
homogeneous isotropic universe filled with a perfect fluid.

%*******************************************************************

We first consider the components of the gauge invariant part
$\stackrel{(1)}{{\cal U}_{a}}$ of the perturbative four-velocity
of the fluid.
Taking into account the perturbation
(\ref{eq:four-vel-norm-first-pert-gauge-inv-2}) of the
normalization condition (\ref{eq:four-velocity-norm-full}), the
components of $\stackrel{(1)}{{\cal U}_{a}}$ are decomposed as 
\begin{equation}
  \stackrel{(1)}{{\cal U}_{a}} = - a \stackrel{(1)}{\Phi} (d\eta)_{a}
  + \left(
    D_{i} \stackrel{(1)}{v} 
    + 
    \stackrel{(1)}{{\cal V}_{i}}
  \right) (dx^{i})_{a}.
  , \quad D^{i}\stackrel{(1)}{{\cal V}_{i}} = 0,
  \label{eq:first-order-four-velocity-components}
\end{equation}
where the $\eta$-component of $\stackrel{(1)}{{\cal U}_{a}}$ are
determined by
Eq.~(\ref{eq:four-vel-norm-first-pert-gauge-inv-2}).
Here, we note that the divergenceless part of the spatial
component of the four-velocity, $\stackrel{(1)}{{\cal V}_{i}}$,
contributes to the vorticity perturbation.
Substituting (\ref{eq:energy-momentum-for-velocity-homogeneous})
and (\ref{eq:first-order-four-velocity-components}) into
(\ref{eq:first-ene-mon-tensor-gauge-inv-def}), the components of
the gauge invariant part ${}^{(1)}\!{\cal T}_{a}^{\;\;b}$ of the
first-order perturbation of the energy-momentum tensor are
obtained as 
\begin{eqnarray}
  {}^{(1)}\!{\cal T}_{\eta}^{\;\;\eta}
  &=& - \stackrel{(1)}{{\cal E}}
  \label{eq:first-order-ene-mon-fluid-eta-eta}
  , \\
  {}^{(1)}\!{\cal T}_{\eta}^{\;\;i}
  &=& \left(\epsilon + p\right) \left\{
    \stackrel{(1)\;\;}{\nu^{i}} - 
    \left(D^{i}\stackrel{(1)}{v} + \stackrel{(1)\;\;}{{\cal V}^{i}} \right)
  \right\}
  \label{eq:first-order-ene-mon-fluid-eta-i}
  , \\
  {}^{(1)}\!{\cal T}_{i}^{\;\;\eta}
  &=&
  \left(\epsilon + p\right) \left(
    D_{i}\stackrel{(1)}{v} + \stackrel{(1)\;\;}{{\cal V}_{i}}
  \right)
  \label{eq:first-order-ene-mon-fluid-i-eta}
  , \\
  {}^{(1)}\!{\cal T}_{i}^{\;\;j}
  &=&
  \stackrel{(1)}{{\cal P}} \gamma_{i}^{\;\;j}.
  \label{eq:first-order-ene-mon-fluid-i-j}
\end{eqnarray}
Through
Eqs.~(\ref{eq:kouchan-10.120})--(\ref{eq:kouchan-10.123}) and
(\ref{eq:first-order-ene-mon-fluid-eta-eta})--(\ref{eq:first-order-ene-mon-fluid-i-j}),
the linearized Einstein equations
(\ref{eq:linear-order-Einstein-equation}) are found to be 
\begin{eqnarray}
  &&
  4 \pi G a^{2} \stackrel{(1)}{{\cal E}}
  =
  \left(
    - 3 {\cal H} \partial_{\eta}
    + \Delta
    + 3 K
  \right) \stackrel{(1)}{\Psi}
  - 3 {\cal H}^{2} \stackrel{(1)}{\Phi}
  \label{eq:linearized-Einstein-eta-eta}
  , \\
  &&
  4 \pi G a^{2} \left(\epsilon + p\right) \left(
    D_{i}\stackrel{(1)}{v} + \stackrel{(1)\;\;}{{\cal V}_{i}}
  \right)
  =
  - \partial_{\eta} D_{i} \stackrel{(1)}{\Psi}
  - {\cal H} D_{i} \stackrel{(1)}{\Phi} 
  + \frac{1}{4} \left(
    \Delta
    + 2 K
  \right)
  \stackrel{(1)\;\;}{\nu_{i}}  
  \label{eq:linearized-Einstein-i-eta}
  , \\
  &&
  4\pi G a^{2} \stackrel{(1)}{{\cal P}} \gamma_{i}^{\;\;j}
  = 
  \frac{1}{2} D_{i} D^{j} \left(
    \stackrel{(1)}{\Psi} - \stackrel{(1)}{\Phi}
  \right)
  \nonumber\\
  && \quad\quad\quad\quad\quad\quad
  + 
  \left\{
    \left(
        \partial_{\eta}^{2} 
      + 2 {\cal H} \partial_{\eta}
      - K
      - \frac{1}{2} \Delta
    \right)
    \stackrel{(1)}{\Psi}
    + \left(
        {\cal H} \partial_{\eta}
      + 2 \partial_{\eta}{\cal H}
      + {\cal H}^{2}
      + \frac{1}{2} \Delta
    \right)
    \stackrel{(1)}{\Phi}
  \right\}
  \gamma_{i}^{\;\;j}
  \nonumber\\
  && \quad\quad\quad\quad\quad\quad
  - \frac{1}{4a^{2}} \partial_{\eta}
  \left\{ 
    a^{2} \left( 
      D_{i} \stackrel{(1)\;\;}{\nu^{j}} + D^{j} \stackrel{(1)\;\;}{\nu_{i}}
    \right)
  \right\}
  \nonumber\\
  && \quad\quad\quad\quad\quad\quad
  + \frac{1}{4} \left(
    \partial_{\eta}^{2}
    + 2 {\cal H} \partial_{\eta}
    + 2 K
    - \Delta
  \right) \stackrel{(1)\;\;\;\;}{\chi_{i}^{\;\;j}}
  ,
  \label{eq:linearized-Einstein-i-j}
\end{eqnarray}
where the component 
${}^{(1)}\!{\cal G}_{\eta}^{\;\;i}\left[{\cal H}\right]=8\pi
G{}^{(1)}\!{\cal T}_{\eta}^{\;\;i}$ is identical to 
Eq.~(\ref{eq:linearized-Einstein-i-eta}) by virtue of the
background Einstein equation 
(\ref{eq:background-Einstein-equations-2}).

%*******************************************************************

We decompose
Eqs.~(\ref{eq:linearized-Einstein-eta-eta})--(\ref{eq:linearized-Einstein-i-j})
similarly to
Eqs.~(\ref{eq:htaui-decomp})--(\ref{eq:hTij-decomp}) for the
metric perturbation $h_{ab}$, whose inverse relations are given by
Eqs.~(\ref{eq:hetai-decomp-inv-1})--(\ref{eq:hTij-decomp-inv-5}).
Then, we obtain the equations for the scalar mode perturbations as
\begin{eqnarray}
  &&
  4 \pi G a^{2} \stackrel{(1)}{{\cal E}}
  =
  \left(
    - 3 {\cal H} \partial_{\eta}
    + \Delta
    + 3 K
  \right) \stackrel{(1)}{\Psi}
  - 3 {\cal H}^{2} \stackrel{(1)}{\Phi}
  \label{eq:linearized-Einstein-eta-eta-scalar}
  , \\
  &&
  4 \pi G a^{2} \left(\epsilon + p\right) D_{i}\stackrel{(1)}{v}
  =
  - \partial_{\eta} D_{i} \stackrel{(1)}{\Psi}
  - {\cal H} D_{i} \stackrel{(1)}{\Phi} 
  \label{eq:linearized-Einstein-i-eta-scalar}
  , \\
  &&
  4 \pi G a^{2} \stackrel{(1)}{{\cal P}}
  = 
  \left(
      \partial_{\eta}^{2} 
    + 2 {\cal H} \partial_{\eta}
    - K
    - \frac{1}{3} \Delta
  \right)
  \stackrel{(1)}{\Psi}
  \nonumber\\
  && \quad\quad\quad\quad\quad\quad
  + \left(
      {\cal H} \partial_{\eta}
    + 2 \partial_{\eta}{\cal H}
    + {\cal H}^{2}
    + \frac{1}{3} \Delta
  \right)
  \stackrel{(1)}{\Phi}
  \label{eq:linearized-Einstein-i-j-trace}
  , \\
  &&
  \frac{1}{3} \Delta \left( \Delta + 3 K \right) 
  \left(\stackrel{(1)}{\Psi} - \stackrel{(1)}{\Phi}\right)
  = 
  0,
  \label{eq:linearized-Einstein-i-j-traceless-scalar}
\end{eqnarray}
the equations for the vector-mode perturbation as 
\begin{eqnarray}
  &&
  4 \pi G a^{2} \left(\epsilon + p\right) \stackrel{(1)\;\;}{{\cal V}_{i}}
  =
  \frac{1}{4} \left(
    \Delta
    + 2 K
  \right)
  \stackrel{(1)\;\;}{\nu_{i}}  
  \label{eq:linearized-Einstein-i-eta-vector}
  , \\
  &&
  \partial_{\eta} \left\{ a^{2} \left( \Delta + 2 K \right) 
    \stackrel{(1)\;\;}{\nu^{j}} \right\}
  =
  0
  \label{eq:linearized-Einstein-i-j-traceless-vector}
  ,
\end{eqnarray}
and the equation for the tensor-mode perturbation as 
\begin{eqnarray}
  \left(
    \partial_{\eta}^{2}
    + 2 {\cal H} \partial_{\eta}
    + 2 K
    - \Delta
  \right) \stackrel{(1)\;\;}{\chi_{ij}}
  = 0
  .
  \label{eq:linearized-Einstein-i-j-traceless-tensor}
\end{eqnarray}
Equation
(\ref{eq:linearized-Einstein-i-j-traceless-tensor}) describes
the evolution of gravitational waves.

%*******************************************************************

Equation (\ref{eq:linearized-Einstein-i-j-traceless-scalar})
yields
\begin{eqnarray}
  \stackrel{(1)}{\Psi} = \stackrel{(1)}{\Phi}.
  \label{eq:absence-of-anisotropic-stress-Einstein-i-j-traceless-scalar}
\end{eqnarray}
From this equation, the energy density perturbation
$\stackrel{(1)}{{\cal E}}$, the velocity perturbation
$D_{i}\stackrel{(1)}{v}$, and the pressure perturbation
$\stackrel{(1)}{{\cal P}}$ are found to satisfy
\begin{eqnarray}
  &&
  4 \pi G a^{2} \stackrel{(1)}{{\cal E}}
  =
  \left(
     \Delta
    - 3 {\cal H} \partial_{\eta}
    - 3 \left( {\cal H}^{2} - K \right)
  \right) \stackrel{(1)}{\Phi}
  \label{eq:linearized-Einstein-energy-perturbation}
  , \\
  &&
  4 \pi G a^{2} \left(\epsilon + p\right) D_{i}\stackrel{(1)}{v}
  =
  - \partial_{\eta} D_{i} \stackrel{(1)}{\Phi}
  - {\cal H} D_{i} \stackrel{(1)}{\Phi} 
  \label{eq:linearized-Einstein-scalar-velocity-perturbation}
  , \\
  &&
  4 \pi G a^{2} \stackrel{(1)}{{\cal P}}
  = 
  \left(
      \partial_{\eta}^{2} 
    + 3 {\cal H} \partial_{\eta}
    + 2 \partial_{\eta}{\cal H}
    + {\cal H}^{2}
    - K
  \right)
  \stackrel{(1)}{\Phi}.
  \label{eq:linearized-Einstein-perssure-perturbation}
\end{eqnarray}
In the Newtonian limit,
Eq.~(\ref{eq:linearized-Einstein-energy-perturbation}) reduces
to the usual Poisson equation for the gravitational potential
$\stackrel{(1)}{\Phi}$ induced by the energy-density
perturbation $\stackrel{(1)}{{\cal E}}$.
This supports the interpretation of $\stackrel{(1)}{\Phi}$ as
the relativistic generalization of the Newtonian gravitational
potential.
Equation (\ref{eq:linearized-Einstein-energy-perturbation}) is
the generalized form of the Poisson equation obtained by taking
into account the expansion of the universe.

%*******************************************************************

Next, we apply the equation of state
(\ref{eq:first-order-equation-of-state-gauge-inv}) for the first
order perturbation.
Then, from Eqs.~(\ref{eq:linearized-Einstein-eta-eta-scalar})
and (\ref{eq:linearized-Einstein-i-j-trace}), we obtain the
well-known master equation for the scalar mode
perturbation\cite{Kodama-Sasaki-1984,Mukhanov-Feldman-Brandenberger-1992}:
\begin{eqnarray}
  &&
  \left\{
    \partial_{\eta}^{2} 
    + 3 {\cal H} (1 + c_{s}^{2}) \partial_{\eta}
    - c_{s}^{2} \Delta
    + 2 \partial_{\eta}{\cal H}
    + (1 + 3 c_{s}^{2}) ({\cal H}^{2} - K)
  \right\}
  \stackrel{(1)}{\Phi}
  \nonumber\\
  &=&
  4 \pi G a^{2}\tau\stackrel{(1)}{{\cal S}}.
  \label{eq:linearized-Einstein-scalar-master-eq}
\end{eqnarray}
The scalar mode perturbations are completely determined by this
master equation (\ref{eq:linearized-Einstein-scalar-master-eq}).
If we obtain the solution $\stackrel{(1)}{\Phi}$ to
Eq.~(\ref{eq:linearized-Einstein-scalar-master-eq}), we can
obtain another scalar perturbation $\stackrel{(1)}{\Psi}$
through
Eq.~(\ref{eq:absence-of-anisotropic-stress-Einstein-i-j-traceless-scalar}),
and the energy density perturbation $\stackrel{(1)}{{\cal E}}$,
the velocity perturbation $D_{i}\stackrel{(1)}{v}$, and the
pressure perturbation $\stackrel{(1)}{{\cal P}}$ are obtained
from Eqs.~(\ref{eq:linearized-Einstein-energy-perturbation}), 
(\ref{eq:linearized-Einstein-scalar-velocity-perturbation}), and
(\ref{eq:linearized-Einstein-perssure-perturbation}),
respectively.
It is also well-known that
Eq.~(\ref{eq:linearized-Einstein-scalar-master-eq}) is reduced
to simpler equation through a change of
variables\cite{Mukhanov-Feldman-Brandenberger-1992}.

%*******************************************************************

Here, we comment on the contribution of the anisotropic stress,
which is ignored in the above derivation of the linearized
Einstein equation.
If an anisotropic stress exists in the linear-order
energy-momentum tensor, these can be formally decomposed into
scalar, vector and tensor types in forms similar to 
Eqs.~(\ref{eq:hij-decomp}) and (\ref{eq:hTij-decomp}).
If these exist, the anisotropic stress of scalar type
contributes to the scalar mode of the perturbation and will
appear on the right hand side of
Eq.~(\ref{eq:linearized-Einstein-i-j-traceless-scalar}).
In this case, the equation
(\ref{eq:absence-of-anisotropic-stress-Einstein-i-j-traceless-scalar})
is no longer valid.
Instead,
$\stackrel{(1)}{\Psi}-\stackrel{(1)}{\Phi}$ is proportional to
the anisotropic stress of scalar type. 
As a result, the master equation
(\ref{eq:linearized-Einstein-scalar-master-eq}) will have a
source term which is proportional to the anisotropic stress, in
addition to the entropy perturbation.
The anisotropic stress of vector type contributes as the
source term of
Eq.~(\ref{eq:linearized-Einstein-i-j-traceless-vector}), if it
exists at linear order.
Though the solution $\stackrel{(1)\;\;}{\nu_{i}}$ to
Eq.~(\ref{eq:linearized-Einstein-i-j-traceless-vector}) is only
the decaying mode in the absence of the anisotropic stress of
vector type, the vector perturbation
$\stackrel{(1)\;\;}{\nu_{i}}$ of the metric is generated by
the anisotropic stress of vector type, if it exits, and the
resulting vector perturbation, $\stackrel{(1)\;\;}{\nu_{i}}$, of
the metric directly generates the divergenceless part,
$\stackrel{(1)\;\;}{{\cal V}_{i}}$, of the four-velocity of the
fluid, which may contribute to the vorticity of the fluid. 
Finally, if the anisotropic stress of tensor type exists, it
appears as the source term in
Eq.~(\ref{eq:linearized-Einstein-i-j-traceless-tensor}), and it 
generates gravitational waves,
$\stackrel{(1)\;\;\;\;}{\chi_{ij}}$.
Thus, the anisotropic stress may generate all types of
perturbations.
The situation is simialr in the case of the second-order
perturbations, as shown below.

%*******************************************************************

%%%%%%%%%%%%%%%%%%%%%%%%%%%%%%%%%%%%%%%%%%%%%%%%%%%%%%%%%%%%%%%%%%%%%
\subsection{Scalar field case}
\label{sec:Scalar-feild-case}

%*******************************************************************

Here, we consider the linearized Einstein equation of a
homogeneous isotropic universe filled with a single scalar
field.

%*******************************************************************

We note that the background scalar field $\varphi$ is
homogeneous, i.e., $\varphi = \varphi(\eta)$, where $\eta$ is
the conformal time.
Thus, the components of the gauge invariant part of the
first-order energy-momentum tensor
${}^{(1)}\!{\cal T}_{a}^{\;\;b}$ are given by
\begin{eqnarray}
  {}^{(1)}\!{\cal T}_{\eta}^{\;\;\eta}
  &=& 
  - \frac{1}{a^{2}} \left(
      \partial_{\eta}\varphi_{1}\partial_{\eta}\varphi
    - \stackrel{(1)}{\Phi} (\partial_{\eta}\varphi)^{2}
    + a^{2}\frac{dV}{d\varphi}\varphi_{1}
  \right)
  \label{eq:kouchan-10.161}
  \\
  {}^{(1)}\!{\cal T}_{i}^{\;\;\eta}
  &=& 
  - \frac{1}{a^{2}}
  D_{i}\varphi_{1}\partial_{\eta}\varphi,
  \label{eq:kouchan-10.162}
  \\
  {}^{(1)}\!{\cal T}_{\eta}^{\;\;i}
  &=& 
  \frac{1}{a^{2}} \partial_{\eta}\varphi 
  \left(
    D^{i}\varphi_{1}
    + (\partial_{\eta}\varphi) \stackrel{(1)\;\;}{\nu^{i}}
  \right)
  ,
  \label{eq:kouchan-10.163}
  \\
  {}^{(1)}\!{\cal T}_{i}^{\;\;j}
  &=& 
  \frac{1}{a^{2}} \gamma_{i}^{\;\;j}\left(
      \partial_{\eta}\varphi_{1} \partial_{\eta}\varphi
    - \stackrel{(1)}{\Phi} (\partial_{\eta}\varphi)^{2}
    - a^{2} \frac{dV}{d\varphi} \varphi_{1}
  \right)
  .
  \label{eq:kouchan-10.165}
\end{eqnarray}
Equation (\ref{eq:kouchan-10.165}) shows that there is no
anisotropic stress in the energy-momentum tensor of the single
scalar field.
Then, as in the case of a perfect fluid, we obtain
Eq.~(\ref{eq:absence-of-anisotropic-stress-Einstein-i-j-traceless-scalar}).
From
Eqs.~(\ref{eq:kouchan-10.120})--(\ref{eq:kouchan-10.123}),
(\ref{eq:absence-of-anisotropic-stress-Einstein-i-j-traceless-scalar}),
and (\ref{eq:kouchan-10.161})--(\ref{eq:kouchan-10.165}), the
components of the linearized Einstein equation
(\ref{eq:linear-order-Einstein-equation}) are obtained
as\cite{Mukhanov-Feldman-Brandenberger-1992}
\begin{eqnarray}
  \left(
    + \Delta
    - 3 {\cal H} \partial_{\eta}
    + 4 K
    - \partial_{\eta}{\cal H}
    - 2 {\cal H}^{2}
  \right) \stackrel{(1)}{\Phi}
  &=& 
  4 \pi G \left(
    \partial_{\eta}\varphi_{1} \partial_{\eta}\varphi
    + a^{2}\frac{dV}{d\varphi}\varphi_{1}
  \right)
  \label{eq:kouchan-18.185}
  , \\
  \partial_{\eta}\stackrel{(1)}{\Phi} + {\cal H} \stackrel{(1)}{\Phi}
  &=&
  4 \pi G \varphi_{1} \partial_{\eta}\varphi
  \label{eq:kouchan-18.186}
  , \\
  \left(
    + \partial_{\eta}^{2} 
    + 3 {\cal H} \partial_{\eta}
    +   \partial_{\eta}{\cal H}
    + 2 {\cal H}^{2}
  \right)
  \stackrel{(1)}{\Phi}
  &=&
  4 \pi G
  \left(
    \partial_{\eta}\varphi_{1} \partial_{\eta}\varphi
    - a^{2} \frac{dV}{d\varphi} \varphi_{1}
  \right)
  \label{eq:kouchan-18.187}
  .
\end{eqnarray}
In the derivation of
Eqs.~(\ref{eq:kouchan-18.185})--(\ref{eq:kouchan-18.187}), we
have used Eq.~(\ref{eq:background-Einstein-equations-scalar-3}).
We also note that only two of these equations are independent.
Further, the vector part of the component 
${}^{(1)}\!{\cal G}_{i}^{\;\;\eta}\left[{\cal H}\right]=8\pi G{}^{(1)}\!{\cal T}_{i}^{\;\;\eta}$ 
of the Einstein equations shows that there is no vector mode as
an initial value constraint.
The equation for the tensor mode $\stackrel{(1)\;\;}{\chi_{ij}}$
is identical to
Eq.(\ref{eq:linearized-Einstein-i-j-traceless-tensor}).

%*******************************************************************

Combining Eqs.~(\ref{eq:kouchan-18.185}) and
(\ref{eq:kouchan-18.187}), we eliminate the potential term of
the scalar field and thereby obtain
\begin{eqnarray}
  \left(
        \partial_{\eta}^{2} 
    +   \Delta
    + 4 K
  \right) \stackrel{(1)}{\Phi}
  = 
  8 \pi G \partial_{\eta}\varphi_{1} \partial_{\eta}\varphi.
  \label{eq:scalar-linearized-Einstein-scalar-master-eq-pre}
\end{eqnarray}
Further, using Eq.~(\ref{eq:kouchan-18.186}) to express
$\partial_{\eta}\varphi_{1}$ in terms of
$\partial_{\eta}\stackrel{(1)}{\Phi}$ and
$\stackrel{(1)}{\Phi}$, we also eliminate
$\partial_{\eta}\varphi_{1}$ in
Eq.~(\ref{eq:scalar-linearized-Einstein-scalar-master-eq-pre}).
Hence, we have
\begin{eqnarray}
  \label{eq:scalar-linearized-Einstein-scalar-master-eq}
  \left(
    \partial_{\eta}^{2}
    + 2 \left(
      {\cal H}
      - \frac{2\partial_{\eta}^{2}\varphi}{\partial_{\eta}\varphi}
    \right) \partial_{\eta}
    - \Delta
    - 4 K
    + 2 
    \left( \partial_{\eta}{\cal H}
      - \frac{{\cal H}\partial_{\eta}^{2}\varphi}{\partial_{\eta}\varphi}
    \right)
  \right) \stackrel{(1)}{\Phi}
  = 
  0.
\end{eqnarray}
This is the master equation for the scalar mode perturbation of
the cosmological perturbation in universe filled with a single
scalar field.
It is also known that
Eq.~(\ref{eq:scalar-linearized-Einstein-scalar-master-eq})
reduces to a simple equation through a change of variables, and
this equation has the same form as 
Eq.~(\ref{eq:linearized-Einstein-scalar-master-eq}) with
$c_{s}=1$\cite{Mukhanov-Feldman-Brandenberger-1992}.

%*******************************************************************

%%%%%%%%%%%%%%%%%%%%%%%%%%%%%%%%%%%%%%%%%%%%%%%%%%%%%%%%%%%%%%%%%%%%%
\section{Second-order Einstein equations}
\label{sec:Secnd-order-Einstein-equations}
%%%%%%%%%%%%%%%%%%%%%%%%%%%%%%%%%%%%%%%%%%%%%%%%%%%%%%%%%%%%%%%%%%%%%

In this section, we derive the second-order perturbation of the
Einstein equation in the context of cosmological perturbations.
In the generic case, the second order perturbation of the
Einstein equation is given by
Eq.~(\ref{eq:second-order-Einstein-equation}).
This equation is for the gauge invariant second-order metric
perturbation ${\cal L}_{ab}$, whose components are given in
Eq.~(\ref{eq:second-order-gauge-inv-metrc-pert-components}).
To derive the equations for each component of ${\cal L}_{ab}$,
we have to evaluate 
${}^{(1)}{\cal G}_{a}^{\;\;b}\left[{\cal L}\right]$,
${}^{(2)}{\cal G}_{a}^{\;\;b}\left[{\cal H},{\cal H}\right]$,
and ${}^{(2)}{\cal T}_{a}^{\;\;b}$.

%*******************************************************************

In this paper, we consider the simple situation in which the
first-order vector and tensor modes are negligible:
\begin{eqnarray}
  \stackrel{(1)\;\;}{\nu_{i}} = 0, \quad
  \stackrel{(1)\;\;\;\;}{\chi_{ij}} = 0.
  \label{eq:simplified-condition}
\end{eqnarray}
In the linear-order perturbation, the vector mode
$\stackrel{(1)\;\;}{\nu_{i}}$ decays as $a^{-2}$, which becomes
smaller than the scalar perturbation.
Further, as seen in \S\ref{sec:Scalar-feild-case}, the vector
mode $\stackrel{(1)\;\;}{\nu_{i}}$ is not generated in an
inflationary universe driven by a single scalar field. 
For these reasions, it is reasonable to omit the vector mode
$\stackrel{(1)\;\;}{\nu_{i}}$ for a wide class of scenarios of
the evolution of the universe. 
By contrast, the tensor mode
$\stackrel{(1)\;\;\;\;}{\chi_{ij}}$, i.e., gravitational waves,  
may be generated by quantum fluctuations during the inflation.
The amplitude of these stochastic gravitational waves depends on
the scenario of the inflation.
However, from the observational result of CMB\cite{WMAP}, scalar
mode fluctuations should be dominant, with the scalar-tensor
ratio being less than unity. 
We can thus assume that the dominant contribution to the
fluctuations in the universe is that of scalar type.
Hence, we ignore the first-order tensor mode
$\stackrel{(1)\;\;\;\;}{\chi_{ij}}$ in this paper.
Of course, it is possible to extend our formulation by taking
into account of the vector- and tensor-mode contributions to
the second-order perturbations, but we only consider the main
contribution to the second-order fluctuations in this paper.

%*******************************************************************

Because the components of the second-order gauge invariant metric
perturbation ${\cal L}_{ab}$ are obtained through the replacements
\begin{eqnarray}
  \label{eq:replacement-from-calHab-to-calLab}
  \stackrel{(1)}{\Phi} \rightarrow \stackrel{(2)}{\Phi}, \quad
  \stackrel{(1)\;\;}{\nu_{i}} \rightarrow \stackrel{(2)\;\;}{\nu}_{i}, \quad
  \stackrel{(1)}{\Psi} \rightarrow \stackrel{(2)}{\Psi}, \quad
  \stackrel{(1)\;\;\;\;}{\chi_{ij}} \rightarrow
  \stackrel{(2)\;\;\;\;}{\chi_{ij}} 
\end{eqnarray}
of the variables in the first-order gauge invariant metric
perturbation ${\cal H}_{ab}$, the evaluation of 
${}^{(1)}{\cal G}_{a}^{\;\;b}\left[{\cal L}\right]$ is
accomplished through these replacements in the first-order gauge
invariant Einstein tensor 
${}^{(1)}{\cal G}_{a}^{\;\;b}\left[{\cal H}\right]$, which
appear in Eqs.~(\ref{eq:kouchan-10.120})--(\ref{eq:kouchan-10.123}).
Then, the components of 
${}^{(1)}{\cal G}_{a}^{\;\;b}\left[{\cal L}\right]$ are given by
\begin{eqnarray}
  {}^{(1)}\!{\cal G}_{\eta}^{\;\;\eta} \left[{\cal L}\right]
  &=&
  - \frac{1}{a^{2}} \left\{
    \left(
      - 6 {\cal H} \partial_{\eta}
      + 2 \Delta
      + 6 K
    \right) \stackrel{(2)}{\Psi}
    - 6 {\cal H}^{2} \stackrel{(2)}{\Phi}
  \right\}
  \label{eq:kouchan-15.32}
  , \\
  {}^{(1)}\!{\cal G}_{i}^{\;\;\eta} \left[{\cal L}\right]
  &=&
  - \frac{1}{a^{2}}
  \left(
    2 \partial_{\eta} D_{i} \stackrel{(2)}{\Psi}
    + 2 {\cal H} D_{i} \stackrel{(2)}{\Phi}
    - \frac{1}{2} \left(
      \Delta
      + 2 K
    \right)
    \stackrel{(2)\;\;}{\nu_{i}}
  \right)
  \label{eq:kouchan-15.33}
  , \\
  {}^{(1)}\!{\cal G}_{\eta}^{\;\;i} \left[{\cal L}\right]
  &=&
  \frac{1}{a^{2}} \left\{
    2 \partial_{\eta} D^{i} \stackrel{(2)}{\Psi}
    + 2 {\cal H} D^{i} \stackrel{(2)}{\Phi}
    + \frac{1}{2} \left(
      - \Delta
      + 2 K
      + 4 {\cal H}^{2}
      - 4 \partial_{\eta}{\cal H}
    \right)
    \stackrel{(2)\;\;}{\nu^{i}}
  \right\}
  \label{eq:kouchan-15.34}
  , \\
  {}^{(1)}\!{\cal G}_{i}^{\;\;j} \left[{\cal L}\right]
  &=& 
  \frac{1}{a^{2}} \left[
    D_{i} D^{j} \left(\stackrel{(2)}{\Psi} - \stackrel{(2)}{\Phi}\right)
  \right.
  \nonumber\\
  &&
  \left.
    + 
    \left\{
      \left(
          2 \partial_{\eta}^{2} 
        -   \Delta
        + 4 {\cal H} \partial_{\eta}
        - 2 K
      \right)
      \stackrel{(2)}{\Psi}
      + \left(
          2 {\cal H} \partial_{\eta}
        + 4 \partial_{\eta}{\cal H}
        + 2 {\cal H}^{2}
        + \Delta
      \right)
      \stackrel{(2)}{\Phi}
    \right\}
    \gamma_{i}^{\;\;j}
  \right.
  \nonumber\\
  &&
  \left.
    - \frac{1}{2a^{2}} \partial_{\eta} \left\{
      a^{2} \left( 
        D_{i} \stackrel{(2)\;\;}{\nu^{j}} + D^{j} \stackrel{(2)\;\;}{\nu_{i}}
      \right)
    \right\}
  \right.
  \nonumber\\
  &&
  \left.
    + \frac{1}{2} \left(
      \partial_{\eta}^{2}
      + 2 {\cal H} \partial_{\eta}
      + 2 K
      - \Delta
    \right) \stackrel{(2)\;\;\;\;}{\chi_{i}^{\;\;j}}
  \right]
  .
  \label{eq:kouchan-15.35}
\end{eqnarray}

%*******************************************************************

In the simple situation descirbed by
Eq.~(\ref{eq:simplified-condition}), the components of the 
quadratic term 
${}^{(2)}{\cal G}_{a}^{\;\;b}\left[{\cal H},{\cal H}\right]$ 
of the linear order perturbations, which are
defined by Eq.~(\ref{eq:cal-G-def-second}), are given by 
\begin{eqnarray}
  {}^{(2)}\!{\cal G}_{\eta}^{\;\;\eta}
  &=&
  \frac{2}{a^{2}} 
  \left\{
      12   {\cal H} \partial_{\eta}\stackrel{(1)}{\Psi} \left(
        \stackrel{(1)}{\Psi} - \stackrel{(1)}{\Phi}
      \right)
    - 12 \left(
      K \left(\stackrel{(1)}{\Psi}\right)^{2}
      + {\cal H}^{2} \left(\stackrel{(1)}{\Phi}\right)^{2}
    \right)
  \right.
  \nonumber\\
  && \quad\quad\quad
  \left.
    -  3 \left(
      D_{k}\stackrel{(1)}{\Psi} D^{k}\stackrel{(1)}{\Psi}
      + \left(\partial_{\eta}\stackrel{(1)}{\Psi}\right)^{2}\right)
    -          8  \stackrel{(1)}{\Psi} \Delta \stackrel{(1)}{\Psi}
  \right\}
  \label{eq:kouchan-16.17}
  , \\
  {}^{(2)}\!{\cal G}_{\eta}^{\;\;i}
  &=&
  \frac{4}{a^{2}}
  \left\{
      2 {\cal H} D^{i}\stackrel{(1)}{\Phi} \left(
        \stackrel{(1)}{\Psi} - \stackrel{(1)}{\Phi}
    \right)
    +   \partial_{\eta}\stackrel{(1)}{\Psi} D^{i} \left(
      2 \stackrel{(1)}{\Psi} - \stackrel{(1)}{\Phi}
    \right)
    + 4 \stackrel{(1)}{\Psi} \partial_{\eta}D^{i}\stackrel{(1)}{\Psi}
  \right\}
  \label{eq:kouchan-16.18}
  , \\
  {}^{(2)}\!{\cal G}_{i}^{\;\;\eta}
  &=&
  \frac{4}{a^{2}}
  \left\{
      4 {\cal H} \stackrel{(1)}{\Phi} D_{i}\stackrel{(1)}{\Phi}
    -   \partial_{\eta}\stackrel{(1)}{\Psi} D_{i}\left(
      2 \stackrel{(1)}{\Psi} - \stackrel{(1)}{\Phi}
    \right)
    - 2 \partial_{\eta}D_{i}\stackrel{(1)}{\Psi} \left(
      \stackrel{(1)}{\Psi} - \stackrel{(1)}{\Phi}
    \right)
  \right\}
  \label{eq:kouchan-16.19}
  , \\
  {}^{(2)}\!{\cal G}_{i}^{\;\;j}
  &=&
  \frac{2}{a^{2}}
  \left[
        D_{i}\stackrel{(1)}{\Phi} D^{j}\left(
          \stackrel{(1)}{\Phi} - \stackrel{(1)}{\Psi}
    \right)
    -   D_{i}\stackrel{(1)}{\Psi} D^{j}\left(
      \stackrel{(1)}{\Phi} - 3 \stackrel{(1)}{\Psi}
    \right)
  \right.
  \nonumber\\
  && \quad\quad\quad
  \left.
    + 2 D_{i}D^{j}\stackrel{(1)}{\Phi} \left(
      \stackrel{(1)}{\Phi} - \stackrel{(1)}{\Psi}
    \right)
    + 4 \stackrel{(1)}{\Psi} D_{i}D^{j}\stackrel{(1)}{\Psi}
  \right.
  \nonumber\\
  && \quad\quad\quad
  \left.
    + \left\{
          \left(\partial_{\eta}\stackrel{(1)}{\Psi}\right)^{2}
      - 2 \partial_{\eta}\stackrel{(1)}{\Psi}
          \partial_{\eta}\stackrel{(1)}{\Phi}
      + 4 \partial_{\eta}^{2}\stackrel{(1)}{\Psi} \left(
        \stackrel{(1)}{\Psi} - \stackrel{(1)}{\Phi}
      \right)
      - 8 {\cal H} \stackrel{(1)}{\Phi} \partial_{\eta}\stackrel{(1)}{\Phi}
    \right.
  \right.
  \nonumber\\
  && \quad\quad\quad\quad\quad\quad
  \left.
    \left.
      + 8 {\cal H} \partial_{\eta}\stackrel{(1)}{\Psi} \left(
        \stackrel{(1)}{\Psi} - \stackrel{(1)}{\Phi}
      \right)
      - 4 \left(2\partial_{\eta}{\cal H}+{\cal H}^{2}\right)
          \left(\stackrel{(1)}{\Phi}\right)^{2}
      - 4 K \left(\stackrel{(1)}{\Psi}\right)^{2}
    \right.
  \right.
  \nonumber\\
  && \quad\quad\quad\quad\quad\quad
  \left.
    \left.
      - 2 D_{k}\stackrel{(1)}{\Psi} D^{k}\stackrel{(1)}{\Psi}
      -   D_{k}\stackrel{(1)}{\Phi} D^{k}\stackrel{(1)}{\Phi}
      + 2 \left(\stackrel{(1)}{\Psi}-\stackrel{(1)}{\Phi}\right) \Delta
          \stackrel{(1)}{\Phi} 
    \right.
  \right.
  \nonumber\\
  && \quad\quad\quad\quad\quad\quad
  \left.
    \left.
      - 4 \stackrel{(1)}{\Psi} \Delta \stackrel{(1)}{\Psi}
    \right\}
    \gamma_{i}^{\;\;j}
  \right]
  .
  \label{eq:kouchan-16.20}
\end{eqnarray}
Through the components
(\ref{eq:einstein-conformal-coordiante-calH}) of the background
Einstein tensor $G_{a}^{\;\;b}$, the components
(\ref{eq:kouchan-10.34})--(\ref{eq:kouchan-10.39}) of the tensor
field $H_{ab}^{\;\;\;\;\;c}$, and the components
(\ref{eq:kouchan-10.120})--(\ref{eq:kouchan-10.123}) of the
gauge invariant part 
${}^{(1)}{\cal G}_{a}^{\;\;b}\left[{\cal H}\right]$ of the
linear-order perturbation of the Einstein tensor, it is
straightforward to confirm that 
${}^{(2)}\!{\cal G}_{a}^{\;\;b}\left[{\cal H},{\cal H}\right]$
of the components (\ref{eq:kouchan-16.17})--(\ref{eq:kouchan-16.20})
satisfies the second-order perturbation of the Bianchi identity 
(\ref{eq:second-div-of-calGab-1,1}), i.e.,
\begin{eqnarray}
  \nabla_{a}{}^{(2)}{\cal G}_{b}^{\;\;a}\left[{\cal H},{\cal H}\right] 
  &=& 
  - 2 H_{ca}^{\;\;\;\;a}\left[{\cal H}\right]
  \; {}^{(1)}\!{\cal G}_{b}^{\;\;c}\left[{\cal H}\right]
  + 2 H_{ba}^{\;\;\;\;e}\left[{\cal H} \right]
  \; {}^{(1)}\!{\cal G}_{e}^{\;\;a}\left[{\cal H}\right]
  \nonumber\\
  &&
  - 2 H_{bad}\left[{\cal H}\right]\; {\cal H}^{dc} G_{c}^{\;\;a}
  + 2 H_{cad}\left[{\cal H}\right]\; {\cal H}^{ad} G_{b}^{\;\;c}.
  \label{eq:second-div-of-calGab-2,0}
\end{eqnarray}
This implies that we have consistently derived the components
(\ref{eq:kouchan-16.17})--(\ref{eq:kouchan-16.20}).

%*******************************************************************

Further, we impose the relation
(\ref{eq:absence-of-anisotropic-stress-Einstein-i-j-traceless-scalar}),
which can always be derived from the first-order Einstein
equation when the anisotropic stress of scalar type is
negligible.
Then, the gauge invariant part of the first-order metric
perturbation is given by
\begin{eqnarray}
  {\cal H}_{ab} = 
  - 2 a^{2} \stackrel{(1)}{\Phi} (d\eta)_{a}(d\eta)_{b}
  - 2 a^{2} \stackrel{(1)}{\Phi} \gamma_{ij} (dx^{i})_{a}(dx^{j})_{b},
\end{eqnarray}
and the components
(\ref{eq:kouchan-16.17})--(\ref{eq:kouchan-16.20}) of the tensor
${}^{(2)}\!{\cal G}_{a}^{\;\;b}\left[{\cal H},{\cal H}\right]$
are reduced as follows: 
\begin{eqnarray}
  {}^{(2)}\!{\cal G}_{\eta}^{\;\;\eta}
  &=&
  - \frac{2}{a^{2}} 
  \left\{
      3  D_{k}\stackrel{(1)}{\Phi} D^{k}\stackrel{(1)}{\Phi}
    + 3 \left(\partial_{\eta}\stackrel{(1)}{\Phi}\right)^{2}
    +  8 \stackrel{(1)}{\Phi} \Delta \stackrel{(1)}{\Phi}
  \right.
  \nonumber\\
  && \quad\quad\quad
  \left.
    + 12 \left(K + {\cal H}^{2}\right) \left(\stackrel{(1)}{\Phi}\right)^{2}
  \right\}
  ,
  \label{eq:second-cal-G-eta-eta-simple-2}
  \\
  {}^{(2)}\!{\cal G}_{i}^{\;\;\eta}
  &=&
  \frac{4}{a^{2}}
  \left(
      4 {\cal H} \stackrel{(1)}{\Phi} D_{i}\stackrel{(1)}{\Phi}
    -   \partial_{\eta}\stackrel{(1)}{\Phi} D_{i}\stackrel{(1)}{\Phi}
  \right)
  ,
  \label{eq:second-cal-G-i-eta-simple-2}
  \\
  {}^{(2)}\!{\cal G}_{\eta}^{\;\;i}
  &=&
  \frac{4}{a^{2}}
  \left(
        \partial_{\eta}\stackrel{(1)}{\Phi} D^{i}\stackrel{(1)}{\Phi}
    + 4 \stackrel{(1)}{\Phi} \partial_{\eta}D^{i}\stackrel{(1)}{\Phi}
  \right)
  ,
  \label{eq:second-cal-G-eta-i-simple-2}
  \\
  {}^{(2)}\!{\cal G}_{i}^{\;\;j}
  &=&
  \frac{2}{a^{2}}
  \left[
      2 D_{i}\stackrel{(1)}{\Phi} D^{j}\stackrel{(1)}{\Phi}
    + 4 \stackrel{(1)}{\Phi} D_{i}D^{j}\stackrel{(1)}{\Phi}
  \right.
  \nonumber\\
  && \quad\quad
  \left.
    - \left(
        3 D_{k}\stackrel{(1)}{\Phi} D^{k}\stackrel{(1)}{\Phi}
      + 4 \stackrel{(1)}{\Phi} \Delta \stackrel{(1)}{\Phi}
      +   \left(\partial_{\eta}\stackrel{(1)}{\Phi}\right)^{2}
    \right.
  \right.
  \nonumber\\
  && \quad\quad\quad\quad
  \left.
    \left.
      + 8 {\cal H} \stackrel{(1)}{\Phi} \partial_{\eta}\stackrel{(1)}{\Phi}
      + 4 \left(
        2 \partial_{\eta}{\cal H} + K + {\cal H}^{2}
      \right) \left(\stackrel{(1)}{\Phi}\right)^{2}
    \right)
    \gamma_{i}^{\;\;j}
  \right].
  \label{eq:second-cal-G-i-j-simple-2}
\end{eqnarray}
Thus, we have evaluated 
${}^{(1)}{\cal G}_{a}^{\;\;b}\left[{\cal L}\right]$
and ${}^{(2)}{\cal G}_{a}^{\;\;b}\left[{\cal H},{\cal H}\right]$
in the second-order Einstein equation
(\ref{eq:second-order-Einstein-equation}).
Next, we evaluate ${}^{(2)}{\cal T}_{a}^{\;\;b}$ separately in
two cases, that of a universe filled with a single perfect fluid
and that of the universe filled with a single scalar field.

%*******************************************************************

%%%%%%%%%%%%%%%%%%%%%%%%%%%%%%%%%%%%%%%%%%%%%%%%%%%%%%%%%%%%%%%%%%%%%
\subsection{Perfect fluid case}
\label{sec:Perfect-fluid-case-second-order}

%*******************************************************************

Here, we consider the second-order perturbation of the Einstein
equation in the case of a universe filled with a
single-component perfect fluid.

%*******************************************************************

Because we concentrate only on the case in which the vector and
tensor modes are negligible, as in
Eq.~(\ref{eq:simplified-condition}), we should ignore the 
divergenceless part of the spatial velocity of the fluid, setting
\begin{eqnarray}
  \stackrel{(1)\;\;}{{\cal V}_{i}} = 0,
  \label{eq:first-order-vorticity-negligible}
\end{eqnarray}
in accordance with the first-order Einstein equation
(\ref{eq:linearized-Einstein-i-eta-vector}).
Then, the components of the gauge invariant first-order
perturbation of the fluid four-velocity are given by
\begin{eqnarray}
  \stackrel{(1)}{{\cal U}_{a}} 
  = 
  - a \stackrel{(1)}{\Phi} (d\eta)_{a}
  + a D_{i}\stackrel{(1)}{v} (dx^{i})_{a}
\end{eqnarray}
In the simple situation that
Eqs.~(\ref{eq:simplified-condition}) and
(\ref{eq:first-order-vorticity-negligible}) are satisfied, the
normalization condition
(\ref{eq:four-vel-norm-second-pert-gauge-inv-2}) of the
second-order perturbation $\stackrel{(2)}{{\cal U}_{a}}$ of the
fluid four-velocity is given by 
\begin{eqnarray}
  u^{a} \stackrel{(2)}{{\cal U}_{a}} 
  &=&
  \left(\stackrel{(1)}{\Phi}\right)^{2}
  - D_{i}\stackrel{(1)}{v} D^{i}\stackrel{(1)}{v}
  - \stackrel{(2)}{\Phi}
  ,
\end{eqnarray}
and the components of $\stackrel{(2)}{{\cal U}_{a}}$ are
decomposed as
\begin{eqnarray}
  \stackrel{(2)}{{\cal U}_{a}}
  =
  a \left(
    \left(\stackrel{(1)}{\Phi}\right)^{2}
    - D_{i}\stackrel{(1)}{v} D^{i}\stackrel{(1)}{v}
    - \stackrel{(2)}{\Phi}
  \right) (d\eta)_{a}
  +
  a \left(
    D_{i} \stackrel{(2)}{v}
    +
    \stackrel{(2)}{{\cal V}_{i}}
  \right) (dx^{i})_{a},
\end{eqnarray}
where
\begin{eqnarray}
  D^{i} \stackrel{(2)}{{\cal V}_{i}} = 0.
\end{eqnarray}
Then, the components of the gauge invariant part
${}^{(2)}\!{\cal T}_{a}^{\;\;b}$ of the second-order
perturbation of the energy-momentum tensor are given by 
\begin{eqnarray}
  {}^{(2)}\!{\cal T}_{\eta}^{\;\;\eta}
  &=&
  - 2 (\epsilon + p) D^{i}\stackrel{(1)}{v} D_{i}\stackrel{(1)}{v} 
  - \stackrel{(2)}{{\cal E}}
  , \\
  {}^{(2)}\!{\cal T}_{i}^{\;\;\eta}
  &=&
  (\epsilon + p)\left(
    D_{i}\stackrel{(2)}{v} 
    + \stackrel{(2)}{{\cal V}}_{i}
    - 2 \stackrel{(1)}{\Phi} D_{i}\stackrel{(1)}{v}
  \right)
  + 2 \left(
    \stackrel{(1)}{{\cal E}}+\stackrel{(1)}{{\cal P}}
  \right) D_{i}\stackrel{(1)}{v}
  , \\
  {}^{(2)}\!{\cal T}_{\eta}^{\;\;i}
  &=&
  (\epsilon + p) \left(
    \stackrel{(2)}{\nu^{i}}
    - D^{i}\stackrel{(2)}{v} 
    - \stackrel{(2)}{{\cal V}^{i}}
    - 6 \stackrel{(1)}{\Phi} D^{i}\stackrel{(1)}{v}
  \right)
  - 2 \left(
    \stackrel{(1)}{{\cal E}} + \stackrel{(1)}{{\cal P}}
  \right) D^{i}\stackrel{(1)}{v}
  , \\
  {}^{(2)}\!{\cal T}_{i}^{\;\;j}
  &=&
  2 (\epsilon+p) D_{i}\stackrel{(1)}{v} D^{j}\stackrel{(1)}{v}
  + \stackrel{(2)}{{\cal P}} \gamma_{i}^{\;\;j}
  .
\end{eqnarray}
Hence, through the background Einstein equations
(\ref{eq:background-Einstein-equations}) and
(\ref{eq:background-Einstein-equations-2}), and the first-order
perturbation of the Einstein equations
(\ref{eq:linearized-Einstein-energy-perturbation})--(\ref{eq:linearized-Einstein-perssure-perturbation}),
the Einstein equations for the second-order perturbations are
obtained as
\begin{eqnarray}
  &&
  \left(
    - 3 {\cal H} \partial_{\eta}
    +   \Delta
    + 3 K
  \right) \stackrel{(2)}{\Psi}
  - 3 {\cal H}^{2} \stackrel{(2)}{\Phi}
  - 4\pi G a^{2} \stackrel{(2)}{{\cal E}}
  =
  \Gamma_{0}
  ,
  \label{eq:kouchan-18.53}
  \\
  &&
    2 \partial_{\eta}D_{i}\stackrel{(2)}{\Psi}
  + 2 {\cal H} D_{i}\stackrel{(2)}{\Phi}
  - \frac{1}{2} \left(
    \Delta
    + 2 K
  \right)
  \stackrel{(2)\;\;}{\nu_{i}}
  \nonumber\\
  && \quad\quad\quad\quad
  + 8\pi G a^{2} (\epsilon + p)
  \left(
    D_{i}\stackrel{(2)}{v} 
    + \stackrel{(2)\;\;}{{\cal V}}_{i}
  \right)
  =
  \Gamma_{i},
  \label{eq:kouchan-18.55}
  \\
  && 
  D_{i} D^{j} \left(\stackrel{(2)}{\Psi} - \stackrel{(2)}{\Phi}\right)
  \nonumber\\
  && \quad\quad
  + 
  \left\{
    \left(
        2 \partial_{\eta}^{2} 
      -   \Delta
      + 4 {\cal H} \partial_{\eta}
      - 2 K
    \right)
    \stackrel{(2)}{\Psi}
    + \left(
      2 {\cal H} \partial_{\eta}
      + 4 \partial_{\eta}{\cal H}
      + 2 {\cal H}^{2}
      + \Delta
    \right)
    \stackrel{(2)}{\Phi}
  \right\}
  \gamma_{i}^{\;\;j}
  \nonumber\\
  && \quad\quad
  - \frac{1}{2a^{2}} \partial_{\eta} \left\{
    a^{2} \left( 
      D_{i} \stackrel{(2)\;\;}{\nu^{j}} + D^{j} \stackrel{(2)\;\;}{\nu_{i}}
    \right)
  \right\}
  \nonumber\\
  && \quad\quad
  + \frac{1}{2} \left(
    \partial_{\eta}^{2}
    + 2 {\cal H} \partial_{\eta}
    + 2 K
    - \Delta
  \right) \stackrel{(2)\;\;\;\;}{\chi_{i}^{\;\;j}}
  \nonumber\\
  && \quad\quad
  - 8\pi G a^{2} \stackrel{(2)}{{\cal P}} \gamma_{i}^{\;\;j}
  =
  \Gamma_{i}^{\;\;j}
  ,
  \label{eq:kouchan-18.57}
\end{eqnarray}
where
\begin{eqnarray}
  \Gamma_{0}
  &:=&
  8\pi G a^{2} (\epsilon + p) D^{i}\stackrel{(1)}{v} D_{i}\stackrel{(1)}{v} 
  \nonumber\\
  && 
  - 3 D_{k}\stackrel{(1)}{\Phi} D^{k}\stackrel{(1)}{\Phi}
  - 3 \left(\partial_{\eta}\stackrel{(1)}{\Phi}\right)^{2}
  -  8 \stackrel{(1)}{\Phi} \Delta \stackrel{(1)}{\Phi}
  - 12 \left( K + {\cal H}^{2} \right) \left(\stackrel{(1)}{\Phi}\right)^{2}
  ,
  \label{eq:kouchan-18.54}
  \\
  \Gamma_{i}
  &:=&
  - 16 \pi G a^{2} \left(
     \stackrel{(1)}{{\cal E}} + \stackrel{(1)}{{\cal P}}
   \right)D_{i}\stackrel{(1)}{v}
  \nonumber\\
  && 
  +         12  {\cal H} \stackrel{(1)}{\Phi} D_{i}\stackrel{(1)}{\Phi}
  -          4  \stackrel{(1)}{\Phi} \partial_{\eta}D_{i}\stackrel{(1)}{\Phi}
  -          4  \partial_{\eta}\stackrel{(1)}{\Phi} D_{i}\stackrel{(1)}{\Phi}
  ,
  \label{eq:kouchan-18.56}
  \\
  \Gamma_{i}^{\;\;j}
  &:=&
    16 \pi G a^{2} (\epsilon+p) D_{i}\stackrel{(1)}{v} D^{j}\stackrel{(1)}{v}
  -  4  D_{i}\stackrel{(1)}{\Phi} D^{j}\stackrel{(1)}{\Phi}
  -  8  \stackrel{(1)}{\Phi} D_{i}D^{j}\stackrel{(1)}{\Phi}
  \nonumber\\
  && 
  + 2 \left(
      3 D_{k}\stackrel{(1)}{\Phi} D^{k}\stackrel{(1)}{\Phi}
    + 4 \stackrel{(1)}{\Phi} \Delta \stackrel{(1)}{\Phi}
    +   \left(\partial_{\eta}\stackrel{(1)}{\Phi}\right)^{2}
  \right.
  \nonumber\\
  && \quad\quad\quad
  \left.
    + 4 \left(
      2 \partial_{\eta}{\cal H} + K + {\cal H}^{2}
    \right) \left(\stackrel{(1)}{\Phi}\right)^{2}
    + 8 {\cal H} \stackrel{(1)}{\Phi} \partial_{\eta}\stackrel{(1)}{\Phi}
  \right)
  \gamma_{i}^{\;\;j}
    .
  \label{eq:kouchan-18.58}
\end{eqnarray}

%*******************************************************************

Equation (\ref{eq:kouchan-18.55}) can be decomposed into scalar
and vector parts.
Taking the divergence of (\ref{eq:kouchan-18.55}), we obtain 
\begin{eqnarray}
  8\pi G a^{2} (\epsilon + p) D_{i}\stackrel{(2)}{v} 
  =
  - 2 \partial_{\eta}D_{i}\stackrel{(2)}{\Psi}
  - 2 {\cal H} D_{i}\stackrel{(2)}{\Phi}
  + D_{i} \Delta^{-1} D^{k}\Gamma_{k}
  .
  \label{eq:kouchan-18.59}
\end{eqnarray}
Then subtracting Eq.~(\ref{eq:kouchan-18.59}) from
Eq.~(\ref{eq:kouchan-18.55}), we obtain
\begin{eqnarray}
  8\pi G a^{2} (\epsilon + p) \stackrel{(2)}{{\cal V}}_{i}
  =
  \frac{1}{2} \left(
    \Delta
    + 2 K
  \right)
  \stackrel{(2)}{\nu_{i}}
  +\left(
    \Gamma_{i}
    -
    D_{i} \Delta^{-1} D^{k} \Gamma_{k}
  \right).
  \label{eq:kouchan-18.59-vorticity}
\end{eqnarray}

%*******************************************************************

Equation (\ref{eq:kouchan-18.57}) can be decomposed into the
trace part, and the traceless part.
This traceless part of Eq.~(\ref{eq:kouchan-18.57}) can be
decomposed into the scalar, vector, and tensor parts.  
The trace part of (\ref{eq:kouchan-18.57}) is given by 
\begin{eqnarray}
  && 
  \left(
      \partial_{\eta}^{2} 
    + 2 {\cal H} \partial_{\eta}
    - \frac{1}{3} \Delta
    - K
  \right)
  \stackrel{(2)}{\Psi}
  + \left(
    {\cal H} \partial_{\eta}
    + 2 \partial_{\eta}{\cal H}
    + {\cal H}^{2}
    + \frac{1}{3} \Delta
  \right)
  \stackrel{(2)}{\Phi}
  \nonumber\\
  && \quad\quad
  - 4 \pi G a^{2} \stackrel{(2)}{{\cal P}}
  =
  \frac{1}{6} \Gamma_{k}^{\;\;k}
  .
  \label{eq:trace-spatial-second-order-Einstein-fluid}
\end{eqnarray}
The traceless part of Eq.~(\ref{eq:kouchan-18.57}) is given by
\begin{eqnarray}
  && 
  \left(
    D_{i}D_{j} - \frac{1}{3} \gamma_{ij} \Delta
  \right) 
  \left( \stackrel{(2)}{\Psi} - \stackrel{(2)}{\Phi} \right)
  - \frac{1}{a^{2}} \partial_{\eta} \left(
    a^{2} D_{(i} \stackrel{(2)\;\;}{\nu_{j)}}
  \right)
  \nonumber\\
  && \quad\quad\quad
  + \frac{1}{2} \left(
    \partial_{\eta}^{2}
    + 2 {\cal H} \partial_{\eta}
    + 2 K
    - \Delta
  \right) \stackrel{(2)\;\;\;\;}{\chi_{ij}}
  =
  \Gamma_{ij}
  - \frac{1}{3} \gamma_{ij} \Gamma_{k}^{\;\;k}
  ,
  \label{eq:kouchan-18.62}
\end{eqnarray}
where $\Gamma_{ij} = \gamma_{jk}\Gamma_{i}^{\;\;k}$.
Taking the divergence of Eq.~(\ref{eq:kouchan-18.62}), we
obtain 
\begin{eqnarray}
  &&
  \left(
    \frac{2}{3} D_{i}\Delta + 2 K D_{i}
  \right) 
  \left( \stackrel{(2)}{\Psi} - \stackrel{(2)}{\Phi} \right)
  - \frac{1}{2a^{2}} \partial_{\eta} \left\{
    a^{2} \left( \Delta + 2 K \right) \stackrel{(2)}{\nu_{i}}
  \right\}
  \nonumber\\
  &=&
  D_{j}\Gamma_{i}^{\;\;j} - \frac{1}{3} D_{i}\Gamma_{k}^{\;\;k}
  .
  \label{eq:kouchan-18.63}
\end{eqnarray}
Further, taking the divergence of Eq.~(\ref{eq:kouchan-18.63}),
we have 
\begin{eqnarray}
  \frac{2}{3} \left( \Delta + 3 K \right) \Delta
  \left( \stackrel{(2)}{\Psi} - \stackrel{(2)}{\Phi} \right)
  =
  D^{i}D_{j}\Gamma_{i}^{\;\;j} - \frac{1}{3} \Delta\Gamma_{k}^{\;\;k}
  .
  \label{eq:kouchan-18.64}
\end{eqnarray}
Thus, we have extracted the scalar part in traceless part
(\ref{eq:kouchan-18.62}) of Eq.~(\ref{eq:kouchan-18.57}), which
is given by 
\begin{eqnarray}
  \stackrel{(2)}{\Psi} - \stackrel{(2)}{\Phi}
  =
  \frac{3}{2}
  \left( \Delta + 3 K \right)^{-1}
  \left(
    \Delta^{-1} D^{i}D_{j}\Gamma_{i}^{\;\;j}
    - \frac{1}{3} \Gamma_{k}^{\;\;k}
  \right)
  .
  \label{eq:kouchan-18.65}
\end{eqnarray}
Substituting (\ref{eq:kouchan-18.65}) into
(\ref{eq:kouchan-18.63}), we obtain the vector part of
Eq.~(\ref{eq:kouchan-18.57}),
\begin{eqnarray}
  \partial_{\eta} \left(a^{2}\stackrel{(2)}{\nu_{i}}\right)
  =
  2a^{2}
  \left( \Delta + 2 K \right)^{-1}
  \left\{
    D_{i} \Delta^{-1} D^{k}D_{l}\Gamma_{k}^{\;\;l}
    - D_{k}\Gamma_{i}^{\;\;k}
  \right\}
  ,
  \label{eq:kouchan-18.66}
\end{eqnarray}
and substituting Eqs.~(\ref{eq:kouchan-18.65}) and
(\ref{eq:kouchan-18.66}), we obtain the tensor part of
Eq.~(\ref{eq:kouchan-18.57}),
\begin{eqnarray}
  && 
  \left(
    \partial_{\eta}^{2} + 2 {\cal H} \partial_{\eta} + 2 K  - \Delta
  \right)
  \stackrel{(2)\;\;\;\;}{\chi_{ij}}
  \nonumber\\
  &=&
  2 \Gamma_{ij}
  - \frac{2}{3} \gamma_{ij} \Gamma_{k}^{\;\;k}
  - 3
  \left(
    D_{i}D_{j} - \frac{1}{3} \gamma_{ij} \Delta
  \right) 
  \left( \Delta + 3 K \right)^{-1}
  \left(
    \Delta^{-1} D^{k}D_{l}\Gamma_{k}^{\;\;l}
    - \frac{1}{3} \Gamma_{k}^{\;\;k}
  \right)
  \nonumber\\
  &&
  + 4
  \left( 
    D_{(i}\left( \Delta + 2 K \right)^{-1}
    D_{j)}\Delta^{-1}D^{l}D_{k}\Gamma_{l}^{\;\;k}
    - D_{(i} \left( \Delta + 2 K \right)^{-1} D^{k}\Gamma_{j)k}
  \right)
  .
  \label{eq:kouchan-18.68}
\end{eqnarray}
Equations (\ref{eq:kouchan-18.66}) and (\ref{eq:kouchan-18.68})
imply that the second-order vector and tensor modes may be
generated by the scalar-scalar mode coupling of the first-order
perturbation if accidental cancellation in the source term does
not occur.

%*******************************************************************

Further, the equations of the scalar mode perturbations
(\ref{eq:kouchan-18.53}) and
(\ref{eq:trace-spatial-second-order-Einstein-fluid}) are reduced
to a single equation for $\stackrel{(2)}{\Phi}$ as follows.
Substituting (\ref{eq:kouchan-18.65}) into
(\ref{eq:kouchan-18.53}), (\ref{eq:kouchan-18.59}), and
(\ref{eq:trace-spatial-second-order-Einstein-fluid}), the
second-order perturbation of the energy density, the scalar part
of the 
spatial velocity, and the pressure of the fluid are given by
\begin{eqnarray}
  4\pi G a^{2} \stackrel{(2)}{{\cal E}}
  &=&
  \left(
    - 3 {\cal H} \partial_{\eta}
    +   \Delta
    + 3 K
    - 3 {\cal H}^{2}
  \right)
  \stackrel{(2)}{\Phi}
  - \Gamma_{0}
  \nonumber\\
  &&
  +
  \frac{3}{2}
  \left(
    \Delta^{-1} D^{i}D_{j}\Gamma_{i}^{\;\;j}
    - \frac{1}{3} \Gamma_{k}^{\;\;k}
  \right)
  \nonumber\\
  && 
  -
  \frac{9}{2}
  {\cal H} \partial_{\eta}
  \left( \Delta + 3 K \right)^{-1}
  \left(
    \Delta^{-1} D^{i}D_{j}\Gamma_{i}^{\;\;j}
    - \frac{1}{3} \Gamma_{k}^{\;\;k}
  \right)
  \label{eq:kouchan-18.79}
  , \\
  8\pi G a^{2} (\epsilon + p) D_{i}\stackrel{(2)}{v} 
  &=&
  - 2 \partial_{\eta}D_{i}\stackrel{(2)}{\Phi}
  - 2 {\cal H} D_{i}\stackrel{(2)}{\Phi}
  + D_{i} \Delta^{-1} D^{k}\Gamma_{k}
  \nonumber\\
  &&
  - 3 \partial_{\eta}D_{i}
  \left( \Delta + 3 K \right)^{-1}
  \left(
    \Delta^{-1} D^{i}D_{j}\Gamma_{i}^{\;\;j}
    - \frac{1}{3} \Gamma_{k}^{\;\;k}
  \right)
  \label{eq:second-velocity-scalar-part-Einstein}
  , \\
  4 \pi G a^{2} \stackrel{(2)}{{\cal P}}
  &=&
  \left(
      \partial_{\eta}^{2} 
    + 3{\cal H} \partial_{\eta}
    - K
    + 2\partial_{\eta}{\cal H}
    + {\cal H}^{2}
  \right)
  \stackrel{(2)}{\Phi}
  \nonumber\\
  &&
  +
  \frac{3}{2}
  \left(
        \partial_{\eta}^{2} 
    + 2 {\cal H} \partial_{\eta}
  \right)
  \left( \Delta + 3 K \right)^{-1}
  \left(
    \Delta^{-1} D^{i}D_{j}\Gamma_{i}^{\;\;j}
    - \frac{1}{3} \Gamma_{k}^{\;\;k}
  \right)
  \nonumber\\
  &&
  - \frac{1}{2} \Delta^{-1} D^{i}D_{j}\Gamma_{i}^{\;\;j}
  .
  \label{eq:kouchan-18.80}
\end{eqnarray}
Through the second-order perturbation
(\ref{eq:second-order-equation-of-state-gauge-inv}) of the
equation of state, Eqs.~(\ref{eq:kouchan-18.79}) and
(\ref{eq:kouchan-18.80}) yield a single equation for
$\stackrel{(2)}{\Phi}$:
\begin{eqnarray}
  &&
  \left\{
      \partial_{\eta}^{2} 
    + 3 {\cal H} (1 + c_{s}^{2}) \partial_{\eta}
    -   c_{s}^{2} \Delta
    + 2 \partial_{\eta}{\cal H}
    + (1 + 3 c_{s}^{2}) ({\cal H}^{2} - K)
  \right\}
  \stackrel{(2)}{\Phi}
  \nonumber\\
  &=&
  4\pi G a^{2} \left\{
    \tau \stackrel{(2)}{{\cal S}}
    + \frac{\partial c_{s}^{2}}{\partial\epsilon}
    \left(\stackrel{(1)}{{\cal E}}\right)^{2}
    + 2 \frac{\partial c_{s}^{2}}{\partial S}
    \stackrel{(1)}{{\cal E}}
    \stackrel{(1)}{{\cal S}}
    + \frac{\partial\tau}{\partial S}
    \left(\stackrel{(1)}{{\cal S}}\right)^{2}
  \right\}
  \nonumber\\
  && \quad
  - c_{s}^{2} \left(
    \Gamma_{0} + \frac{1}{2} \Gamma_{k}^{\;\;k}
  \right)
  +
  \frac{3}{2}
  \left(c_{s}^{2} + \frac{1}{3}\right)
  \Delta^{-1} D^{i}D_{j}\Gamma_{i}^{\;\;j}
  \nonumber\\
  && \quad
  -
  \frac{3}{2}
  \left(
      \partial_{\eta}^{2} 
    + \left( 2 + 3 c_{s}^{2} \right) {\cal H} \partial_{\eta}
  \right)
  \left( \Delta + 3 K \right)^{-1}
  \left(
    \Delta^{-1} D^{i}D_{j}\Gamma_{i}^{\;\;j}
    - \frac{1}{3} \Gamma_{k}^{\;\;k}
  \right)
  .
  \label{eq:second-order-Einstein-scalar-master-eq}
\end{eqnarray}
This is the second-order extension of the master equation
(\ref{eq:linearized-Einstein-scalar-master-eq}) of the scalar
mode perturbations in the case of a universe filled with a
perfect fluid.
Actually, if the quadratic terms of the first order
perturbations on the right-hand side of
Eq.~(\ref{eq:second-order-Einstein-scalar-master-eq}) are
absent, this equation coincides with
Eq.~(\ref{eq:linearized-Einstein-scalar-master-eq}).

%*******************************************************************

To solve the system of second-order perturbations of the
Einstein tensor, we have to carry out the following process.
First, we solve the first-order master equation
(\ref{eq:linearized-Einstein-scalar-master-eq}) for the
perturbations.
The solution to
Eq.~(\ref{eq:linearized-Einstein-scalar-master-eq}) gives the
energy density, the velocity, and the pressure perturbation of
the fluid through
Eqs.~(\ref{eq:linearized-Einstein-energy-perturbation})--(\ref{eq:linearized-Einstein-perssure-perturbation}).
Next, we evaluate the source term of
(\ref{eq:second-order-Einstein-scalar-master-eq}) and solve this
equation.
Since the homogeneous solutions to
Eq.~(\ref{eq:second-order-Einstein-scalar-master-eq}) coincide
with the solutions to
Eq.~(\ref{eq:linearized-Einstein-scalar-master-eq}), which are
known as the growing and decaying modes of linear
perturbation theory, the general solution to
Eq.~(\ref{eq:second-order-Einstein-scalar-master-eq}) is given
by an inhomogeneous solution to
Eq.~(\ref{eq:second-order-Einstein-scalar-master-eq}), together
with the growing and decaying modes of the linear-order scalar 
mode perturbation $\stackrel{(1)}{\Phi}$ with arbitrary
coefficients.
Once we obtain the solution to
Eq.~(\ref{eq:second-order-Einstein-scalar-master-eq}), we can
also obtain the energy density, velocity, pressure perturbation
at second order through
Eqs.~(\ref{eq:kouchan-18.79})--(\ref{eq:kouchan-18.80}).

%*******************************************************************

Further, we have equations for the vector and tensor modes
(\ref{eq:kouchan-18.59-vorticity}), (\ref{eq:kouchan-18.66}),
and (\ref{eq:kouchan-18.68}).
Once we obtain the solution to the linearized Einstein equations,
(\ref{eq:linearized-Einstein-scalar-master-eq}) and
(\ref{eq:linearized-Einstein-energy-perturbation})--(\ref{eq:linearized-Einstein-perssure-perturbation}),
we can evaluate the quadratic terms of the linear-order
perturbations in Eqs.~(\ref{eq:kouchan-18.59-vorticity}),
(\ref{eq:kouchan-18.66}), and (\ref{eq:kouchan-18.68}).
The evolution of the vector mode of the second-order metric
perturbation is determined by Eq.~(\ref{eq:kouchan-18.66}), and
the rotational part $\stackrel{(2)}{{\cal V}}_{i}$ of the
spatial velocity of the fluid is determined by
Eq.~(\ref{eq:kouchan-18.59-vorticity}).
The tensor mode, i.e., the gravitational wave mode, at second
order is determined by Eq.~(\ref{eq:kouchan-18.68}). 
Since the homogeneous solutions to Eq.~(\ref{eq:kouchan-18.68})
coincide with the solutions to
Eq.~(\ref{eq:linearized-Einstein-i-j-traceless-tensor}), the
general solution to Eq.~(\ref{eq:kouchan-18.68}) is also given
by an inhomogeneous solution to Eq.~(\ref{eq:kouchan-18.68}),
together with two independent solutions to
(\ref{eq:linearized-Einstein-i-j-traceless-tensor}) of linear
order with arbitrary coefficients.

%*******************************************************************

Of course, we need the additional information concerning the
entropy perturbations $\stackrel{(1)}{{\cal S}}$ and
$\stackrel{(2)}{{\cal S}}$ at each order to determine the first- 
and second-order perturbation\cite{Kodama-Sasaki-IJMPA}.
Once we obtain this information, all modes of the second-order
perturbation are determined by the above second-order
perturbation equations, 
(\ref{eq:kouchan-18.59-vorticity}),
(\ref{eq:kouchan-18.65})--(\ref{eq:second-order-Einstein-scalar-master-eq})
of the Einstein equation.
This is one of the main results of this paper.

%**************************************************************

%%%%%%%%%%%%%%%%%%%%%%%%%%%%%%%%%%%%%%%%%%%%%%%%%%%%%%%%%%%%%%%%%%%%%
\subsection{Scalar field case}
%%%%%%%%%%%%%%%%%%%%%%%%%%%%%%%%%%%%%%%%%%%%%%%%%%%%%%%%%%%%%%%%%%%%%

%**************************************************************

In the simple situation in which the first-order vector and
tensor modes are negligible, the components of the second-order
perturbation
(\ref{eq:second-order-energy-momentum-scalar-gauge-inv}) of the
energy-momentum tensor for a single scalar field are
given by 
\begin{eqnarray}
  {}^{(2)}\!{\cal T}_{\eta}^{\;\;\eta}
  &=&
  -   \frac{1}{a^{2}} 
  \left( 
        \partial_{\eta}\varphi \partial_{\eta}\varphi_{2} 
    - 4 \partial_{\eta}\varphi \stackrel{(1)}{\Phi} \partial_{\eta}\varphi_{1}
    + 4 (\partial_{\eta}\varphi)^{2} \left(\stackrel{(1)}{\Phi}\right)^{2}
    -   (\partial_{\eta}\varphi)^{2} \stackrel{(2)}{\Phi}
    +   (\partial_{\eta}\varphi_{1})^{2}
  \right.
  \nonumber\\
  && \quad\quad\quad\quad
  \left.
    +   D_{i}\varphi_{1} D^{i}\varphi_{1}
    +   a^{2} \varphi_{2} \frac{\partial V}{\partial\varphi}
    +   a^{2} (\varphi_{1})^{2} \frac{\partial^{2}V}{\partial\varphi^{2}}
  \right)
  , \\
  {}^{(2)}\!{\cal T}_{i}^{\;\;\eta}
  &=&
  - \frac{1}{a^{2}}
  \left(
        D_{i}\varphi_{2} \partial_{\eta}\varphi  
    + 2 D_{i}\varphi_{1} \partial_{\eta}\varphi_{1}
    - 4 D_{i}\varphi_{1} \stackrel{(1)}{\Phi} \partial_{\eta}\varphi 
  \right)
  , \\
  {}^{(2)}\!{\cal T}_{\eta}^{\;\;i}
  &=&
  \frac{1}{a^{2}}
  \left(
        \partial_{\eta}\varphi D^{i}\varphi_{2} 
    + 4 \partial_{\eta}\varphi \stackrel{(1)}{\Psi} D^{i}\varphi_{1} 
    +   (\partial_{\eta}\varphi)^{2} \stackrel{(2)}{\nu^{i}}
    + 2 \partial_{\eta}\varphi_{1} D^{i}\varphi_{1}
  \right)
  , \\
  {}^{(2)}\!{\cal T}_{i}^{\;\;j}
  &=&
    2 \frac{1}{a^{2}} D_{i}\varphi_{1} D^{j}\varphi_{1}
  \nonumber\\
  &&
  +   \frac{1}{a^{2}} \gamma_{i}^{\;\;j}
  \left(
    +   \partial_{\eta}\varphi \partial_{\eta}\varphi_{2}
    - 4 \partial_{\eta}\varphi \stackrel{(1)}{\Phi} \partial_{\eta}\varphi_{1}
    + 4 (\partial_{\eta}\varphi)^{2} \left(\stackrel{(1)}{\Phi}\right)^{2}
    -   (\partial_{\eta}\varphi)^{2} \stackrel{(2)}{\Phi}
  \right.
  \nonumber\\
  && \quad\quad\quad\quad\quad\quad
  \left.
    +   (\partial_{\eta}\varphi_{1})^{2}
    -   D_{k}\varphi_{1} D^{k}\varphi_{1} 
    -   a^{2} \varphi_{2}\frac{\partial V}{\partial\varphi}
    -   a^{2} \varphi_{1}^{2}\frac{\partial^{2}V}{\partial\varphi^{2}}
  \right)
  .
\end{eqnarray}

%**************************************************************

Through Eqs.~(\ref{eq:background-Einstein-equations-scalar-3})
and (\ref{eq:kouchan-18.186}), the components 
${}^{(1)}{\cal G}_{\eta}^{\;\;i}\left[{\cal L}\right]+{}^{(2)}{\cal G}_{\eta}^{\;\;i}\left[{\cal H},{\cal H}\right]=8\pi G\;\;{}^{(2)}{\cal T}_{\eta}^{\;\;i}$
and
${}^{(1)}{\cal G}_{i}^{\;\;\eta}\left[{\cal L}\right]+{}^{(2)}{\cal G}_{i}^{\;\;\eta}\left[{\cal H},{\cal H}\right]=8\pi G\;\;{}^{(2)}{\cal T}_{i}^{\;\;\eta}$
of the second-order Einstein equation
(\ref{eq:second-order-Einstein-equation}) give the single equation 
\begin{eqnarray}
  &&
    2 \partial_{\eta} D_{i} \stackrel{(2)}{\Psi}
  + 2 {\cal H} D_{i} \stackrel{(2)}{\Phi}
  - \frac{1}{2} \left(
    \Delta
    + 2 K
  \right)
  \stackrel{(2)}{\nu_{i}}
  -
  8\pi G D_{i}\varphi_{2} \partial_{\eta}\varphi  
  = 
  \Gamma_{i}
  ,
  \label{eq:kouchan-18.199}
\end{eqnarray}
where
\begin{eqnarray}
  \Gamma_{i}
  &:=&
  -          4  \partial_{\eta}\stackrel{(1)}{\Phi} D_{i}\stackrel{(1)}{\Phi}
  +          8  {\cal H} \stackrel{(1)}{\Phi} D_{i}\stackrel{(1)}{\Phi}
  -          8  \stackrel{(1)}{\Phi} \partial_{\eta} D_{i}\stackrel{(1)}{\Phi}
  +         16 \pi G D_{i}\varphi_{1} \partial_{\eta}\varphi_{1} 
  .
  \label{eq:kouchan-18.200}
\end{eqnarray}
Taking the divergence of Eq.~(\ref{eq:kouchan-18.199}), we
obtain the scalar part of Eq.~(\ref{eq:kouchan-18.199}),
\begin{eqnarray}
  &&
    2 \partial_{\eta} \stackrel{(2)}{\Psi}
  + 2 {\cal H} \stackrel{(2)}{\Phi}
  -
  8\pi G \varphi_{2} \partial_{\eta}\varphi  
  = 
  \Delta^{-1} D^{k} \Gamma_{k}.
  \label{eq:kouchan-18.199-2}
\end{eqnarray}
Subtracting Eq.~(\ref{eq:kouchan-18.199-2}) from
Eq.~(\ref{eq:kouchan-18.199}), we obtain the vector part of
Eq.(\ref{eq:kouchan-18.199}),
\begin{eqnarray}
  &&
  \stackrel{(2)}{\nu_{i}}
  = 
  2 ( \Delta + 2 K )^{-1}
  \left\{
    D_{i} \Delta^{-1} D^{k} \Gamma_{k}
    - \Gamma_{i}
  \right\}
  \label{eq:kouchan-18.199-3}
  .
\end{eqnarray}

%**************************************************************

Through Eqs.~(\ref{eq:background-Einstein-equations-scalar-3})
and (\ref{eq:scalar-linearized-Einstein-scalar-master-eq-pre}),
the component 
${}^{(1)}{\cal G}_{\eta}^{\;\;\eta}\left[{\cal L}\right]+{}^{(2)}{\cal G}_{\eta}^{\;\;\eta}\left[{\cal H},{\cal H}\right]=8\pi G \;\; {}^{(2)}{\cal T}_{\eta}^{\;\;\eta}$
of the second-order Einstein equation
(\ref{eq:second-order-Einstein-equation}) gives
\begin{eqnarray}
  &&
  \left(
    - 3 {\cal H} \partial_{\eta}
    +   \Delta
    + 3 K
  \right) \stackrel{(2)}{\Psi}
  +
  \left(
    -   \partial_{\eta}{\cal H}
    - 2 {\cal H}^{2}
    +   K
  \right)
  \stackrel{(2)}{\Phi} 
  \nonumber\\
  && \quad\quad\quad\quad\quad\quad\quad\quad
  - 4 \pi G 
  \left(
        \partial_{\eta}\varphi \partial_{\eta}\varphi_{2} 
    +   a^{2} \varphi_{2} \frac{\partial V}{\partial\varphi}
  \right)
  =
  \Gamma_{0}
  ,
  \label{eq:kouchan-18.218}
\end{eqnarray}
where
\begin{eqnarray}
  &&
  \Gamma_{0}
  :=
  -          2  \stackrel{(1)}{\Phi} \partial_{\eta}^{2} \stackrel{(1)}{\Phi} 
  -          3  \left(\partial_{\eta}\stackrel{(1)}{\Phi}\right)^{2}
  -          3  D_{k}\stackrel{(1)}{\Phi} D^{k}\stackrel{(1)}{\Phi}
  -         10  \stackrel{(1)}{\Phi} \Delta \stackrel{(1)}{\Phi}
  \nonumber\\
  && \quad\quad\quad\quad\quad
  -          4  \left( \partial_{\eta}{\cal H} + 4 K + 2 {\cal H}^{2} \right)
                \left(\stackrel{(1)}{\Phi}\right)^{2}
  \nonumber\\
  && \quad\quad\quad\quad\quad
  + 4 \pi G 
  \left(
        (\partial_{\eta}\varphi_{1})^{2}
    +   D_{k}\varphi_{1} D^{k}\varphi_{1}
    +   a^{2} \varphi_{1}^{2} \frac{\partial^{2}V}{\partial\varphi^{2}}
  \right)
  .
\end{eqnarray}

%*******************************************************************

Similarly, through
Eqs.~(\ref{eq:background-Einstein-equations-scalar-3}) and
(\ref{eq:scalar-linearized-Einstein-scalar-master-eq-pre}), the
component 
${}^{(1)}{\cal G}_{i}^{\;\;j}\left[{\cal L}\right]+{}^{(2)}{\cal G}_{i}^{\;\;j}\left[{\cal H},{\cal H}\right]=8\pi G \;\;{}^{(2)}{\cal T}_{i}^{\;\;j}$
of the second-order Einstein equation
(\ref{eq:second-order-Einstein-equation}) yields  
\begin{eqnarray}
  && 
  D_{i} D_{j} \left( \stackrel{(2)}{\Psi} - \stackrel{(2)}{\Phi} \right)
  \nonumber\\
  && 
  + 
  \left\{
    \left(
      -   \Delta
      + 2 \partial_{\eta}^{2} 
      + 4 {\cal H} \partial_{\eta}
      - 2 K
    \right)
    \stackrel{(2)}{\Psi}
    + \left(
        2 {\cal H} \partial_{\eta}
      + 2 \partial_{\eta}{\cal H}
      + 4 {\cal H}^{2}
      + \Delta
      + 2 K
    \right)
    \stackrel{(2)}{\Phi}
  \right\}
  \gamma_{ij}
  \nonumber\\
  &&
  - \frac{1}{a^{2}} \partial_{\eta} \left(
    a^{2} D_{(i} \stackrel{(2)}{\nu_{j)}}
  \right)
  \nonumber\\
  &&
  + \frac{1}{2} \left(
    \partial_{\eta}^{2}
    + 2 {\cal H} \partial_{\eta}
    + 2 K
    - \Delta
  \right) \stackrel{(2)}{\chi}_{ij}
  \nonumber\\
  &&
  - 8 \pi G \left(
    \partial_{\eta}\varphi\partial_{\eta}\varphi_{2}
    - a^{2} \varphi_{2}\frac{\partial V}{\partial\varphi}(\varphi)
  \right) \gamma_{ij} = \Gamma_{ij}
  \label{eq:kouchan-18.207}
  ,
\end{eqnarray}
where
\begin{eqnarray}
  \Gamma_{ij}
  &:=&
  -          4  D_{i}\stackrel{(1)}{\Phi} D_{j}\stackrel{(1)}{\Phi}
  -          8  \stackrel{(1)}{\Phi} D_{i}D_{j}\stackrel{(1)}{\Phi}
  + 16 \pi G D_{i}\varphi_{1} D_{j}\varphi_{1}
  \nonumber\\
  && 
  + 2 \left(
       8 {\cal H} \stackrel{(1)}{\Phi} \partial_{\eta}\stackrel{(1)}{\Phi}
    -  2  \stackrel{(1)}{\Phi} \partial_{\eta}^{2}\stackrel{(1)}{\Phi}
    +     \left(\partial_{\eta}\stackrel{(1)}{\Phi}\right)^{2}
    +  3  D_{k}\stackrel{(1)}{\Phi} D^{k}\stackrel{(1)}{\Phi}
  \right.
  \nonumber\\
  && \quad\quad
  \left.
    +  2  \stackrel{(1)}{\Phi} \Delta \stackrel{(1)}{\Phi}
    +  4  \left( \partial_{\eta}{\cal H} + 2 {\cal H}^{2} \right)
                  \left(\stackrel{(1)}{\Phi}\right)^{2}
  \right.
  \nonumber\\
  && \quad\quad
  \left.
    + 4 \pi G
    \left(
        (\partial_{\eta}\varphi_{1})^{2}
      - D_{k}\varphi_{1} D^{k}\varphi_{1} 
      - a^{2} (\varphi_{1})^{2}\frac{\partial^{2}V}{\partial\varphi^{2}}
    \right)
  \right)
  \gamma_{ij}
  .
  \label{eq:kouchan-18.208}
\end{eqnarray}
As seen in the case of a perfect fluid in
\S\ref{sec:Perfect-fluid-case-second-order},
Eq.~(\ref{eq:kouchan-18.207}) can be decomposed into the trace
and the traceless parts. 
Further, the traceless part Eq.~(\ref{eq:kouchan-18.207}) is 
decomposed into the scalar, vector, and tensor parts.
The trace part of Eq.~(\ref{eq:kouchan-18.207}) is given by
\begin{eqnarray}
  && 
  \left(
                  \partial_{\eta}^{2} 
    +          2  {\cal H} \partial_{\eta}
    - \frac{1}{3} \Delta
    -             K
  \right)
  \stackrel{(2)}{\Psi}
  + \left(
                  {\cal H} \partial_{\eta}
    +             \partial_{\eta}{\cal H}
    +          2  {\cal H}^{2}
    + \frac{1}{3} \Delta
    +             K
  \right)
  \stackrel{(2)}{\Phi}
  \nonumber\\
  && \quad\quad\quad\quad\quad\quad\quad\quad\quad\quad\quad\quad
  - 4 \pi G \left(
    \partial_{\eta}\varphi\partial_{\eta}\varphi_{2}
    - a^{2} \varphi_{2}\frac{\partial V}{\partial\varphi}(\varphi)
  \right)
  = \frac{1}{6} \Gamma_{k}^{\;\;k}
  ,
  \label{eq:kouchan-18.221}
\end{eqnarray}
where $\Gamma_{k}^{\;\;k} = \gamma^{ij}\Gamma_{ij}$.
The traceless scalar part of Eq.~(\ref{eq:kouchan-18.207}) is
given by 
\begin{eqnarray}
  \stackrel{(2)}{\Psi} - \stackrel{(2)}{\Phi}
  = 
  \frac{3}{2}
  (\Delta + 3 K)^{-1}
  \left\{
    \Delta^{-1} D^{i}D_{j}\Gamma_{i}^{\;\;j} - \frac{1}{3} \Gamma_{k}^{\;\;k}
  \right\},
  \label{eq:kouchan-18.213}
\end{eqnarray}
where $\Gamma_{i}^{\;\;j} := \gamma^{kj}\Gamma_{ik}$.
The vector part of Eq.~(\ref{eq:kouchan-18.207}) is given by 
\begin{eqnarray}
  \partial_{\eta}
  \left(
    a^{2} \stackrel{(2)}{\nu_{i}}
  \right)
  =
  2 a^{2}
  (\Delta + 2 K)^{-1} 
  \left\{
    D_{i}\Delta^{-1} D^{k}D_{l}\Gamma_{k}^{\;\;l}
    - D_{k}\Gamma_{i}^{\;\;k}
  \right\}
  .
  \label{eq:kouchan-18.214}
\end{eqnarray}
Finally, the tensor part of Eq.~(\ref{eq:kouchan-18.207}) is
given by 
\begin{eqnarray}
  && 
  \left(
    \partial_{\eta}^{2} + 2 {\cal H} \partial_{\eta} + 2 K  - \Delta
  \right)
  \stackrel{(2)\;\;\;\;}{\chi_{ij}}
  \nonumber\\
  &=&
  2 \Gamma_{ij}
  - \frac{2}{3} \gamma_{ij} \Gamma_{k}^{\;\;k}
  - 3
  \left(
    D_{i}D_{j} - \frac{1}{3} \gamma_{ij} \Delta
  \right) 
  \left( \Delta + 3 K \right)^{-1}
  \left(
    \Delta^{-1} D^{k}D_{l}\Gamma_{k}^{\;\;l}
    - \frac{1}{3} \Gamma_{k}^{\;\;k}
  \right)
  \nonumber\\
  &&
  + 4
  \left\{ 
      D_{(i} (\Delta+2K)^{-1} D_{j)}\Delta^{-1}D^{l}D_{k}\Gamma_{l}^{\;\;k}
    - D_{(i}(\Delta+2K)^{-1}D^{k}\Gamma_{j)k}
  \right\}
  .
  \label{eq:kouchan-18.215}
\end{eqnarray}

%*******************************************************************

From Eqs.~(\ref{eq:kouchan-18.199-2}),
(\ref{eq:kouchan-18.218}), (\ref{eq:kouchan-18.221}), and
(\ref{eq:kouchan-18.213}), we obtain a single equation for the
second-order perturbation $\stackrel{(2)}{\Phi}$ as
Eq.~(\ref{eq:second-order-Einstein-scalar-master-eq}) in the
case of a perfect fluid considered in
\S\ref{sec:Perfect-fluid-case-second-order}.
Combining Eqs.~(\ref{eq:kouchan-18.218}) and
(\ref{eq:kouchan-18.221}), we have
\begin{eqnarray}
  && 
  \left(
                  \partial_{\eta}^{2} 
    -             {\cal H} \partial_{\eta}
    + \frac{2}{3} \Delta
    +          2  K
  \right)
  \stackrel{(2)}{\Psi}
  + \left(
                  {\cal H} \partial_{\eta}
    + \frac{1}{3} \Delta
    +          2  K
  \right)
  \stackrel{(2)}{\Phi}
  \nonumber\\
  && \quad\quad\quad\quad\quad\quad\quad\quad\quad\quad\quad\quad\quad\quad
  - 8 \pi G \partial_{\eta}\varphi \partial_{\eta}\varphi_{2}
  = \Gamma_{0} + \frac{1}{6} \Gamma_{k}^{\;\;k},
  \label{eq:kouchan-18.228}
\end{eqnarray}
and 
\begin{eqnarray}
  &&
  \left(
    -   \partial_{\eta}^{2} 
    - 5 {\cal H} \partial_{\eta}
    + \frac{4}{3} \Delta
    + 4 K
  \right) \stackrel{(2)}{\Psi}
  -
  \left(
      2 \partial_{\eta}{\cal H}
    +   {\cal H} \partial_{\eta}
    + 4 {\cal H}^{2}
    +   \frac{1}{3} \Delta
  \right)
  \stackrel{(2)}{\Phi} 
  \nonumber\\
  && \quad\quad\quad\quad\quad\quad\quad\quad\quad\quad\quad\quad\quad\quad\quad\quad\quad
  - 8 \pi G a^{2} \varphi_{2} \frac{\partial V}{\partial\varphi}
  =
  \Gamma_{0} - \frac{1}{6} \Gamma_{k}^{\;\;k}
  .
  \label{eq:kouchan-18.228-2}
\end{eqnarray}
Further, substituting Eq.~(\ref{eq:kouchan-18.213}) into
Eq.~(\ref{eq:kouchan-18.228}), we obtain
\begin{eqnarray}
  && 
  \left(
                  \partial_{\eta}^{2} 
    +             \Delta
    +          4  K
  \right)
  \stackrel{(2)}{\Phi}
  - 8 \pi G \partial_{\eta}\varphi \partial_{\eta}\varphi_{2}
  \nonumber\\
  &=& 
  \Gamma_{0}
  + \frac{1}{6} \Gamma_{k}^{\;\;k}
  -
  \Delta^{-1} D^{i}D_{j}\Gamma_{i}^{\;\;j} + \frac{1}{3} \Gamma_{k}^{\;\;k}
  \nonumber\\
  &&
  -
  \frac{3}{2}
  \left(
    \partial_{\eta}^{2} 
    - {\cal H} \partial_{\eta}
  \right)
  (\Delta + 3 K)^{-1}
  \left\{
    \Delta^{-1} D^{i}D_{j}\Gamma_{i}^{\;\;j} - \frac{1}{3} \Gamma_{k}^{\;\;k}
  \right\}
  \label{eq:kouchan-18.230}
\end{eqnarray}
On the other hand, differentiating
Eq.~(\ref{eq:kouchan-18.199-2}) with respect to the conformal
time $\eta$, we obtain
\begin{eqnarray}
  - 8\pi G \partial_{\eta}\varphi_{2} \partial_{\eta}\varphi
  &=& 
  \left(
    \frac{2\partial_{\eta}^{2}\varphi}{\partial_{\eta}\varphi}\partial_{\eta}
    - 2 \partial_{\eta}^{2}
  \right)
  \stackrel{(2)}{\Psi}
  + 
  \left(
    \frac{2\partial_{\eta}^{2}\varphi}{\partial_{\eta}\varphi} {\cal H}
    - 2 \partial_{\eta}{\cal H}
    - 2 {\cal H} \partial_{\eta}
  \right)
  \stackrel{(2)}{\Phi} 
  \nonumber\\
  &&
  - \frac{\partial_{\eta}^{2}\varphi}{\partial_{\eta}\varphi}
  \Delta^{-1}D^{k}\Gamma_{k} 
  + \partial_{\eta}\Delta^{-1}D^{k}\Gamma_{k}
  \nonumber\\
  &=& 
  \left(
    - 2 \partial_{\eta}^{2}
    - 2 {\cal H} \partial_{\eta}
    + \frac{2\partial_{\eta}^{2}\varphi}{\partial_{\eta}\varphi}\partial_{\eta}
    + \frac{2\partial_{\eta}^{2}\varphi}{\partial_{\eta}\varphi} {\cal H}
    - 2 \partial_{\eta}{\cal H}
  \right)
  \stackrel{(2)}{\Phi}
  \nonumber\\
  &&
  + \frac{3}{2}
  \left(
    \frac{2\partial_{\eta}^{2}\varphi}{\partial_{\eta}\varphi}\partial_{\eta}
    - 2 \partial_{\eta}^{2}
  \right)
  (\Delta + 3 K)^{-1}
  \left\{
    \Delta^{-1} D^{i}D_{j}\Gamma_{i}^{\;\;j} - \frac{1}{3} \Gamma_{k}^{\;\;k}
  \right\}
  \nonumber\\
  &&
  - \frac{\partial_{\eta}^{2}\varphi}{\partial_{\eta}\varphi}
  \Delta^{-1}D^{k}\Gamma_{k} 
  + \partial_{\eta}\Delta^{-1}D^{k}\Gamma_{k}
  ,
  \label{eq:kouchan-18.232}
\end{eqnarray}
where we have again used Eq.~(\ref{eq:kouchan-18.213}).
Substituting (\ref{eq:kouchan-18.232}) into
(\ref{eq:kouchan-18.230}), we obtain the master equation:
\begin{eqnarray}
  && 
  \left\{
    \partial_{\eta}^{2}
    + 2 \left(
      {\cal H}
      - \frac{\partial_{\eta}^{2}\varphi}{\partial_{\eta}\varphi}
    \right)
    \partial_{\eta}
    -             \Delta
    -          4  K
    + 2 \left(
      \partial_{\eta}{\cal H}
      - \frac{\partial_{\eta}^{2}\varphi}{\partial_{\eta}\varphi} {\cal H}
    \right)
  \right\}
  \stackrel{(2)}{\Phi}
  \nonumber\\
  &=& 
  - \Gamma_{0}
  - \frac{1}{2} \Gamma_{k}^{\;\;k}
  +
  \Delta^{-1} D^{i}D_{j}\Gamma_{i}^{\;\;j}
  + \left(
    \partial_{\eta}
    - \frac{\partial_{\eta}^{2}\varphi}{\partial_{\eta}\varphi}
  \right)
  \Delta^{-1}D^{k}\Gamma_{k}
  \nonumber\\
  && 
  -
  \frac{3}{2}
  \left\{
    \partial_{\eta}^{2}
    - \left(
      \frac{2\partial_{\eta}^{2}\varphi}{\partial_{\eta}\varphi} - {\cal H}
    \right)
    \partial_{\eta}
  \right\}
  (\Delta + 3 K)^{-1}
  \left\{
    \Delta^{-1} D^{i}D_{j}\Gamma_{i}^{\;\;j} - \frac{1}{3} \Gamma_{k}^{\;\;k}
  \right\}.
  \label{eq:kouchan-18.233}
\end{eqnarray}
This is the second-order extension of
Eq.~(\ref{eq:scalar-linearized-Einstein-scalar-master-eq}),
which is the master equation of scalar mode of the second-order
cosmological perturbation in a universe filled with a single
scalar field.
This equation would coincide with
Eq.~(\ref{eq:scalar-linearized-Einstein-scalar-master-eq}) if
the quadratic terms of the linear-order perturbations were
absent.

%*******************************************************************

Thus, we have a set of ten equations for the second-order
perturbations of a universe filled with a single scalar field,
Eqs.~(\ref{eq:kouchan-18.199-2}),
(\ref{eq:kouchan-18.199-3}),
(\ref{eq:kouchan-18.213})--(\ref{eq:kouchan-18.215}),
(\ref{eq:kouchan-18.228-2}), and (\ref{eq:kouchan-18.233}).
To solve this system, we have to solve the linear-order system
first.
Next, we evaluate the quadratic terms, $\Gamma_{0}$,
$\Gamma_{i}$ and $\Gamma_{ij}$, of the linear-order
perturbations. 
Then, using the information of the quadratic terms of the
linear-order perturbation, we estimate the source term in 
Eq.~(\ref{eq:kouchan-18.233}).
The general solution to Eq.~(\ref{eq:kouchan-18.233}) is given
by an inhomogeneous solution to Eq.~(\ref{eq:kouchan-18.233})
and two independent homogeneous solutions of this equation with
arbitrary constants.
Since Eq.~(\ref{eq:kouchan-18.233}) is the same as
Eq.~(\ref{eq:scalar-linearized-Einstein-scalar-master-eq}),
except for the source term, which consists of the quadratic terms
of the linear-order perturbations, the homogeneous solutions to
(\ref{eq:kouchan-18.233}) coincide with the solutions to the
linear-order perturbations, $\stackrel{(1)}{\Phi}$.
Hence, we can construct the general solution to
Eq.~(\ref{eq:kouchan-18.233}) if we obtain a special solution of
this equation.
After constructing the solution $\stackrel{(2)}{\Phi}$ to
Eq.~(\ref{eq:kouchan-18.233}) , we can obtain the second-order
metric perturbation $\stackrel{(2)}{\Psi}$ through
Eq.~(\ref{eq:kouchan-18.213}).
Thus, we have obtained the second-order gauge invariant
perturbation $\varphi_{2}$ of the scalar field through
Eq.~(\ref{eq:kouchan-18.199-2}).
Equation (\ref{eq:kouchan-18.228-2}) is then used to check the
consistency of the second-order perturbation of the Klein Gordon
equation,
\begin{eqnarray}
  \bar{\nabla}^{a}\bar{\nabla}_{a}\bar{\varphi}
  + \frac{\partial V}{\partial\bar{\varphi}} = 0.
\end{eqnarray}

%*******************************************************************

Evaluating the source terms in Eq.~(\ref{eq:kouchan-18.215})
through the evaluation of the quadratic terms $\Gamma_{0}$,
$\Gamma_{i}$, and $\Gamma_{ij}$ of the linear-order
perturbations, we can solve Eq.~(\ref{eq:kouchan-18.215}).
We also note that Eq.~(\ref{eq:kouchan-18.215}) is identical to
Eq.~(\ref{eq:kouchan-18.68}), except for the definition 
of the quadratic terms.
As in the case of a perfect fluid, this equation implies that a
scalar mode perturbation of linear order may generat the
second-order tensor mode, i.e., the second-order gravitational
waves if accidental cancellation in the source term does not
occur.

%*******************************************************************

For the vector-mode perturbation $\stackrel{(1)\;\;}{\nu_{i}}$,
the situation is different from that in the case of a perfect
fluid.
Since there is no rotational component in a single scalar
field system, there is no rotational spatial component of the
velocity of the matter field, i.e., there is no vorticity.
Instead, in addition to the evolution equation
(\ref{eq:kouchan-18.214}) of the vector mode, we have the
initial value constraint (\ref{eq:kouchan-18.199-3}) of the
vector mode.
However, the constraint (\ref{eq:kouchan-18.199-3}) and the
evolution equation (\ref{eq:kouchan-18.214}) also imply that the
second-order vector-mode perturbation may be generated by the
scalar-scalar mode coupling of the linear order perturbations.

%*******************************************************************

Thus, we have formulated a second-order perturbation theory
of a universe filled with a single scalar field.
All modes of the second-order perturbation are determined by the
above second-order perturbation equations,
(\ref{eq:kouchan-18.199-2}), (\ref{eq:kouchan-18.199-3}),
(\ref{eq:kouchan-18.213})--(\ref{eq:kouchan-18.215}),
(\ref{eq:kouchan-18.228-2}), and (\ref{eq:kouchan-18.233}).
This and the result in the case of a perfect fluid are the main
results of this paper.

%*******************************************************************

%%%%%%%%%%%%%%%%%%%%%%%%%%%%%%%%%%%%%%%%%%%%%%%%%%%%%%%%%%%%%%%%%%%%%%
\section{Summary and discussions}
\label{sec:summary}
%%%%%%%%%%%%%%%%%%%%%%%%%%%%%%%%%%%%%%%%%%%%%%%%%%%%%%%%%%%%%%%%%%%%%%

%*******************************************************************

In summary, we have confirmed that the general formulation of
the second-order perturbation theory developed in
KN2003\cite{kouchan-gauge-inv} and KN2005\cite{kouchan-second} 
is applicable to cosmological perturbation theory.
We have shown that the method for finding higher-order gauge
invariant variables proposed in KN2003 does work in
the case of cosmological perturbations.
The key point of our method is the assumption that we already
know the procedure for decomposing the first-order metric
perturbation into gauge invariant and variant parts. 
In particular, to apply this method to higher-order
perturbations, we have to find the gauge variant vector field 
$X_{a}$ in the first-order metric perturbation, which is defined
by Eq.~(\ref{eq:linear-metric-decomp}).
As shown in \S\ref{sec:First-order-metric-perturbations}, the
vector field $X_{a}$ is found by restricting the domain of the
perturbations to the space of functions in which the Green
functions $\Delta^{-1}$, $(\Delta+2K)^{-1}$, and
$(\Delta+3K)^{-1}$ can be defined, where $\Delta$ and $K$ are 
the Laplacian associated with the metric $\gamma_{ij}$ and the
curvature constant of the maximally symmetric three-space,
respectively.
As a result, we found gauge invariant variables for the
first-order metric perturbation that are discussed in the 
literature\cite{Bardeen-1980,Kodama-Sasaki-1984,Mukhanov-Feldman-Brandenberger-1992}.
The resulting gauge invariant metric perturbation has the same
form as the metric perturbation described by the longitudinal
gauge (the Newtonian gauge).
This result for linear metric perturbations was then extended to
second-order perturbations, as proposed in KN2003.
In this way, we obtain second-order gauge invariant metric
perturbations whose components are similar to those of
linear metric perturbations in the longitudinal gauge. 
If we apply the gauge fixing so that $X_{a}=Y_{a}=0$, where
$X_{a}$ and $Y_{a}$ are the first- and second-order gauge
variant parts of the metric perturbations studied in
\S\ref{sec:Gauge-Invariant-Variables-of-Cosmological-Perturbations},
this gauge fixing corresponds to the Poisson gauge in the
literature\cite{Non-Gaussianity-in-CMB}, which is the
higher-order extension of the longitudinal gauge. 
Further, as proposed in KN2003, we can also define the
gauge invariant variables for the perturbations of an arbitrary
field other than the metric, as shown in
Eqs.~(\ref{eq:matter-gauge-inv-decomp-1.0}) and
(\ref{eq:matter-gauge-inv-decomp-2.0}).

%*******************************************************************

We have also shown that the formulae
(\ref{eq:matter-gauge-inv-decomp-1.0}) and
(\ref{eq:matter-gauge-inv-decomp-2.0}) of the decomposition of
the gauge invariant and variant parts for an arbitrary field
other than the metric play crucial roles in the second-order
gauge invariant perturbation theory.
We have seen that the first- and second-order perturbative
energy momentum-tensor of a perfect fluid
[Eqs.~(\ref{eq:first-ene-mon-tensor-decomp})--(\ref{eq:second-ene-mon-tensor-gauge-inv-def})]
and a single scalar field
[Eqs.~(\ref{eq:first-order-energy-momentum-scalar-decomp})--(\ref{eq:second-order-energy-momentum-scalar-gauge-inv})]
can be decomposed into gauge invariant and variant parts in the
same forms as Eqs.~(\ref{eq:matter-gauge-inv-decomp-1.0}) and
(\ref{eq:matter-gauge-inv-decomp-2.0}).
Here again, we note that no background values of the fluid
components, the scalar field, nor the metric were explicitly used
in the derivation of these expressions.
We thus conclude that
Eqs.~(\ref{eq:first-ene-mon-tensor-decomp})--(\ref{eq:second-ene-mon-tensor-gauge-inv-def})
and
Eqs.~(\ref{eq:first-order-energy-momentum-scalar-decomp})--(\ref{eq:second-order-energy-momentum-scalar-gauge-inv})
are valid in the perturbation theory on any background
spacetime.

%*******************************************************************

Further, we have derived the perturbed Einstein equations of
first and second order in terms of the above gauge invariant 
variables.
As shown in KN2005, each order of the Einstein equations can be
written in terms of only the gauge invariant variables, and 
we do not have to consider the gauge degree of freedom when we
treat the perturbed Einstein equations.
Though it is well known that the first-order metric perturbation 
in a homogeneous and isotropic universe can be decomposed into
scalar, vector, and tensor types, we have shown that the
second-order metric perturbation can also be decomposed into
these three types.
In the perturbation theory at linear order, these three types of
perturbations are decoupled.
However, at higher orders, these three types of perturbations
are coupled.
At second order, this mode-mode coupling appears as a source
term which consists of quadratic terms of the linear-order
perturbations. 
Because the scalar mode of the perturbations yields the dominant 
contribution in many cosmological contexts, we consider only the
situation in which the first-order vector and the tensor modes
are negligible.
Even in this simple situation, the second-order Einstein
equations imply that the second-order vector and tensor 
modes may be generated by the scalar-scalar mode coupling.
Further, we have shown that the Einstein equations for the
second-order perturbations have forms similar to those for
the linear-order perturbations, but there are source terms due
to the quadratic terms of the linear-order perturbations.
Since we have developed the perturbation theory order by order,
the Einstein equations for the second-order perturbations can be
solved using techniques for linear-order perturbations.
In particular, we have also seen that the second-order Einstein
equations for the scalar mode perturbations are reduced to
single equations in the cases of both a perfect fluid and a
scalar field.
The resulting equations have forms similar to those for the
linear-order perturbations, but there is a source term which 
consists of the quadratic terms of the linear-order perturbations.

%*******************************************************************

Now, we discuss the definitions of the gauge invariant variables
found in the literature.
It is well known that there are many definitions of the gauge
invariant variables for density 
perturbation\cite{Bardeen-1980,Kodama-Sasaki-1984}.
Thus, there is no uniqueness in the definitions of the gauge
invariant variables.
This results from the fact that we can always construct new
gauge invariant quantities from combinations of other gauge
invariant variables.
In many works, the interpretation of gauge invariant quantities
is based on the coincidence of the perturbative variables in an
appropriate gauge choice. 
For example, 
{\it the gauge invariant variable $\stackrel{(1)}{{\cal E}}$
  defined by Eq.~(\ref{eq:kouchan-016.13}) describes the density
  perturbation, because the variable $\stackrel{(1)}{{\cal E}}$
  coincides with $\stackrel{(1)}{\epsilon}$, defined in
  Eq.~(\ref{eq:kouchan-16.2}) in the gauge choice $X^{\eta}=0$}.
This criterion for the interpretation of a gauge invariant
variable for density perturbations produces many different
definitions of the density perturbations, as pointed out in the
literature\cite{Bardeen-1980,Kodama-Sasaki-1984}.
However, we have to recall that the gauge choice is the point
identification map between the physical spacetime 
${\cal M}_{\lambda}$ and the background spacetime 
${\cal M}_{0}$, as reviewed
\S\ref{sec:General-framework-of-the-gauge-invariant-perturbation-theory}.
The concept of the gauge choice does not exist for the
physical spacetime; it has meaning only if we introduce a
reference manifold, ${\cal M}_{0}$. 
Moreover, because all variables on the physical spacetime, 
${\cal M}_{\lambda}$, are pulled back to the background spacetime,
${\cal M}_{0}$, these pulled-back variables necessarily depend
on the gauge choice, in general.
This gauge dependence is due to the point identification map
between ${\cal M}_{\lambda}$ and ${\cal M}_{0}$, and it is not
due to the nature of ${\cal M}_{\lambda}$ itself.
Thus, the density perturbation $\stackrel{(1)}{\epsilon}$ in
Eq.~(\ref{eq:kouchan-16.2}) is defined by the pull back of the
gauge choice and depends on the gauge choice.
However, this gauge dependence is not due to the nature of the
physical spacetime nor the background spacetime themselves.
We should emphasize that the physical spacetime is described
only by the gauge invariant quantities, and hence physical
density perturbation is the gauge invariant variables
$\stackrel{(1)}{{\cal E}}$, not $\stackrel{(1)}{\epsilon}$.
Thus, it is meaningless to interpret the variables
$\stackrel{(1)}{{\cal E}}$ in terms of its coincidence with
$\stackrel{(1)}{\epsilon}$ in some gauge choice, because the
physical meaning of the variable $\stackrel{(1)}{\epsilon}$ may
depend on the gauge choices, whereas the physical meaning of the
gauge invariant variable $\stackrel{(1)}{{\cal E}}$ does not
depend on this choice.
To understand why the gauge invariant variable
$\stackrel{(1)}{{\cal E}}$ has the meaning of the energy
density, it is enough to point out two facts.
First, note that the gauge variant part $X_{a}$ of the
pulled-back metric perturbation $h_{ab}$ is not a variable on
the physical spacetime, which arises from the gauge choice,
i.e., teh point identification map between the
physical spacetime ${\cal M}_{\lambda}$ and the background
spacetime ${\cal M}_{0}$.
The information regarding the gauge choice is provided by the
gauge variant part, $X_{a}$, of the metric perturbation, but
there is no information regarding the physical spacetime in the
gauge variant part of the metric perturbation.
Second, $\stackrel{(1)}{{\cal E}}$ consists of only the
variables of the energy density, its perturbation, and the
vector field $X_{a}$ of the gauge choice.
Hence, the only possible physical meaning of
$\stackrel{(1)}{{\cal E}}$ is that it is the energy density
perturbation.
Similarly, and more generally, the gauge invariant variables
defined by Eqs.~(\ref{eq:matter-gauge-inv-def-1.0}) and
(\ref{eq:matter-gauge-inv-def-2.0}) for the first- and
second-order perturbations of an arbitrary matter field $Q$ have
the physical meaning of the first- and second-order perturbation
of the physical variable $Q$, respectively.

%****************************************************************

Of course, we can ask which variable is useful when we have many
definitions of the gauge invariant variables. 
This question is a different from the main point discussed
above.
To answer this question, we have to specify the problem which
we want to clarify, and we have to specify for what the
variables are useful.
A partial answer to this question should be provided by the
correspondence found between the gauge invariant variables and
observables in observations and experiments.
The gauge invariant variables which are useful to study physical
processes should appear in these physical processes. 
Therefore, the question of which variable is useful is reduced to
the question of which variable appears in the physical processes.
Because observations and experiments are constructed from some
physical processes, it would be interesting to investigate the
correspondence between gauge invariant variables and observables
in observations and experiments. 
Through such an investigation we could confirm that our
formulation developed in KN2003, KN2005 and in this paper is
relevant to physical processes.

%*******************************************************************

Ten years ago, Mukhanov et al. proposed a gauge invariant
treatment of second-order cosmological
perturbations\cite{Mukhanov-Abramo-Brandenberger-1997}.
Their aim in investigating such perturbations was to clarify the
back reaction effect on the expansion of the universe due to the
inhomogeneities of the gravitational field. 
They also proposed an averaging procedure.
In their papers, they also discussed the gauge issue in
second-order general relativistic perturbations. 
In their proposal, the gauge transformation of a second-order
perturbation should be given by an exponential map.
From our understanding of the ``gauge'' in general relativistic
perturbations, which was reviewed in 
\S\ref{sec:General-framework-of-the-gauge-invariant-perturbation-theory},
this proposal corresponds to our gauge transformation with
$\xi_{(2)}=0$.
Moreover, in their treatment, the perturbative expansions of the
metric and the matter fields take the form
\begin{equation}
  \bar{Q} = Q_{0} + \lambda \delta Q,
  \label{eq:Abramo-expansion}
\end{equation}
instead of Eq.~(\ref{eq:Bruni-35}) in this paper, which includes
a term of $O(\lambda^{2})$.
They also discussed the back reaction effect due to the
nonlinear effects of the Einstein equation through the
substitution of the expansion (\ref{eq:Abramo-expansion}) into
the Einstein equation, and the evaluation of the quadratic terms
of $\delta Q$.
Thus, their treatment of the second order-perturbations is quite
different from the formulation developed in this paper.
Thus, we have to regard their treatment of second-order
perturbations, including their arguments concerning the gauge,
as being based on a perturbation scheme that is quite different
from that given in this paper.
Further, the correspondence between their works and ours are
highly non-trivial.
At this time, it is not clear that we should clarify the
correspondence between their works and ours, because our
formulation has not yet been constructed to treat the back
reaction effect.
If the formulations of the back reaction effect or some
averaging procedures are formulated on the basis of our
formulation, it will be worthwhile to compare their works and
ours.

%*******************************************************************

Recently, Noh and Hwang studied second-order cosmological
perturbations\cite{Noh-Hwang-2004} on the basis of the ADM
formulation. 
They investigated various gauge fixing methods, gauge invariance,
and the second-order Einstein equations in a complicated manner.
Contrastingly, in our formulation, all gauge invariant
variables for all fields were prepared before the derivation of
the perturbed Einstein equations.
As shown in KN2005, the Einstein equation is necessarily given
in terms of gauge invariant variables only. 
This is shown without assuming an explicit background spacetime
metric.
Therefore, we do not have to consider the gauge degree of freedom
when we study perturbations of the Einstein equation in both the
cosmological perturbation theory and any other general
relativistic perturbation theory, as shown in KN2005.
In this sense, we can conclude that the formulation developed in
this paper is clearer than the formulation of Noh and
Hwang\cite{Noh-Hwang-2004}.
However, it would be interesting to compare their approach and
ours.

%*******************************************************************

In addition to the above works treating the formulation of
second-order perturbation theory, there is a series of papers by
Matarrese and his
co-workers\cite{Non-Gaussianity-in-CMB,Matarrese-Mollerach-Bruni-1998,Mollerach-Harari-Matarrse-2004}
concerning non-Gaussian behavior generated by second-order
general relativistic perturbations.
They also considered gauge invariant variables, but they
concentrated on only the conserved quantities which correspond to
Bardeen's parameter in the linear-order perturbation theory.
By contast, in this paper, we found gauge invariant
variables for the first- and second-order perturbations of
all quantities.
The second-order perturbative Einstein equations on a
homogeneous isotropic background universe were derived in terms
of gauge invariant variables, without any gauge fixing.
This is the main result of this paper.
Hence, with regard to the gauge issue of second-order
perturbation theory, we regard the formulation of second-order
gauge invariant cosmological perturbations to be completed in
this paper.
Many parts of their works are based on the Poisson gauge
explained above, and the Poisson gauge is a complete gauge
fixing method.
Therefore, it would be interesting to follow their physical
arguments based on the gauge invariant perturbation theory
formulated in this paper.

%****************************************************************

Though we have ignore the first-order vector and tensor modes, it
is, in principle, possible to include these modes.
To do this, the long algebraic calculations are needed.
Other than these long calculations, however, there is no
technical problem to include them.
If we include the first-order vector and tensor modes in our
consideration, all types of mode coupling, i.e., scalar-vector,
scalar-tensor, vector-vector, vector-tensor, and tensor-tensor
mode couplings may occur, in addition to the scalar-scalar mode
coupling discussed in this paper.
These additional mode couplings are included in the source terms
$\Gamma_{0}$, $\Gamma_{i}$, and $\Gamma_{ij}$ in the
second-order Einstein equations, which consist of the quadratic
terms of the first-order perturbations, as in the case of the
scalar-scalar mode coupling studied in
\S\ref{sec:Secnd-order-Einstein-equations}. 
Here, we emphasize that even if we consider the simple situation
in which the first-order vector and tensor modes are 
negligible, the second-order vector and tensor modes may be
generated by the scalar-scalar mode coupling if accidental
cancellations in the quadratic terms of the linear perturbations 
do not occur.

%*******************************************************************

Recently, Tomita has extended his pioneering works to a universe
filled with dust and a cosmological constant, and he also
discussed non-Gaussian behavior in CMB due to the nonlinear 
effect of the gravitational perturbations\cite{Tomita-2005}.
His works are based on the synchronous gauge.
He also claimed that there is no vorticity perturbation in this
system even if we take into account the effects of second-order
perturbations.
Contrastingly, in our analyses, the divergenceless part of the
spatial velocity of the fluid may be generated by the
non-linear effects.
The divergenceless part of the spatial velocity of the fluid
contributes to the vorticity.
Of course, in the model of a universe filled with a single
scalar field, there is no vorticity, because the matter current
of the scalar field is proportional to the gradient of the
scalar field.
In this case, the Einstein equation gives the constraint
equation for the vector-mode metric perturbations.
Because the Einstein equations constitute a first-class
constrained system, the initial value constraint should be
consistent with the evolution equation of the vector-mode metric
perturbations.
In this sense, the initial value constraint
(\ref{eq:kouchan-18.199-3}) for the vector mode should be
consistent with the evolution equation
(\ref{eq:kouchan-18.214}).
We can easily understand the reason for the absence of vorticity
in a universe filled with a single scalar field, but the absence
of vorticity in a universe filled with a dust field is not so
trivial.
Since the vorticity in the early universe is related to the
generation of the magnetic field in the early universe through
the Harrison
mechanism\cite{Matarrese-etal-2005,Takahashi-etal-2005}, and to 
the generation of the B-mode polarization in CMB
anisotropy\cite{Mollerach-Harari-Matarrse-2004}, the existence
of vorticity in the early universe is very important in
cosmology.
Because vorticity perturbations of the fluid velocity related
to vector-mode perturbations, we conclude that the generation
of the vector-mode perturbations is an important issue in the
cosmological context.

%*******************************************************************

In addition to the generation of vector-mode perturbations,
the generation of tensor modes, which corresponds to
gravitational waves, is also interesting for cosmology. 
The upper limit of the amplitude of the vorticity perturbations
and the gravitational wave perturbations is constrained by the
observational data of CMB anisotropy\cite{WMAP}.
However, it is known that the fluctuations of the scalar mode
perturbations do exist in the early universe from the anisotropy
of the CMB.
Hence, the generation of  vector and tensor modes due to the
scalar-scalar mode coupling in second-order perturbations
will give a lower limits on the vorticity and gravitational
waves in the early universe from a theoretical point of view.

%*******************************************************************

From the above discussion, we see that there are many issue
which should be clarified using second-order cosmological
perturbations.
These are quite interesting not only from the theoretical point
of view but also from the observational point of view.
We have to clarify these issues one by one.
To carry this out, the gauge invariant formulation developed in
this paper should provide very powerful theoretical tools, and
we hope that these issues will be clarified in terms of the
gauge invariant variables defined in this paper.
We leave these issues as future works.

%************************************************

%%%%%%%%%%%%%%%%%%%%%%%%%%%%%%%%%%%%%%%%%%%%%%%%%%%%%%%%%%%%%%%%%%%%%%
\section*{Acknowledgements}
%%%%%%%%%%%%%%%%%%%%%%%%%%%%%%%%%%%%%%%%%%%%%%%%%%%%%%%%%%%%%%%%%%%%%%
The author thanks participants in the international workshop
``Black holes, spacetime singularities, and cosmic censorship'', 
which was held at TIFR in India in March, 2006, for valuable
discussions.
In particular, he is grateful to Prof. P.~S.~Joshi and his
colleagues for their hospitality during the workshop.
The author also thanks to Prof. N.~Dadhich and his colleagues
for their hospitality during his visit to IUCAA in India, and
Prof. Tomita for his valuable comments.
The author is deeply grateful to the members of Division of
Theoretical Astronomy at NAOJ and his family for their
continuous encouragement and taking care of his health when he
was writing this paper.

%****************************************************************

%%%%%%%%%%%%%%%%%%%%%%%%%%%%%%%%%%%%%%%%%%%%%%%%%%%%%%%%%%%%%%%%%%%%%%
\appendix
%%%%%%%%%%%%%%%%%%%%%%%%%%%%%%%%%%%%%%%%%%%%%%%%%%%%%%%%%%%%%%%%%%%%%%
\section{The components of $H_{abc}\left[{\cal H}\right]$,
  $H_{a}^{\;\;bc}\left[{\cal H}\right]$,
  $H_{a\;\;c}^{\;\;b}\left[{\cal H}\right]$, and
  $H^{abc}\left[{\cal H}\right]$}
\label{sec:components-of-Habcs}
%%%%%%%%%%%%%%%%%%%%%%%%%%%%%%%%%%%%%%%%%%%%%%%%%%%%%%%%%%%%%%%%%%%%%%

In the derivation of the gauge invariant part of the perturbed
Einstein tensor, the components $H_{abc}\left[{\cal H}\right]$,
$H_{a}^{\;\;bc}\left[{\cal H}\right]$,
$H_{a\;\;c}^{\;\;b}\left[{\cal H}\right]$, and
$H^{abc}\left[{\cal H}\right]$ defined by
Eqs.~(\ref{eq:Habc-def-1}) and (\ref{eq:Habc-def-2}) are
useful.
These components are given below.
\begin{itemize}
\item Components of $H_{abc}\left[{\cal H}\right]$ :
  \begin{eqnarray}
    H_{\eta\eta\eta}\left[{\cal H}\right]
    &=&
    - a^{2} \partial_{\eta}\Phi
    , \\
    H_{i\eta\eta}\left[{\cal H}\right]
    &=&
    - a^{2}
    \left(
      D_{i} \Phi
      + {\cal H} \nu_{i}
    \right)
    , \\
    H_{ij\eta}\left[{\cal H}\right]
    &=&
    a^{2} 
    \left\{
      \left(
        2 {\cal H} \left(\Psi + \Phi\right)
        + \partial_{\eta} \Psi
      \right)
      \gamma_{ij}
      +
      D_{(i} \nu_{j)}
      - \frac{1}{2} \left(
        \partial_{\eta}
        + 2 {\cal H}
      \right) \chi_{ij}
    \right\}
    , \\
    H_{\eta\eta i}\left[{\cal H}\right]
    &=&
    a^{2} \left\{
      D_{i} \Phi
      + \left( \partial_{\eta} + {\cal H} \right) \nu_{i}
    \right\}
    , \\
    H_{j\eta i}\left[{\cal H}\right]
    &=&
    a^{2} \left\{
      - \partial_{\eta} \Psi \gamma_{ij} 
      - D_{[i} \nu_{j]} 
      + \frac{1}{2} \partial_{\eta} \chi_{ij} 
    \right\}
    , \\
    H_{jki}\left[{\cal H}\right]
    &=&
    a^{2}
    \left\{
      D_{i} \Psi \gamma_{kj}
      - 2 \gamma_{i(k} D_{j)} \Psi
      - {\cal H} \gamma_{kj} \nu_{i}
      + D_{(j} \chi_{k)i} 
      - \frac{1}{2} D_{i} \chi_{kj} 
    \right\}.
  \end{eqnarray}
\item Components of $H_{a}^{\;\;bc}\left[{\cal H}\right]$ :
  \begin{eqnarray}
    H_{\eta}^{\;\;\eta\eta}\left[{\cal H}\right]
    &=&
    - \frac{1}{a^{2}} \partial_{\eta}\Phi
    \label{eq:kouchan-10.40}
    , \\
    H_{i}^{\;\;\eta\eta}\left[{\cal H}\right]
    &=&
    - \frac{1}{a^{2}}\left(
      D_{i} \Phi + {\cal H} \nu_{i}
    \right)
    \label{eq:kouchan-10.41}
    , \\
    H_{\eta}^{\;\;i\eta}\left[{\cal H}\right]
    &=&
    \frac{1}{a^{2}} \left\{
      D^{i} \Phi + {\cal H} \nu^{i}
    \right\}
    \label{eq:kouchan-10.42}
    , \\
    H_{i}^{\;\;j\eta}\left[{\cal H}\right]
    &=&
    -
    \frac{1}{a^{2}}
    \left\{
      \left(
        2 {\cal H} \left(\Psi + \Phi\right)
        + \partial_{\eta} \Psi
      \right)
      \gamma_{i}^{\;\;j}
      + \frac{1}{2} \left(D_{i} \nu^{j} + D^{j} \nu_{i}\right) 
    \right.
    \nonumber\\
    && \quad\quad\quad\quad
    \left.
      - \frac{1}{2} \left(
        \partial_{\eta}
        + 2 {\cal H}
      \right) \chi_{i}^{\;\;j}
    \right\}
    \label{eq:kouchan-10.43}
    , \\
    H_{\eta}^{\;\;\eta i}\left[{\cal H}\right]
    &=&
    - \frac{1}{a^{2}}
    \left\{
        D^{i} \Phi
      + \left( \partial_{\eta} + {\cal H} \right) \nu^{i}
    \right\}
    \label{eq:kouchan-10.44}
    , \\
    H_{j}^{\;\;\eta i}\left[{\cal H}\right]
    &=&
    \frac{1}{a^{2}} \left\{
      \partial_{\eta} \Psi \gamma_{j}^{\;\;i} 
      +
      \frac{1}{2} \left(
        D^{i} \nu_{j} - D_{j} \nu^{i} 
      \right)
      - \frac{1}{2} \partial_{\eta} \chi_{j}^{\;\;i} 
    \right\}
    \label{eq:kouchan-10.45}
    , \\
    H_{\eta}^{\;\;ij}\left[{\cal H}\right]
    &=&
    \frac{1}{a^{2}}
    \left\{
      - \partial_{\eta} \Psi \gamma^{ji} 
      + D^{[i} \nu^{j]}
      + \frac{1}{2} \partial_{\eta} \chi^{ji} 
    \right\}
    \label{eq:kouchan-10.46}
    , \\
    H_{j}^{\;\;ki}\left[{\cal H}\right]
    &=&
    \frac{1}{a^{2}} \left\{
      - \gamma^{ik} D_{j} \Psi
      + 2 \gamma^{[k}_{\;\;j} D^{i]} \Psi
      - {\cal H} \gamma^{k}_{\;\;j} \nu^{i}
    \right.
    \nonumber\\
    && \quad\quad\quad\quad
    \left.
      + \frac{1}{2} D_{j} \chi^{ki} 
      + D^{[k} \chi_{j}^{\;\;i]} 
    \right\}
    .
    \label{eq:kouchan-10.47}
  \end{eqnarray}
\item Components of $H_{a\;\;c}^{\;\;b}\left[{\cal H}\right]$ :
  \begin{eqnarray}
    H_{\eta\;\;\eta}^{\;\;\eta}\left[{\cal H}\right]
    &=&
    \partial_{\eta}\Phi
    , \\
    H_{i\;\;\eta}^{\;\;\eta}\left[{\cal H}\right]
    &=&
    D_{i} \Phi + {\cal H} \nu_{i}
    , \\
    H_{\eta\;\;\eta}^{\;\;i}\left[{\cal H}\right]
    &=&
    - D^{i} \Phi - {\cal H} \nu^{i}
    , \\
    H_{i\;\;\eta}^{\;\;j}\left[{\cal H}\right]
    &=&
    \frac{1}{2} \left(D_{i} \nu^{j} + D^{j} \nu_{i}\right) 
    + 
    \left(
      2 {\cal H} \left(\Psi + \Phi\right)
      + \partial_{\eta} \Psi
    \right)
    \gamma_{i}^{\;\;j}
    \nonumber\\
    && \quad
    - \frac{1}{2} \left(
      \partial_{\eta}
      + 2 {\cal H}
    \right) \chi_{i}^{\;\;j}
    , \\
    H_{\eta\;\;i}^{\;\;\eta}\left[{\cal H}\right]
    &=&
    - D_{i} \Phi
    - \left( \partial_{\eta} + {\cal H} \right) \nu_{i}
    , \\
    H_{j\;\;i}^{\;\;\eta}\left[{\cal H}\right]
    &=&
    \partial_{\eta} \Psi \gamma_{ji}
    + D_{[i} \nu_{j]}
    - \frac{1}{2} \partial_{\eta} \chi_{ji}
    , \\
    H_{\eta\;\;j}^{\;\;i}\left[{\cal H}\right]
    &=&
    - \partial_{\eta} \Psi \gamma_{j}^{\;\;i} 
    + \frac{1}{2} \left( D^{i} \nu_{j} - D_{j} \nu^{i} \right)
    + \frac{1}{2} \partial_{\eta} \chi_{j}^{\;\;i} 
    , \\
    H_{j\;\;i}^{\;\;k}\left[{\cal H}\right]
    &=&
    - \gamma_{ij} D^{k} \Psi
    + 2 \gamma^{k}_{\;\;[j} D_{i]} \Psi
    - {\cal H} \gamma^{k}_{\;\;j} \nu_{i}
    + \frac{1}{2} D^{k} \chi_{ji}
    + D_{[j} \chi^{k}_{\;\;i]}  
    .
  \end{eqnarray}
\item Components of $H^{abc}\left[{\cal H}\right]$ :
  \begin{eqnarray}
    H^{\eta\eta\eta}\left[{\cal H}\right]
    &=&
    \frac{1}{a^{4}} \partial_{\eta}\Phi
    , \\
    H^{i\eta\eta}\left[{\cal H}\right]
    &=&
    -
    \frac{1}{a^{4}}\left(
      D^{i} \Phi + {\cal H} \nu^{i}
    \right)
    , \\
    H^{ij\eta}\left[{\cal H}\right]
    &=&
    -
    \frac{1}{a^{4}}
    \left\{
      \left(
        2 {\cal H} \left(\Psi + \Phi\right)
        + \partial_{\eta} \Psi
      \right)
      \gamma^{ij}
      + D^{(i} \nu^{j)}
    \right.
    \nonumber\\
    && \quad\quad\quad\quad
    \left.
      - \frac{1}{2} \left(
        \partial_{\eta}
        + 2 {\cal H}
      \right) \chi^{ij}
    \right\}
    , \\
    H^{\eta\eta i}\left[{\cal H}\right]
    &=&
    \frac{1}{a^{4}}
    \left\{
      D^{i} \Phi
      + \left( \partial_{\eta} + {\cal H} \right) \nu^{i}
    \right\}
    , \\
    H^{i\eta j}\left[{\cal H}\right]
    &=&
    \frac{1}{a^{4}} \left\{
        \partial_{\eta} \Psi \gamma^{ij} 
      - D^{[i} \nu^{j]}
      - \frac{1}{2} \partial_{\eta} \chi^{ij} 
    \right\}
    , \\
    H^{jki}\left[{\cal H}\right]
    &=&
    \frac{1}{a^{4}} \left\{
      - \gamma^{ik} D^{j} \Psi
      + 2 \gamma^{j[k} D^{i]} \Psi
      - {\cal H} \gamma^{jk} \nu^{i}
    \right.
    \nonumber\\
    && \quad\quad\quad\quad
    \left.
      + D^{(j} \chi^{k)i} 
      - \frac{1}{2} D^{i} \chi^{jk}
    \right\}.
  \end{eqnarray}
\end{itemize}

%*********************************************************************

%%%%%%%%%%%%%%%%%%%%%%%%%%%%%%%%%%%%%%%%%%%%%%%%%%%%%%%%%%%%%
%%%%%%%%%%%%%%%%%%%%%%%%%%%%%%%%%%%%%%%%%%%%%%%%%%%%%%%%%%%%%

\end{document}